\documentclass[%
reprint,
 amsmath,amssymb,
rmp,
]{revtex4-2}

\usepackage{graphicx}
\usepackage{dcolumn}
\usepackage{bm}
\usepackage{hyperref}
\usepackage{xcolor} 
\usepackage{braket}
\usepackage{siunitx}
\usepackage{amsmath} 
\usepackage[english]{babel}

\newcommand{\cc}{\text{c.c.}}

\newcommand{\Isat}{I_{\text{sat}}}

\begin{document}

\title{Paraxial fluids of light}

\author{Quentin Glorieux}%
 \email{quentin.glorieux@lkb.upmc.fr}
  \author{Clara Piekarski}
  \author{Quentin Schibler}
  \author{Tangui Aladjidi}
  \author{Myrann Baker-Rasooli} 
 
\affiliation{%
 Laboratoire Kastler Brossel, Sorbonne Université, CNRS, ENS-PSL, Collège de France. 4 Place Jussieu, Paris, France.
}%

\date{\today}

\begin{abstract}
Paraxial fluids of light are a promising platform for exploring collective phenomena in a highly tunable environment.
These systems, which map the propagation of light through nonlinear media onto the wavefunction of effective 2D quantum fluids, offer a complementary approach to traditional platforms such as cold atomic gases or superfluid helium.
In this review, we present a detailed overview of the theoretical framework underlying paraxial fluids of light, including the nonlinear Schrödinger equation (NLSE) and its mapping to the 2D+1 Gross-Pitaevskii equation (GPE).
We explore the hydrodynamic formulation of these systems and we provide a comparative analysis of fluids of light and cold atomic gases, examining key parameters and figures of merit.

We then review the recent experimental advances and the experimental platforms currently used to realize paraxial fluids of light, including hot atomic vapors, photorefractive crystals, and thermo-optic media. Additionally, we question the geometry of the system extending the analogy from 2D+1 to lower or higher dimensions.

Looking forward, we outline the potential future directions for the field, including the use of laser cooled atoms as nonlinear media, the study of two-component mixtures, and the exploration of quantum effects beyond the mean-field approximation. These developments promise to deepen our understanding of quantum fluids and potentially contribute to advances in quantum technologies.
\end{abstract}

\maketitle
\tableofcontents

\section{Introduction and historical overview}
The Nonlinear Schrödinger Equation (NLSE) is one of the most broadly applicable nonlinear models in physics \citep{fibich2015nonlinear}. 
Its relevance extends across numerous fields, such as condensed matter and ultracold atoms \citep{dalfovo1999theory, pethick2008bose}, plasma physics \citep{zakharov1972collapse}, nonlinear optics \citep{hasegawa1973transmission,ablowitz1991solitons,kivshar2003optical}, laser physics \citep{lugiato1987spatial}, fluid mechanics \citep{whitham2011linear}, turbulence \citep{zakharov2012kolmogorov}, phase transitions \citep{cross1993pattern}, biophysics \citep{davydov1985solitons}, and even astrophysics \citep{shukla2007nonlinear}. 
In all of these areas, the NLSE serves as a ``universal" framework, capturing the nonlinear processes in order to understand the complex behavior that arises in these diverse physical systems.

In particular, in the context of optics, the NLSE describes the evolution of the complex envelope of an optical field traveling through a Kerr nonlinear medium. 
This equation describes different configurations such as the propagation through optical fibers \citep{hasegawa1973transmission} (where pulses propagate along a single spatial axis and evolve in time), and the transverse evolution during propagation through bulk nonlinear media \citep{carusotto2014superfluid}.

In each case, the governing equation is virtually the same as the one used for describing the temporal evolution of quantum gases such as dilute Bose–Einstein condensates (BECs), leading to the name of ``quantum fluids of light" \citep{gross1963hydrodynamics,pitaevskii1961vortex}.
This one-to-one correspondence means that optical experiments can serve as highly controllable platforms for studying phenomena usually associated with ultracold atomic quantum gases or superfluid helium, such as vortex formation, superfluidity, quantum turbulence, and quantum many-body physics. 
On the other hand, quantum gases could serve as an inspiration to imagine new nonlinear or quantum optics experiments, leading to novel photonic applications. 
In this way, paraxial fluids of light not only broaden the application spectrum of the NLSE but also establish a strong connection between nonlinear optics, quantum optics and BEC physics, allowing to share ideas and techniques between the fields.

\subsection{Early developments: NLSE in nonlinear optics}
The importance of the NLSE in optics has been recognized since the early development of laser physics and the advent of nonlinear optics.
\cite{chiao1964self} and \cite{talanov1965focusing,talanov1970focusing} independently derived an envelope equation describing how an optical beam could self-focus in a Kerr medium (with a cubic nonlinearity). 
While the term ``nonlinear Schrödinger equation” was not explicitly used at the time, the equation obtained by \cite{chiao1964self} is mathematically what we recognize today as the (2D+1) focusing NLSE in the paraxial limit and is the main topic of this review.

In the years immediately following, \cite{hasegawa1973transmission,hasegawa1973transmission2} in the context of optical solitons in fibers, \cite{wagner1968large} for paraxial propagation and \cite{shabat1972exact} for integrable systems, adopted and popularized the NLSE terminology. 
By the early 1980s, references to the ``nonlinear Schrödinger equation” in nonlinear optics became standard, especially after that \cite{mollenauer1980experimental} observed experimentally solitons in optical fibers.

\begin{figure}[h]
    \centering
    \includegraphics[width=0.85\linewidth]{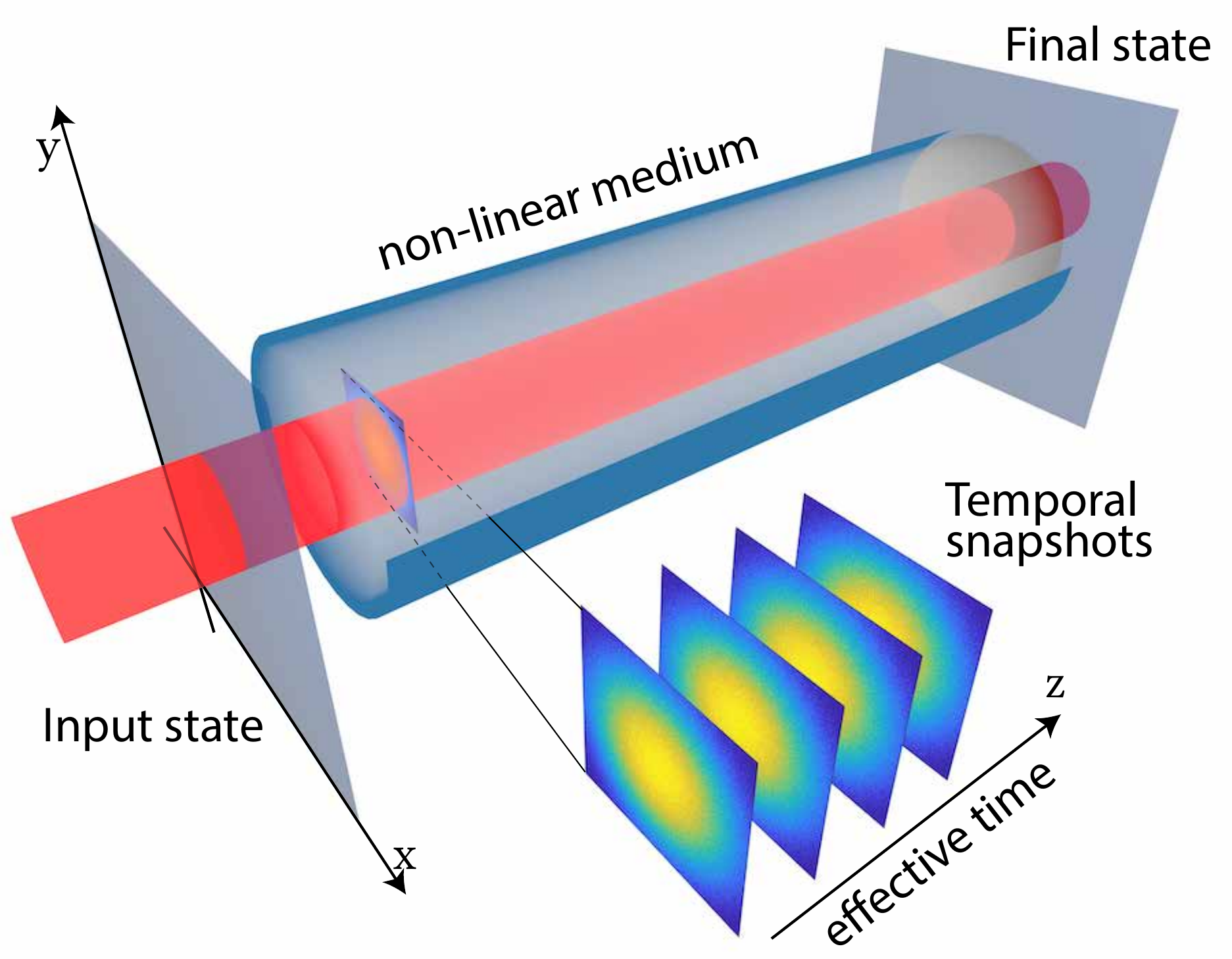}
    \caption{Sketch of the 2D+1 paraxial fluid of light. A laser propagates along $z$ in a nonlinear medium. Each transverse plane $(x,y)$ is equivalent to a temporal snapshot. Input state is user-defined and the final state is measured experimentally.}
    \label{fig:photon-fluid}
\end{figure}

Two distinct regimes of the NLSE phenomenology were rapidly recognized: the focusing (attractive nonlinearity) and defocusing (repulsive nonlinearity) cases. 
In the focusing regime, nonlinearity counteracts dispersion or diffraction, enabling the formation of localized, stable structures known as bright solitons. 
In the defocusing regime, they work together to support dark solitons, characterized by intensity dips in a continuous background. 
The dimensionality of the system further divides the NLSE phenomenology into two major categories, each tied to distinct experimental setups, theoretical considerations, and scientific communities.
In a one-dimensional configuration (often referred to as 1D+1), such as in optical fibers, the NLSE governs the evolution of the wave envelope along the propagation direction $z$ playing the role of time, while the physical time $t$ is analogous to a spatial coordinate. 
Temporal dispersion introduced an effective mass (due to the group velocity dispersion) and interacts with nonlinearity.
This (1D+1) framework has been extensively studied due to its direct relevance to optical communication systems \citep{sulem2007nonlinear}.
In contrast, a two-dimensional (in space) plus one propagation direction (2D+1) applies to paraxial beams propagation in bulk media.
Here, the NLSE describes how the wave envelope evolves in the transverse plane $(x, y)$ as it travels along $z$ as shown in Fig. \ref{fig:photon-fluid}.
Instead of dispersion, diffraction serves as the primary linear effect at the origin of the effective mass, balanced by focusing or defocusing nonlinearities.

\noindent These distinctions (focusing versus defocusing, and (1D+1) versus (2D+1)) have significant implications for both fundamental research and practical applications.
On the one hand, the (1D+1) configuration has been primarily associated with soliton dynamics, particularly in optical fibers, and has paved the way for soliton-based optical communication technologies.
On the other hand, the paraxial geometry has been closely linked to classical hydrodynamics \citep{MATTAR19811} and to  quantum gases and quantum hydrodynamics  \citep{pomeau1993diffraction}, leading to the term ``quantum fluids of light” \citep{carusotto2013quantum}.
This connection has triggered the development of a broad research field dedicated to exploring fundamental phenomena such as superfluidity, wave condensation, topology, turbulence, and, more generally, quantum simulation using photonic systems, which we will explore in this review.

\subsection{Cavity systems: the emergence of fluids of light}
Before going forward on the review of propagating systems, we mention a broad class of experimental platforms that study fluids of light in confined geometries.
A foundational step in the study of confined photonic system was provided by the pioneering theoretical work of \cite{lugiato1987spatial}. 
They introduced the so-called Lugiato-Lefever equation, which describes the spatiotemporal dynamics of light in a driven-dissipative Kerr cavity. 
This equation established one of the first bridges between laser physics and hydrodynamics, and these cavity-based systems have historically supported the emergence of the concept of fluids of light.

Actually, one of the decisive conceptual advances in this direction came from a pioneer of the NLSE mapping, Raymond Chiao \citep{chiao1964self}. 
In 1999, \cite{chiao1999bogoliubov} proposed the possibility of superfluid-like behavior in light using the formal analogy between quantum gases and the evolution of an electromagnetic field within a nonlinear cavity.
Drawing analogies between nonlinear optics and the hydrodynamics of superfluids, \cite{chiao2000bogoliubov} suggested that light propagating in nonlinear media could exhibit a Bogoliubov dispersion relation and proposed several experimental implementations using hot rubidium vapors or Rydberg atoms in a microwave cavity \citep{chiao2004effective}.
While these proposals initially faced experimental challenges, they opened the way to modern fluids of light research and to the best of our knowledge they invented the expression ``fluids of light".

The next significant breakthrough came with the advent of exciton polariton systems.
Exciton-polaritons are hybrid quasiparticles formed from the strong coupling of photons with excitations in semiconductor microcavities. 
These systems inherently possess a driven-dissipative nature, where external pumping and losses play an essential role in their dynamics. 
Despite this complexity, experimental observations of Bose-Einstein condensation (BEC) \citep{deng2002condensation,kasprzak2006bose} and superfluidity in exciton-polariton systems \citep{amo2009superfluidity,stepanov2019dispersion, claude2022high} played an important role in the study of quantum fluids of light \citep{amo2009collective}. 
Polaritons demonstrated not only superfluid properties but also the ability to support exotic nonlinear phenomena, such as solitons \citep{amo2011polariton,maitre2020dark,claude2020taming,lerario2020parallel}, vortices \citep{lagoudakis2008quantized,boulier2015vortex,boulier2016injection,koniakhin2019stationary}, and more recently, topological excitations \citep{st2017lasing,klembt2018exciton,solnyshkov2021microcavity} and KPZ universality class \citep{fontaine2022kardar}. 
These advancements highlighted the richness of driven-dissipative systems as an experimental platform for exploring quantum hydrodynamics.

Simultaneously, a parallel line of research emerged in dye-filled microcavities, where  Bose-Einstein condensation of photons was demonstrated \citep{klaers2010bose}. 
Unlike exciton-polariton systems, where the quantum fluid arises from the hybrid nature of polaritons, photon BEC relies on thermalization of photons via their interaction with a dye medium \citep{klaers2010thermalization}. 
This platform has since been extended to explore the collective dynamics of photon fluids, including the emergence of phase coherence, thermalization, and beyond-mean-field effects in photonic condensates \citep{klaers2012statistical,busley2022compressibility,karkihalli2024dimensional}.

\subsection{Removing the cavity: the paraxial fluids of light}

Paraxial fluids of light build on this historical background to form a distinctive branch in the study of quantum fluids of light.
Unlike the driven-dissipative dynamics of exciton-polariton condensates or photon BECs in cavities, paraxial systems evolve as nearly closed systems, remaining effectively conservative if linear losses (absorption) are negligible.
Therefore, one key advantage of paraxial systems lies in their ability to replicate quantum hydrodynamic phenomena without the complexities of cavity-based dissipation and external driving fields, allowing direct exploration of fundamental quantum fluid effects.

Early experiments in this field explored soliton instabilities \citep{swartzlander1992optical,tikhonenko1996observation}, vortex nucleation \citep{tikhonenko1996excitation,tikhonenko1995spiraling}, and pattern formation \citep{petrossian1992transverse}.
Soon after the vocabulary of fluid-like motion was used by \cite{PhysRevLett.79.3399} to describe the rotation of two same sign vortices in free space.
Even though this work studies the propagation in a linear medium, the authors emphasized that both fluid flow and diffraction of light may be described using potential theory, and therefore they expected similar phenomena to occur in both systems.

Thereafter, one of the pioneering groups in the study of paraxial fluids of light, J. Fleischer and colleagues at Princeton, have made significant experimental contributions using photorefractive crystals.
In an early work, \cite{fleischer2003observation} demonstrated the formation of optical patterns and nonlinear self-organization in a fluid of light. 
In particular, they explored how spatial solitons interact to create complex structures such as optical lattices and localized wave packets, effectively mimicking fluid-like phenomena and bridging the gap between nonlinear optics and fluid dynamics.
Following this approach of classical nonlinear dynamics, they obtained pioneered results on dispersive shock waves \citep{wan2007dispersive}.
An intense theoretical activity has followed, targeting in particular cubic-quintic nonlinearity, where a fifth order nonlinear term is present to counter balance the third order one \citep{paz2005superfluidlike,novoa2009pressure,kozyreff2010capillary}. 
These works are widely inspired by a seminal theoretical work by \cite{josserand1995cavitation} on vortex nucleation in a superfluid model.

Interest in paraxial fluids of light has steadily grown, driven in part by the theoretical work of \cite{larre2015propagation}, which advanced the analogy between atomic ultracold gases and fluids of light, particularly through the derivation of a generalized quantum theory of paraxial light propagation.
As pioneered by \cite{lai1989quantuma,lai1989quantum} for optical fibers, \cite{larre2015propagation} proposed a theoretical approach with a quantized electromagnetic field that is able to describe in its full generality the dynamics of interacting photons propagating in the paraxial approximation. 
Interestingly this formalism allows for many-body phenomena to be investigated with the reconstruction of the quantum state of light and its statistics after propagation using typical quantum-optics techniques.
While this quantum-optics approach for paraxial fluids of light is still missing for clear experimental demonstrations and therefore is not at the core of this review, this direction is truly promising and will be described in Section~\ref{sec:quantum}.

In the meantime, we give a general overview of the paraxial fluids of light formalism in Section~\ref{sec:framework}.
In particular, we explicit the link between fluids of light and ultracold atomic quantum gases which has been very fruitful in the last decade.
The question of the dimensionality of the system is discussed in Section~\ref{sec:dimension}.
In Section~\ref{comparison}, we precise the analogy with a dictionary of terms between the two fields, to help bridging the gap between them. 
Moreover, we will compare typical experimental parameters and experimental techniques to see where fluids of light could bring novel opportunities to study quantum gases physics.
For fluids of light, several experimental platforms have emerged: atomic clouds, photorefractive crystals and thermo-optic liquids and in Section~\ref{sec:platforms}, we compare these systems with common figures of merits to evidence the advantages of each platforms.
We describe specific experimental and numerical techniques in Section~\ref{sec:techniques}.
In Section~\ref{sec:expt}, we review the recent experiments in the field, following three directions: quantum hydrodynamics (solitons and vortices), out-of-equilibrium physics and superfluidity.
We  conclude this review with our vision on the future directions of the field in Section~\ref{sec:future}.

\section{Detailed theoretical framework}
\label{sec:framework}

To describe the phenomena involving fluids of light, we first establish the formal description of the system, and then link the physics of light propagating in a nonlinear medium with the physics of cold gases.

\subsection{Nonlinear optics and the $\chi^{(3)}$ nonlinearity}
\label{sec:nlse_derivation}
The propagation of light in a nonlinear medium is described by  an expression of the medium electric polarization $\mathbf{P}$ as a function of the incoming electric field $\mathbf{E}$.
From the Maxwell equations, we solve the propagation of an electromagnetic field $\mathbf{E}$ in a medium as follows \citep{boyd_nl_optics}:
\begin{equation}
    \nabla^2\mathbf{E}-\frac{1}{c^2}\frac{\partial^2\mathbf{E}}{\partial t^2}=\frac{1}{\varepsilon_0 c^2}\frac{\partial^2\mathbf{P}}{\partial t^2},
    \label{eq:general_propag}
\end{equation}
where $c$ is the speed of light in vacuum, $\varepsilon_0$ is the vacuum electric permittivity and $\mathbf{P}$ is the electric polarization in the medium.
The electric susceptibility $\chi$ links these two quantities as $\mathbf{P}(t)=\varepsilon_0\chi\mathbf{E}(t)$.
In the case of nonlinear media, the susceptibility also depends on the electric field.
The effect of the susceptibility is thus described as an expansion in powers of the electric field: 
\begin{equation}
\mathbf{P}=\varepsilon_0[\hat{\chi}^{(1)} . \mathbf{E}+\hat{\chi}^{(2)}:\mathbf{E}\mathbf{E}+\hat{\chi}^{(3)} \vdots  \mathbf{E}\mathbf{E}\mathbf{E}+\dots]
\end{equation}
where $\hat{\chi}^{(n)}$ is the $n$-th order susceptibility, a tensor of rank $n+1$. 
The corresponding tensor products are indicated by the dots notation $.,~:,~\vdots$. 
This expansion can equivalently be written as:
\begin{equation} \label{eq:pola}
    P_i = \varepsilon_0 \chi^{(1)}_{ij}E_j + \varepsilon_0 \chi^{(2)}_{ijk}E_jE_k + \varepsilon_0 \chi^{(3)}_{ijkl}E_jE_kE_l \dots
\end{equation}
where the indices $(i,j,k,l,\dots)$ run over the cartesian coordinates $(x,y,z)$ and a repetition of indices implies summation using the Einstein summation convention.

Since $\mathbf{P}(r) = -\mathbf{P}(-r)$ in a medium with central symmetry, all terms with even powers vanish. Terms of order higher than 3 are neglected. Assuming an isotropic medium and a linearly or circularly polarized light, Eq.~\eqref{eq:pola} becomes a scalar equation \citep{boyd_nl_optics}, meaning that the electric polarization is aligned with the field's polarization:
\begin{align}
    P(t) = \varepsilon_0 \chi^{(1)} E(t) +
     \varepsilon_0 \chi^{(3)} E^3(t).
\end{align}
\noindent The electric field is assumed to be monochromatic:
\begin{equation}
\label{eq:e}
E(t) = \frac{1}{2} E_0 \text{e}^{-\mathrm{i}\omega t} + \mathrm{c.c. }, 
\end{equation}
\noindent  and can be expanded to:
\begin{equation}
    E^3(t) = \frac{1}{8}\left(3E_0 |E_0|^2 e^{-\mathrm{i}\omega t} + E_0^3e^{-3\mathrm{i}\omega t} + \mathrm{c.c.}\right).
\end{equation}
Since the cubic term only contains frequencies at $\omega$ and $3\omega$, the polarization is written:
\begin{equation}
    P(t) = \frac{P_0}{2}e^{-\mathrm{i}\omega t} + \frac{P_1}{2}e^{-3\mathrm{i}\omega t} + \mathrm{c.c.}
\end{equation}
Following \cite{boyd_nl_optics}, the assumption that the susceptibility response is instantaneous is dropped and the different susceptibility orders are defined as:
\begin{align}
    P_0 &= \varepsilon_0 \chi^{(1)}(\omega)E_0 + \frac{3}{4}\varepsilon_0\chi^{(3)}(\omega)|E_0|^2E_0, \\
    P_1 &= \frac{1}{4}\varepsilon_0 \chi^{(3)}(3\omega)E_0^3.
\end{align}
The terms rotating at $3\omega$ are neglected as $\chi^{(3)}(3\omega) \ll \chi^{(3)}(\omega)$ for near resonant excitations, giving the final expression for the polarization:
\begin{equation}
    P(t) = \varepsilon_0 \left(\frac{1}{2}\chi^{(1)}(\omega) + \frac{3}{8}\chi^{(3)}(\omega)|E_0|^2\right)E_0e^{-\mathrm{i}\omega t} + \text{c.c.} ,
\end{equation}
which enables us to define the total susceptibility as:
\begin{equation}
\label{eq:chi}
    \chi = \chi^{(1)} + \frac{3}{4} \chi^{(3)} |E_0|^2.
\end{equation}

We then aim to derive an expression for the refractive indices. Experimentally, it is possible to modify the refractive index locally as explained in Section~\ref{sec:platforms}. This local change is modeled by writing the susceptibility as $\chi(\mathbf{r}) = \chi^{(1)}(\mathbf{r}) + \frac{3}{4} \chi^{(3)} |E_0|^2 = \bar\chi^{(1)} + \delta \chi^{(1)}(\mathbf{r}) + \frac{3}{4} \chi^{(3)} |E_0|^2$,
where $\bar\chi^{(1)}$ denotes the mean value of $\chi^{(1)}(\mathbf{r})$.
We account for this change in the refractive index by writing
$n(\mathbf{r}) = n_0 + \delta n_0(\mathbf{r}) + n_2^E |E_0|^2$
and identifying terms by comparing with
$n(\mathbf{r}) = \sqrt{1 + \chi(\mathbf{r})}$.

Note that the nonlinear refractive index is noted with a superscript E, indicating that it is defined in terms of the electric field's envelope. 
An alternative and commonly used convention expresses it in terms of intensity: $n = n_0 + \delta n_0(\mathbf{r}) + n_2^I I$. 
Since intensity is more readily measurable, the symbol $n_2$ without a superscript will refer to $n_2^I$ throughout this work.\\
As the susceptibility (in this case) is several orders of magnitude smaller than unity, the square root expands to:
\begin{align}
    n &= \sqrt{1 + \bar\chi^{(1)} + \delta \chi^{(1)}(\mathbf{r}) + \frac{3}{4}\chi^{(3)}|E_0|^2} \\
    &= \sqrt{1 + \bar\chi^{(1)}}\sqrt{1 + \frac{\delta \chi^{(1)}(\mathbf{r})}{1 + \bar\chi^{(1)}} + \frac{3}{4}\frac{\chi^{(3)}}{1 + \bar\chi^{(1)}}|E_0|^2} \\
    &\simeq n_0\left(1 + \frac{\delta \chi^{(1)}(\mathbf{r})}{2n_0^2} + \frac{3}{8}\frac{\chi^{(3)}}{n_0^2} |E_0|^2\right) \\
    &= n_0 + \delta n_0(\mathbf{r)} + n_2^E|E_0|^2.
\end{align}
We use the definition of the intensity $I=\frac{1}{2}\varepsilon_0 n_0 c |E_0|^2$ to find the expressions of $n_0$, $\delta n_0(\mathbf{r})$, $n_2^E$and $n_2^I$: 
\begin{align} \label{eq:refractive_indices}
    n_0 &= \sqrt{1 + \bar\chi^{(1)}} = n_0' + i n_0'', \\
    \delta n_0(\mathbf{r}) &= \frac{\delta\chi^{(1)}(\mathbf{r})}{2n_0} , \\
        n_2^E &= \frac{1}{2}\varepsilon_0n_0c\ n_2^I = \frac{3}{8}\frac{\chi^{(3)}}{n_0},\\
    n_2 & \equiv n_2^I = \frac{3\chi^{(3)}}{4\varepsilon_0 n_0^2 c}.
\end{align}

\begin{figure}
    \centering
    \includegraphics[width=0.9\linewidth]{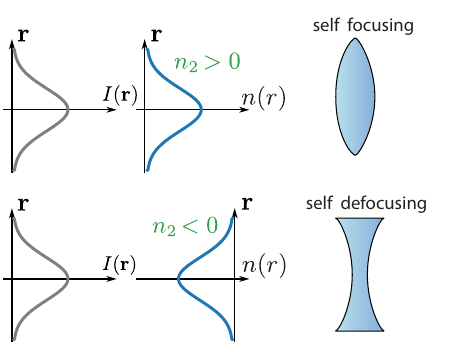}
    \caption{Self-focusing and self-defocusing nonlinearity. Adapted from \cite{aladjidi2023full}.}
    \label{fig:self}
\end{figure}

\noindent The nonlinear index $n_2$ is known as self-Kerr nonlinearity and is at the origin of self-defocusing or self-focusing depending on the sign of $\chi^{(3)}$. 
The intensity of a Gaussian beam is higher at its center than at its edge. 
Then, if $n_2$ is positive (resp. negative), the center of the beam will experience a higher (resp. lower) refractive index than the edge, similarly as in a converging (resp. diverging) lens.
The beam will then self-focus (resp. self-defocus) as shown in Fig. \ref{fig:self}.

\subsection{Nonlinear Schrödinger equation in 2D+1}
\label{sec:NLSE}
We consider a perfectly monochromatic field $E$ written using the convention of Eq.~\eqref{eq:e} propagating in a Kerr medium of susceptibility defined in Eq.~\eqref{eq:chi}.
Injecting this field in Eq.~\eqref{eq:general_propag} yields the propagation equation for the field envelope $E_0$:
\begin{equation}
    \nabla^2 E_0+\dfrac{\omega^2}{c^2}\left(1 + \chi^{(1)}(\mathbf{r})\right)E_0=-\frac{3}{4}\dfrac{\omega^2}{c^2} \chi^{(3)}|E_0|^2 E_0 .
    \label{eq:chi3_propag}
 \end{equation}
Note the $\frac{3}{4}$ prefactor in front of $\chi^{(3)}$ which follows from the definition of $E_0$ in Eq.\eqref{eq:e}.
This is the most commonly used convention, although others are found in the literature \citep{grynberg2010introduction}.
Some authors define $E_0$ as $E = E_0 \text{e}^{-\mathrm{i}\omega t} + \cc$, which leads to a factor of $3$ instead of $\frac{3}{4}$.\\

The field is decomposed as $E_0 = \mathcal{E} \text{e}^{\mathrm{i} k_0 z}$ and the slowly varying envelope approximation is applied, where the envelope $\mathcal{E}$ varies slowly along the propagation axis $z$ compared to the carrier wavelength $k_0 = n_0'\dfrac{\omega}{c}$. 
This approximation implies that:
\begin{align}
    \nabla^2 \mathcal{E} &= \left(\nabla_\perp^2 \mathcal{E} + \frac{\partial^2\mathcal{E}}{\partial z^2}- k_0^2\mathcal{E}+2\mathrm{i}k_0\dfrac{\partial \mathcal{E}}{\partial z} \right) e^{\mathrm{i}k_0z} \\
    & \simeq \left(\nabla_\perp^2 \mathcal{E} - k_0^2\mathcal{E}+2\mathrm{i}k_0\dfrac{\partial \mathcal{E}}{\partial z} \right) e^{\mathrm{i}k_0z}.
\end{align}

Note that this approximation also implies that the field is paraxial, as the envelop can't change significantly over a wavelength, the wavefront must be flat enough to keep the beam near the optical axis. 
With these approximations, we decompose the Laplacian of Eq.~\eqref{eq:chi3_propag} as follows:
\begin{align}
    \nabla^2 \mathcal{E} &= \left(\nabla_\perp^2 \mathcal{E} + \frac{\partial^2\mathcal{E}}{\partial z^2}- k_0^2\mathcal{E}+2\mathrm{i}k_0\dfrac{\partial \mathcal{E}}{\partial z} \right) e^{\mathrm{i}k_0z} \\
    & \simeq \left(\nabla_\perp^2 \mathcal{E} - k_0^2\mathcal{E}+2\mathrm{i}k_0\dfrac{\partial \mathcal{E}}{\partial z} \right) e^{\mathrm{i}k_0z}.
\end{align}
Injecting this expression into Eq.~\eqref{eq:chi3_propag} leads to the paraxial wave equation:
\begin{equation} \label{eq:nlse_paraxial_in_progress}
    \nabla_\perp^2 \mathcal{E} + 2\text{i}k_0\dfrac{\partial \mathcal{E}}{\partial z} - k_0^2\mathcal{E}+ \frac{\omega^2}{c^2}(1 + \chi^{(1)}(\mathbf{r}))\mathcal{E} = -3 k^2 \chi^{(3)}|\mathcal{E}|^2\mathcal{E}.
\end{equation}
The refractive index is split $\sqrt{1 + \chi^{(1)}}$ into real and imaginary parts since they lead to different effects, and we rewrite the linear susceptibility as:
\begin{equation}
    \frac{\omega^2}{c^2}\left(1 + \chi^{(1)}\right) = (k_0 + \mathrm{i}\frac{\alpha}{2})^2 = k_0^2 + \mathrm{i}k_0 \alpha - \frac{\alpha^2}{4} \simeq k_0^2 + \mathrm{i}k_0\alpha,
\end{equation}
where $\alpha= \frac{2\omega n_0''}{c}$ is the linear  loss coefficient for the intensity, hence the factor $\frac{1}{2}$ in front of the electric field. In the following, it is assumed that the medium is transparent, so $n_0'' \ll n_0' \implies \alpha \ll k_0$.
We neglect the term $\alpha^2$ since $\alpha$ is much smaller than $k_0$, but we keep the term $ \mathrm{i}k_0 \alpha$ since $k_0^2$ will cancel out in the propagation equation.
We then find:

\begin{equation} \label{eq:nlse_paraxial_in_progress2}
    \nabla_\perp^2 \mathcal{E} + 2\mathrm{i}k_0\dfrac{\partial \mathcal{E}}{\partial z} + \mathrm{i}k_0\alpha\mathcal{E} + \frac{\omega^2}{c^2}\delta\chi^{(1)}(\mathbf{r})\mathcal{E} = -\frac34 \frac{\omega^2}{c^2} \chi^{(3)}|\mathcal{E}|^2\mathcal{E}.
\end{equation}

\noindent A word of caution here: one should be careful when comparing the coefficients to those obtained in other works. Choosing the convention $E = \mathcal{E} \text{e}^{\mathrm{i} \omega t} + \text{c.c.}$ eliminates the $\frac{1}{4}$ coefficient in front of the nonlinearity. Moreover, adopting a convention for the susceptibility that absorbs factors arising from powers of $E$ causes the factor $3$ in front of the nonlinearity to disappear.

\noindent The susceptibilities are replaced with their respective refractive index expressions of Eq.~\eqref{eq:refractive_indices} and the 2D+1 nonlinear Schrödinger equation \noindent Eq.~\eqref{eq:NLSE_2D} is obtained:
\begin{equation}
   { \mathrm{i}\dfrac{\partial \mathcal{E}}{\partial z} = - \frac{1}{2 k_0}\nabla_\perp^2 \mathcal{E} - \mathrm{i}\frac{\alpha}{2} \mathcal{E} - k_0\frac{\delta n_0(\mathbf{r})}{n_0} \mathcal{E}-k_0\frac{n_2^E}{n_0}|\mathcal{E}|^2\mathcal{E}.}
    \label{eq:NLSE_2D}
\end{equation}
In general the potential term $\frac{\delta n_0(\mathbf{r)}}{n_0}$ and the nonlinear term $\frac{n_2^E}{n_0}$ are complex as the refractive index $n_0$ is complex.
However these terms can be approximated real if: $\Im (k_0 \frac{n_2}{n_0}I) \ll \alpha/2$ and $\Im (k_0\frac{\delta n_0(\mathbf{r})}{n_0}) \ll \alpha/2$.
The imaginary part of the nonlinear term can be rewritten as:
\begin{equation}
    \begin{aligned}
   \Im (k_0 \frac{n_2}{n_0}I) &= \Im \left(k_0\frac{n_2 I}{n_0' + \mathrm{i}n_0''}\right) = -k_0\frac{n_2 I n_0''}{n_0'^2 + n_0''^2} \\
   &\simeq -k_0\frac{n_2 In_0''}{n_0'^2} = -n_2I\frac{\omega}{c}\frac{n_0''}{n_0'}.   
    \end{aligned}
\end{equation}
where the transparent medium approximation has been made, i.e., \( n_0'' \ll n_0' \). From this, we deduce that, in order to safely neglect the imaginary part of the nonlinear term, this condition must hold:
$n_2 I \, \frac{\omega}{c} \, \frac{n_0''}{n_0'} \ll \frac{\alpha}{2} = \frac{\omega}{c} n_0''$  and therefore $ n_2 I \ll n_0'$.
Similarly, the potential term can be considered real if the spatial index variation satisfies \( \delta n_0(\mathbf{r}) \ll n_0' \).

\subsection{Mapping the Gross-Pitaevskii and the nonlinear Schrödinger equations}
\label{sec:mapping}
The concept of fluids of light relies on the mathematical mapping that can be made between the NLSE and the Gross-Pitaevskii equation (GPE), which describes the evolution of the macroscopic wavefunction $\psi$ of a weakly interacting Bose gas, with an additional damping term $\gamma$:
\begin{equation}
   { \mathrm{i}\hbar\dfrac{\partial \psi}{\partial t} = \left( \dfrac{-\hbar^2}{2m}\nabla^2  - \mathrm{i}\dfrac{\gamma}{2} + V(\mathbf{r}) +g|\psi|^2    \right)\psi .}
    \label{eq:GPE}
\end{equation}
In this equation, $g=\dfrac{4\pi\hbar^2a_s}{m}$, with $a_s$ the s-wave scattering length, describes two-body contact interaction \citep{dalfovo1999theory}.
$V(\mathbf{r})$ is a potential energy term, typically a trapping potential.
$\gamma$ is a term modeling atom losses in the system, and has been used e.g. to describe atom lasers \citep{kneerlaser}, but could be ignored in most closed-system BEC experiments.
The comparison between the NLSE and the (damped) GPE enables us to map the propagation of the envelope of the electric field $\mathcal{E}$ to the time evolution of a Bose gas wavefunction $\psi$.

There is a key difference between the two equations though, as the NLSE describes a propagation along the $z$-axis when the GPE describes a temporal evolution.
Consequently, we can map the propagation axis to an effective time $\tau = z/c$.
The spatial dynamics takes place in the transverse $(x,y)$ plane, as the kinetic energy term  involves only the derivatives along this plane. 
Hence, in this derivation for a monochromatic light field, the NLSE describes an analogous two-dimensional photonic Bose gas evolving along an effective time that is the propagation axis.
This is known as paraxial fluids of light or fluid of light in a propagating geometry, or in 2D+1 geometry.
Imaging the beam at the output of the nonlinear medium of length $L$ is equivalent to detecting a 2D Bose gas that evolved during an effective time $L/c$.
A more detailed discussion on the dimensionality of the system will be given in Section~\ref{sec:dimension}.

To complete this formal analogy, we multiply Eq.~\eqref{eq:NLSE_2D} by $\hbar$ and make the variable change $\tau = z/c$.
We obtain:
\begin{equation}
    \mathrm{i}\hbar\dfrac{\partial \mathcal{E}}{\partial \tau} = \hbar c\left[- \frac{1}{2 k_0}\nabla_\perp^2 - \mathrm{i}\frac{\alpha}{2} - k_0\frac{\delta n_0(\mathbf{r})}{n_0} -k_0\frac{n_2^E}{n_0}|\mathcal{E}|^2\right]\mathcal{E}.
   \label{eq:NLSE_2D_rescaled}
\end{equation}
All terms in factor of $\mathcal{E}$ on the right-hand-side of Eq.~\eqref{eq:NLSE_2D_rescaled} have the dimension of an energy, like in the Gross-Pitaevskii equation.
Consequently, we can extract an analogous kinetic, potential and kinetic energy.

\begin{enumerate}
    \item The \textbf{kinetic energy} term is given by the transverse diffraction: 
    $$-\hbar c\ \dfrac{\nabla_\perp^2 \mathcal{E}}{2 k_0} \leftrightarrow \dfrac{-\hbar^2}{2m}\nabla^2\psi$$
    which enables us to define the effective mass of the photons in the transverse plane as:
    $$\bar{m}=\frac{\hbar}{c} k_0.$$
     .
    \item The \textbf{interaction energy} term is given by the nonlinear index:
  
    $$-\hbar c \ k_0\frac{n_2^E}{n_0}|\mathcal{E}|^2 \leftrightarrow g|\psi|^2 $$
    \text{ which defines an effective interaction coefficient}
    $$\bar{g}=-\hbar c\  k_0 \frac{n_2^E}{n_0}. $$
     Bose gases are stable against collapse in the case of repulsive interactions, i.e. when $g>0$. 
     For the effective interaction term in fluids of light, this implies $n_2<0$. 
     In this case, the beam experiences self-defocusing,
     which is consistent with the picture of repulsive photon-photon interactions. 
     By contrast, when $n_2>0$ (and $\bar{g}<0$), the beam experiences self-focusing up to the point of filamentation \citep{askar1974self}, which 
     is a similar phenomenon to the collapse of BECs with attractive interactions \citep{roberts2001controlled}.
     
    \item The \textbf{potential energy} term is given by the spatial modulation of the linear index: 
    $$-\hbar c\  k_0\dfrac{\delta n_0(\mathbf{r})}{n_0} \leftrightarrow V(\mathbf{r}),$$
    which defines  the effective potential
    $${\bar{V}(\mathbf{r})=-\hbar c\  k_0 \dfrac{\delta n_0(\mathbf{r})}{n_0}.}$$
    Interestingly, while a trapping potential is necessary in BECs to not lose the atoms due to gravity, it is obviously not the case for fluids of light.
    However it is important to note that a spatial modulation of the linear index gives the possibility to shape the potential landscape for the fluid. \\
    
    \item The \textbf{loss} term equivalent to  the  linear absorption term: $$\hbar c\ \dfrac{\alpha}{2} \leftrightarrow \dfrac{\gamma}{2} ,$$
\noindent which defines the effective loss term as 
    $$\bar{\gamma} =  \hbar c\  \alpha .$$

\end{enumerate}
To map Eq.~\eqref{eq:NLSE_2D_rescaled} to the GPE with a time-independent potential, we assume that the susceptibility is independent of  $z$ and therefore we set $\mathbf{r}=\mathbf{r_{\perp}}$ in the potential term.
We can now rewrite Eq.~\eqref{eq:NLSE_2D_rescaled} as:
\begin{equation}
    \mathrm{i}\hbar\dfrac{\partial \mathcal{E}}{\partial \tau} = \left[ -\dfrac{\hbar^2}{2 \bar{m}}\nabla_\perp^2 - \mathrm{i}\dfrac{\bar{\gamma}}{2} + \bar{V}(\mathbf{r}_\perp)+\bar{g}|\mathcal{E}|^2 \right]\mathcal{E} .
   \label{eq:NLSE_2D_gpe}
\end{equation}
Eq.~\eqref{eq:NLSE_2D_gpe} is nearly equivalent to the Gross-Pitaevskii equation \eqref{eq:GPE}.
A difference is that the electric field envelope has not been renormalized to a particle density. 
Consequently, even though $\bar{g}|\mathcal{E}|^2$ in the NLSE has the same dimension as $g|\psi|^2$ in the GPE, the separate terms  do not have the same dimension.
Moreover, the gradient operator is applied in two dimensions rather than three, suggesting a correspondence with a 2D GPE. This highlights the need for a proper normalization of the ``wavefunction” to either a 2D or 3D photon density, which would, in turn, affect the effective nonlinear coefficient $\bar{g}$.
This ambiguity in the dimensionality of the system will be addressed in Section~\ref{sec:dimension}. A complete mapping to a 3D GPE will be achieved once the finite spectral linewidth of the laser is taken into account and the field is properly renormalized  (see Section~\ref{sec:dimension}).

For the sake of simplicity, we will for now satisfy ourselves with this ``partial" mapping, as it is sufficient to understand most  of the phenomena studied up to now in fluids of light, as we will see in Section~\ref{sec:dimension}. 

The mapping in Eq.~\eqref{eq:NLSE_2D_gpe} is useful to explain the mathematical mapping, however, when working with fluids of light, we rather keep the native units of the original NLSE given by Eq.~\eqref{eq:NLSE_2D}.
This is equivalent to ignoring the $\hbar$ and $c$ factors in all analogous terms.
The analogous kinetic and interaction energy terms are then expressed in m$^{-1}$ and generally, the interaction coefficient $g=-k_0 n_2^E/n_0$ which  appears in Eq.~\eqref{eq:NLSE_2D}  is preferred over the effective one $\bar{g}=\hbar c g$. The analogous interaction energy is then given by $-k_0 \Delta n$ with: 
\begin{equation}
    \Delta n = \dfrac{n_2^E}{n_0}| \mathcal{E}|^2=\dfrac{n_2}{n_0}I.
    \label{eq:Delta_n_def}
\end{equation}
In Table~1, we compare the optical NLSE and the GPE formalism in the 2D+1 geometry.
   \begin{table}[h]
\centering
\begin{tabular}{p{3cm} p{3cm} p{2cm}}
\hline
\textbf{Quantity} & \textbf{Optics (2D+1)} & \textbf{BEC} \\

\hline
Evolution & $z$ in [m]  & $t$ in [s] \\
\hline
Mass & $k_0$ in [m$^{-1}]$ & $m$ in [kg] \\
 \hline
Energies& in [m$^{-1}$] & in [Hz] $\times \hbar$ \\

Kinetic energy & $E_k = \dfrac{k_{\perp}^2}{2 k_0}$ & $E_k = \dfrac{\hbar^2k^2}{2m}$ \\
Potential energy & $E_p = -k_0 \dfrac{\delta n_0(r)}{n_0}$ & $E_p = V(r)$ \\
Interaction energy & $E_i =  -k_0 \Delta n$ & $E_i = g \rho$ \\
Loss rate & $\alpha$ & $\gamma$ (usually 0) \\
\hline
Healing length  in [m]\\ (see Section~\ref{sec:bogo})   & $\xi = \dfrac{1}{k_0 \sqrt{2|\Delta n|}}$ & $\xi = \dfrac{\hbar}{\sqrt{2 m g \rho}}$ \\
\hline
Sound velocity & $c_s = \sqrt{|\Delta n|}$ & $c_s = \sqrt{\dfrac{g \rho}{m}}$ \\
(see Section~\ref{sec:bogo})& in [m.m$^{-1}$] & in [m.s$^{-1}$]\\
\hline
Non-linear length & $z_{NL} = \dfrac{1}{k_0 |\Delta n|}$ & $t_{NL}=\dfrac{\hbar}{g \rho}$ \\
(see Section~\ref{sec:adim})& in [m] & in [s]\\
\hline
\end{tabular}
\caption{Comparison between NLSE and GPE parameters. The natural experimental units to express the energies are [m$^{-1}]$ in optics and [J]  for atomic BEC (or [Hz], by dividing the expressions in the table by $\hbar$). }

\label{tab:nlse_gpe_comparison}
\end{table}

\subsection{Hydrodynamics equation}
\label{sec:hydro}
\begin{figure}[h]
    \centering
    \includegraphics[width=1\linewidth]{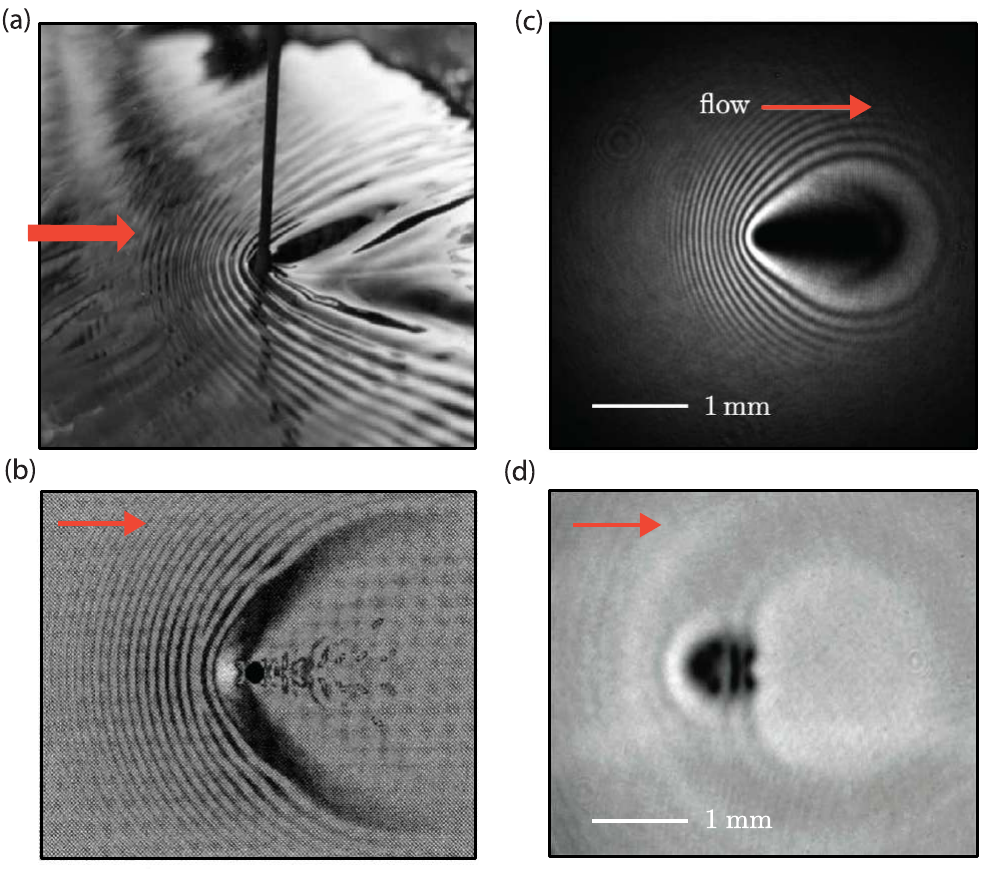}
    \caption{Illustration of the fluids of light. a) A classical fluid (the Loire river) hitting a defect (a wood stick). b) Numerical simulation by \cite{pomeau1993diffraction}. c) A fluid of light  (no interaction) hitting a defect at high velocity (faster than the sound velocity). d) A fluid of light hitting a defect at much slower velocity (0.5 times the sound velocity). Figures c) and d) are done in a rubidium vapor. d) is adapted from \cite{aladjidi2023full}.
    }
    \label{fig:loire}
\end{figure}

To understand the denomination of ``fluid" of light, it useful to move to an hydrodynamics framework.
This can be done through the Madelung transform, where the field envelope $\mathcal{E}$ is expressed  as a function of its density $\rho$ and phase $\phi$: 
\begin{equation}
    \mathcal{E}(\mathbf{r}_\perp,z)=\sqrt{\rho(\mathbf{r}_\perp,z)}e^{\mathrm{i}\phi(\mathbf{r}_\perp,z)}.
\end{equation}
Injecting this expression in Eq.~\eqref{eq:NLSE_2D}, we obtain, similarly as for a classical fluids, a set of Euler equations:

\begin{align}
    &\dfrac{\partial \rho}{\partial z} + \frac{1}{c}\mathbf{\nabla}_\perp \cdot (\rho\mathbf{v})= -\alpha \rho, \label{eq:euler_rho} \\
    & \dfrac{\partial \mathbf{v}}{\partial z} +\dfrac{1}{c}\mathbf{v}\cdot\nabla_\perp \mathbf{v} =c\nabla_\perp \left(  \frac{\delta n_0}{n_0} + \frac{n_2^E}{n_0}\rho + \dfrac{1}{2k_0^2}\dfrac{\nabla^2\sqrt{\rho}}{\sqrt{\rho}} \right) ,
\label{eq:euler_v}
\end{align}
where $\mathbf{v}=\dfrac{c}{k_0}\mathbf{\nabla}_\perp \phi$ is the velocity of the fluid.
Eq.~\eqref{eq:euler_rho} is a continuity equation that describes the dissipation of the photon density due to absorption.
Eq.~\eqref{eq:euler_v} is a convection equation coupling the velocity flow to different source terms on the right-hand side. 
From left to right, these terms represent: a potential term, an interaction term,  and a term analogous to the Bohm quantum potential, commonly referred to as the quantum pressure.
The term has no counterpart in classical fluids, and becomes dominant in case of rapid density fluctuations in the transverse plane.
While \cite{tsang2005metaphoric} have interpreted  the quantum pressure term to play analogous roles to viscosity in the Navier-Stokes equations, the similarity between quantum pressure and viscosity is still an open problem.

\subsection{Bogoliubov perturbation theory }
\label{sec:bogo}
Eq.~\eqref{eq:NLSE_2D} cannot be solved exactly due to the presence of the nonlinear term.
When $g>0$ ($n_2<0$), it is however possible to solve it in a perturbative approach up to first order, which is the Bogoliubov theory \citep{pethick2008bose}.
This will give us the dispersion relation of the fluid, i.e. its response to weak excitations.
We first derive a stationary solution of the NLSE. 
We treat the problem in the absence of  potential ($\delta n(r_\perp)=0$), so we can consider the stationary solution to be uniform in the transverse plane and look for it as $\mathcal{E}^{(0)}=|\mathcal{E}^{(0)}|e^{\mathrm{i}\kappa z}$.
It satisfies
\begin{equation}
    \left(\dfrac{-\nabla_\perp^2}{2k_0} -\mathrm{i}\dfrac{\alpha}{2}-k_0\dfrac{n_2^E}{n_0}|\mathcal{E}^{(0)}|^2 +\kappa\right)\mathcal{E}^{(0)} = 0 .
\end{equation}
Since the field is uniform, $\nabla_\perp\mathcal{E}^{(0)}=0$ and we find $\kappa=k_0\frac{n_2^E}{n_0}|\mathcal{E}^{(0)}|^2+\mathrm{i}\frac{\alpha}{2}$.
\noindent We can then write the unperturbed field envelope as: 
\begin{equation}
    \mathcal{E}^{(0)}=| \mathcal{E}^{(0)}|e^{(\mathrm{i}k_0\Delta n -\frac{\alpha}{2})z},
    \label{eq:nlse_stationnary_solution}
\end{equation}
with $\Delta n = \frac{n_2^E}{n_0}| \mathcal{E}^{(0)}|^2$ as defined in Eq.~\eqref{eq:Delta_n_def}.
$\kappa$ is the equivalent of the chemical potential $\mu=gn$ for atomic BECs, $n$ being the average density of atoms. However in the optics case, $\kappa$ also has a complex part to include the linear losses.
A full mapping to a chemical potential would require a proper renormalization of the electric field to an average density of photons.
We will come back to this normalization in Section~\ref{sec:dimension}.
In addition, 
the system we consider is out-of-equilibrium, due to the initial quench of the interactions at the entrance of the nonlinear medium, consequently, we will avoid using the term of ``chemical potential", as it is defined for systems at equilibrium. \\
We now look for a perturbative $z$-dependent solution around the steady state given by Equation~\eqref{eq:nlse_stationnary_solution} and adding a weak perturbation $\delta \mathcal{E}(\mathbf{r_\perp,z}) \ll \mathcal{E}^{(0)}$:
\begin{equation}
    \mathcal{E}(\mathbf{r_\perp},z) = \mathcal{E}^{(0)} +\delta \mathcal{E}(\mathbf{r_\perp},z).
\end{equation}
We look for solutions of the form:
\begin{equation}
    \delta \mathcal{E}(\mathbf{r_\perp},z)  = \left(u(\mathbf{r_\perp})e^{-\mathrm{i}\Omega_\text{B} z} -v^*(\mathbf{r_\perp})e^{\mathrm{i}\Omega_\text{B} z}\right)e^{(\mathrm{i}k_0\Delta n -\frac{\alpha}{2})z},
\end{equation}
where $\Omega_\text{B}$ is the (real)  eigenenergy associated to the  perturbation.
Injecting this ansatz into the linearized Eq.~\eqref{eq:NLSE_2D}, we obtain two coupled equations by equating the terms in front of $e^{-\mathrm{i}\Omega_\text{B} z}$ and $e^{\mathrm{i}\Omega_\text{B} z}$ to 0:

\begin{align}
    \left[-\dfrac{1}{2k_0}\nabla_\perp^2 -k_0\Delta n -\Omega_\text{B} \right]u(\mathbf{r_\perp}) +[k_0\Delta n] v (\mathbf{r_\perp}) &= 0 \\
        \left[-\dfrac{1}{2k_0}\nabla_\perp^2 -k_0\Delta n + \Omega_\text{B} \right]v(\mathbf{r_\perp})  +[k_0\Delta n]u (\mathbf{r_\perp}) &= 0 .
    \label{eq:bogo_uv}
\end{align}

\noindent We write $u(\mathbf{r_\perp})$ and $v(\mathbf{r_\perp})$ as plane waves in the transverse plane:
$u(\mathbf{r_\perp})=u_{k_\perp}e^{\mathrm{i}\mathbf{k}_\perp \mathbf{r}_\perp}$ and $v(\mathbf{r_\perp})=v_{k_\perp}e^{\mathrm{i}\mathbf{k}_\perp \mathbf{r}_\perp}$.
Then, Eq.~\eqref{eq:bogo_uv} gives a system of two linear equations that are consistent only if the determinant of the corresponding matrix vanishes.
Setting this determinant to 0, we obtain: 
\begin{equation}
    {\Omega_\text{B}(k_\perp) = \sqrt{\dfrac{k_\perp^2}{2k_0} \left(\dfrac{k_\perp^2}{2k_0}+ 2 k_0 |\Delta n|\right)}
    =\sqrt{\left(\dfrac{k_\perp^2}{2k_0}\right)^2 + k_\perp^2 |\Delta n|} 
    \text{ .}}
    \label{eq:bogo_disp}
\end{equation}

\begin{figure}[h]
    \centering
    \includegraphics[width=1\linewidth]{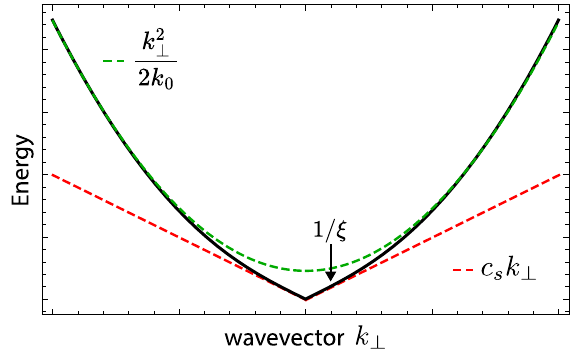}
    \caption{Sketch of the Bogoliubov dispersion in black. Red is the linear limit at low $k_{\perp}$, while green is the large $k_{\perp}$ with a parabolic shape.}
    \label{fig:bogoliubov}
\end{figure}

\noindent Equation \eqref{eq:bogo_disp} is the Bogoliubov dispersion, which is characteristic of weakly interacting Bose gases.
It is here expressed in terms of propagating fluid of light quantities and units, but one can check that it is consistent with 
its usual expression for BECs by introducing the effective terms defined earlier.\\
This dispersion, that we note $\Omega_\text{B}(k_\perp)$ shows two regimes plotted in Fig \ref{fig:bogoliubov}: 
\begin{itemize}
    \item for $k_\perp \ll  k_0\sqrt{|\Delta n|}$, the dispersion is linear: $\Omega_\text{B}(k_\perp) \approx k_\perp \sqrt{\Delta n}$.
    This is a sonic or phononic regime, which enables us to define a speed of sound $c_\text{s}=\sqrt{|\Delta n|}$.
    \item for $k_\perp \gg  k_0\sqrt{|\Delta n|}$, we find the dispersion of free massive particles: $\Omega_\text{B}(k_\perp) \approx \dfrac{k_\perp^2}{2k_0}$.
\end{itemize}

\noindent The transition between those two regimes happens around  $k_\xi=1/\xi$, where $\xi$ is the healing length, which is defined as the length scale which equates the kinetic energy and the interaction energy: $\xi=1/k_0\sqrt{2|\Delta n|}$.
$\xi$ is the characteristic scale of the interactions in the transverse plane. Note that $\xi$ is defined as the inverse of the wavevector $k_\xi$ and therefore should be multiplied by $2\pi$ to obtain the wavelength associated to the wavevector $k_\xi$.

It is well know that this dispersion relation leads to superfluidity in quantum gases \cite{pitaevskij_bose-einstein_2016}.
Similarly, superfluidity of light, i.e. a transverse flow of light without energy dissipation,  can also be observed in the linear region of the dispersion, for $k_\perp < k_\xi$.
\noindent To convince ourselves of this fact, let us recall the Landau criterion for superfluidity, which defines a critical speed above which a flow dissipates energy in the form of density waves. 
In the case of a transverse flow of the background fluid at a velocity $\mathbf{v}$, the energy of the elementary wave is shifted by $\mathbf{k}_\perp.\mathbf{v}$ due to the Doppler effect: 
\begin{equation}
    \Omega_\text{B}(k_\perp) = \mathbf{k}_\perp.\mathbf{v} + \sqrt{\left(\dfrac{k_\perp^2}{2k_0}\right)^2 + k_\perp^2 |\Delta n}|  \text{ .}
    \label{eq:bogo_disp_doppler}
\end{equation}
For the flow to generate density waves, the expression~\eqref{eq:bogo_disp_doppler} has to be negative, to be energetically favorable: 
\begin{equation}
    \Omega_\text{B}(k_\perp)_{\mathbf{v}=0} + \mathbf{k}_\perp.\mathbf{v} < 0 .
\end{equation}
As $\Omega_\text{B}(k_\perp)_{\mathbf{v}=0}>0$, this is possible only if $ \mathbf{k}_\perp.\mathbf{v}<0$, that is if the density wave propagates counter-flow, and if $\Omega_\text{B}(k_\perp)_{\mathbf{v}=0}<|\mathbf{k}_\perp||\mathbf{v}|$ . 
These conditions define a critical speed for the background flow above which density waves are excited:
\begin{equation}
    v_\text{c} = \min_{k_\perp}\left\{\dfrac{\Omega_\text{B}(k_\perp)_{\mathbf{v}=0}}{|\mathbf{k}_\perp|}\right\} .
    \label{eq:landau_criteriion}
\end{equation}

This is the Landau criterion for superfluidity \citep{landau_superfluidity_1941,leggett2001bose}. 
If the dispersion is quadratic in $k_{\perp}$, then $v_\text{c}=0$  and energy will always be dissipated in the form of density waves.
But in the sonic region of the dispersion, $\Omega_\text{B}(k_\perp)=c_sk_\perp$ and we obtain a finite value for the critical velocity which is precisely the speed of sound: $v_\text{c}=c_\text{s}$.

\section{What is the dimension of the system?}
\label{sec:dimension}   

So far in this review, we have assumed a perfectly monochromatic laser beam, neglecting the temporal dimension and resulting in a 2D+1 description for fluids of light.
This is the approach widely used in the literature.
However, the most comprehensive way to describe paraxial fluids of light involves considering a full 3D+1 geometry.
In this framework, the three \textit{spatial} dimensions are composed of the transverse coordinates $(x, y)$ and the physical time coordinate $t$. 
In this section, we derive a 3D+1 evolution equation starting from optics notations and then mapping it to the BEC language.

\subsection{The role of time: 3D+1 NLSE in nonlinear optics} 
We consider a laser with a central frequency $\omega_0$, allowing the field envelope $\mathcal{E}$ to vary with time: $E (\mathbf{r}, t) = \dfrac{1}{2}\mathcal{E}(\mathbf{r}, t)\text{e}^{\mathrm{i} (k_0 z - \omega_0 t)} + \text{c.c.}$.
This definition describes well a pulsed laser, but also applies to continuous wave lasers, where the linewidth is finite and could be taken into account with a slowly varying envelope. 
Actually, adding this temporal dependency in the envelope describes any fluctuations of the laser around the carrier frequency at $\omega_0$.
We write the total field $E(r,t)$  in the frequency domain:
\begin{align}
E(\mathbf{r}, \omega) &= \frac{1}{2}\mathcal{E}(\mathbf{r}, \omega - \omega_0)\text{e}^{\mathrm{i} k_0 z} + \frac{1}{2}\mathcal{E}(\mathbf{r}, \omega + \omega_0)\text{e}^{-\mathrm{i} k_0 z} \label{eq:env3D_1} \\
& \simeq \frac{1}{2}\mathcal{E}(\mathbf{r}, \omega - \omega_0)\text{e}^{\mathrm{i} k_0 z}.
\label{eq:env3D}
\end{align}
In the second line, we have neglected the term at $\omega + \omega_0$ since $\mathcal{E}$ 
is supposed to vary slowly compared to the frequency $\omega_0$. As a result, for frequencies $\omega$ around $\omega_0$, such that $|\omega - \omega_0| \ll \omega_0$, only the first term of Eq.~\eqref{eq:env3D} contributes significantly to the field.

After re-injecting the envelope of Eq.~\eqref{eq:env3D_1} into the paraxial wave equation \eqref{eq:nlse_paraxial_in_progress} and writing $k^2(\omega) = \frac{\omega^2}{c^2}\Re\left[1 + \chi^{(1)}(\omega)\right]$, we find the following equation:
\begin{equation} \label{eq:nlse_paraxial_3d}
      2\text{i}k_0\dfrac{\partial \mathcal{E}}{\partial z}(\omega-\omega_0) = -\nabla_\perp^2\mathcal{E}(\omega-\omega_0) - (k^2 - k_0^2)\mathcal{E}(\omega-\omega_0).
\end{equation}
Here, the losses and the nonlinearity, that are not specific to the introduction of $t$ in the derivation, have been ignored. 
However, following the same approach as in Section~\ref{sec:NLSE}, both could be integrated without difficulties.

As $k$ is typically of the same order as $k_0$ due to the paraxial approximation, we can approximate $k^2-k_0^2 \simeq 2k_0(k-k_0)$. 
We then expand $k$ around the carrier frequency $\omega_0$ up to second order:
\begin{equation}
\begin{aligned}
   k(\omega) &\simeq k(\omega_0) +\dfrac{1}{v_g}\\
   &=\left. \dfrac{\partial k}{\partial \omega}\right|_{\omega_0}(\omega-\omega_0)+\frac12 =\left.\dfrac{\partial^2 k}{\partial \omega^2}\right|_{\omega_0}(\omega-\omega_0)^2 \\
   & = k_0 + \frac{1}{v_g}\delta\omega + \frac{D_0}{2}\delta\omega^2,
   \label{eq:deltaomega}
\end{aligned}
\end{equation}
where we defined $\delta\omega = \omega - \omega_0$, and we introduced two terms: the group velocity $\left. \dfrac{1}{v_g}=\dfrac{\partial k}{\partial \omega}\right|_{\omega_0}$ and the group velocity dispersion $D_0=\left.\dfrac{\partial^2 k}{\partial \omega^2}\right|_{\omega_0}$.

\noindent Injecting the expansion in Eq.~\eqref{eq:nlse_paraxial_3d} we find:

\begin{equation} \label{eq:nlse_paraxial_3d_2}
      \text{i}\dfrac{\partial \mathcal{E}}{\partial z}(\delta\omega) = \left(-\frac{1}{2k_0}\nabla_\perp^2 - \frac{1}{v_g}\delta\omega - \frac{D_0}{2}\delta\omega^2\right)\mathcal{E}(\delta\omega).
\end{equation}
By taking the inverse Fourier transform, and simplifying the global phase offset due to the frequency translation $\omega_0$, we obtain:
\begin{equation}
\begin{aligned} \label{eq:nlse_paraxial_3d_3}
      \text{i}\dfrac{\partial \mathcal{E}}{\partial z} = &\left[-\frac{1}{2k_0}\nabla_\perp^2  + \frac{D_0}{2}\frac{\partial^2}{{\partial t}^2} - \frac{\text{i}}{v_g}\frac{\partial}{\partial t}  \right.\\
      &\left.-\mathrm{i}\frac{\alpha}{2} - k_0\frac{\delta n_0(\mathbf{r})}{n_0} -k_0\frac{n_2^E}{n_0}|\mathcal{E}|^2\right]\mathcal{E},
\end{aligned}
\end{equation}
where we reintroduced the loss, potential (taken independent of $z$), and nonlinear terms, assuming that they do not depend on  $t$.

\noindent The term $\frac{\text{i}}{v_g}\frac{\partial}{\partial t}$ in Eq.~\eqref{eq:nlse_paraxial_3d_3} is known as a rigid drift and can be removed by a change of variable to the co-moving frame: 
$t' = t - \frac{z}{v_g}, \text{ } \mathcal{E}(z, t) = \mathcal{E}(z, t')$.
Note that here (and in later variable changes) we keep the notation $\mathcal{E}$ for $\mathcal{E}(z, t')$, even though it is  a new quantity: the field envelope in the co-moving frame.

\noindent The derivative over $z$ is computed using the chain rule:
\begin{equation}
\left.\frac{\partial}{\partial z}\right|_{t}= \frac{\partial t'}{\partial z} \frac{\partial}{\partial t'} + \left. \frac{\partial}{\partial z}\right|_{t'}
= -\frac{1}{v_g} \frac{\partial}{\partial t'} + \left.\frac{\partial}{\partial z}\right|_{t'},
\end{equation}
and allows to cancel the rigid drift.
This transformation is known in optics as a retarded time frame and leads to:
\begin{equation}
\label{eq:nlse_paraxial_3d_4}
    \begin{aligned}
              \text{i}\dfrac{\partial \mathcal{E} (z,t')}{\partial z} = &\left[-\frac{1}{2k_0}\nabla_\perp^2  + \frac{D_0}{2}\frac{\partial^2}{\partial t'^2}\right. \\
      &\left.- \mathrm{i}\frac{\alpha}{2} - k_0\frac{\delta n_0(\mathbf{r})}{n_0} -k_0\frac{n_2^E}{n_0}|\mathcal{E}|^2\right]\mathcal{E} (z,t').
    \end{aligned}
\end{equation} 

\begin{figure}
    \centering
    \includegraphics[width=0.9\linewidth]{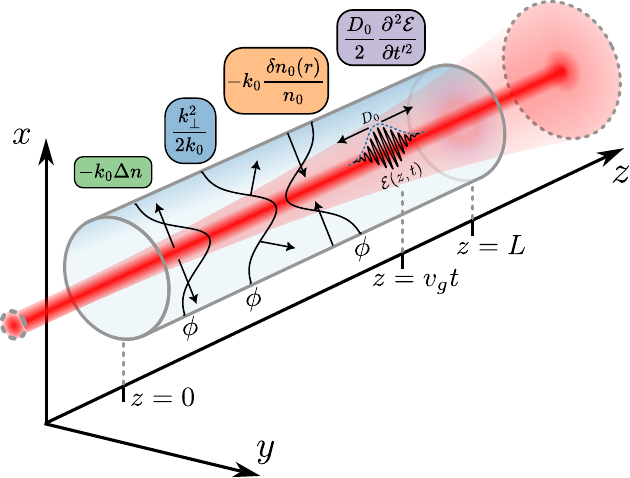}
    \caption{Sketch of the 3D NLSE with all terms contributing the dynamics adapted from \cite{aladjidi2023full}. From left to right: interactions or self-defocusing, kinetic energy along $(x,y)$ or diffraction, potential energy or waveguiding, kinetic energy along $t$ or dispersion, rigid drift or group velocity. This last term disappears in the co-moving frame as explained in the text.} 
    \label{fig:nlse_3d}
\end{figure}

\noindent Conceptually, the fluid of light becomes a 3D+1 system where the propagation axis $z$ still acts as an effective evolution parameter (analogous to time) and the third spatial dimension being the time $t'$ in the reference frame of the pulse.
The mass  $m_{\perp}=k_0$  in the transverse plane is still due to diffraction, and a mass in the temporal direction $m_{t}=\dfrac{-1}{D_0}$ is introduced due to the group velocity dispersion.

\subsection{Mapping the 3D+1 NLSE and the GPE}
We now map the 3D+1 NLSE to the GPE using the same approach as in Section~\ref{sec:mapping}.
We use a similar change of variable  (but in the co-moving frame): $\tau = \dfrac{z}{c}\text{ , } \zeta = v_g t - z \text{ , }  \mathcal{E}(z,t) = \mathcal{E}(\tau, \zeta)$ and we multiply by $\hbar$ to recover the dimension of an energy: 
\begin{equation}
\begin{aligned}
    \mathrm{i}\hbar\dfrac{\partial \mathcal{E}}{\partial \tau} =& \left. [-\dfrac{\hbar^2 }{2 \bar{m}_\perp}\nabla_\perp^2 -\dfrac{\hbar^2 }{2 \bar{m}_\zeta} \frac{\partial^2}{\partial\zeta^2} - \mathrm{i}\dfrac{\bar{\gamma}}{2}\right.\\
    & \left. + \bar{V}(\mathbf{r},\zeta) + \bar{g}(\mathbf{r},\zeta)|\mathcal{E}|^2\right]\mathcal{E}
    \label{eq:nlse_3d_gpe_not_norm1}
    \end{aligned}
\end{equation}
where we defined the analogous mass in the transverse plane as $\bar{m}_{\perp}={\hbar k_0}/{c}$, an the mass along the $\zeta$-axis as $\bar{m}_\zeta={-\hbar }/{cv_g^2D_0}$.
The terms $\bar{V}$, $\bar{\gamma}$ and $\bar{g}$ are defined the same way as in Section~\ref{sec:mapping}.

In Eq.~\ref{eq:nlse_3d_gpe_not_norm1}, the mass is anisotropic.
But since it appears only in the kinetic energy terms, we can rescale the $\zeta$ axis once more to eliminate this anisotropy, at the expense of modifying the definition of momentum along the $\zeta$-axis.
This rescaling is introduced as:
\begin{equation}
\zeta' \equiv \frac{\zeta}{\sqrt{-k_0D_0 v_g^2}},
\end{equation}
leading to an isotropic form for the 3D+1 NLSE:
\begin{equation}
\begin{aligned}  
    \mathrm{i}\hbar\dfrac{\partial }{\partial \tau}\mathcal{E}(x,y,\zeta') =& \left. [-\frac{\hbar^2}{2 \bar m}\nabla^2_{x,y,\zeta'} - \mathrm{i}\dfrac{\bar{\gamma}}{2} + \bar{V}(\mathbf{r}_\perp,\zeta')\right. \\
    &\left. + \bar{g}(\mathbf{r}_\perp,\zeta')|\mathcal{E}|^2 \right]\mathcal{E}(x,y,\zeta'),
    \label{eq:nlse_3d_gpe_not_norm}
    \end{aligned}
\end{equation}
with $\bar m = {\hbar k_0}/{c}$.
Note that in Eq.~\eqref{eq:nlse_3d_gpe_not_norm1} and Eq.~\eqref{eq:nlse_3d_gpe_not_norm}, the terms $\bar{g}$ are proportional but not equal since $|\mathcal{E}|^2$ depends on the rescaling. This will become clear in the next step of the derivation with the proper normalization of $|\mathcal{E}|^2$.

We decomposed these two variable changes for clarity, but the final change from $t$ to $\zeta'$ could have been intuited from our previous remark on the kinetic energy. The overall rescaling is indeed the square root of the ratio of the mass along the time-axis and the mass in the transverse plane:
\begin{equation}
    {\zeta'=\frac{1}{\sqrt{-k_0 D_0}} t'= \sqrt{\frac{m_t}{m_{\perp}}} t'.}
    \label{eq:rescaling}
\end{equation}

\noindent In 3D, there is a final step that can be done to complete the mapping with the GPE.
The electric field envelope $\mathcal{E}(x,y,\zeta')$ can be normalized to be of the same dimension as a wavefunction $\psi(x,y,\zeta')$, in order to obtain the interaction coefficient $\bar{g}_{3D}$ with the same dimension as in the GPE.

\noindent We first convert the electric field $\mathcal{E}(x,y,t')$ into a photon density $|\psi|^2$ along $(x,y,t')$ by dividing  the intensity $I$ by the energy $\hbar \omega_0$ of one photon:
\begin{equation}
\begin{aligned}
     |\mathcal{E }(x,y,t')|^2 &= \frac{2}{\varepsilon_0 n_0 c}I(x,y,t')= \frac{2}{\varepsilon_0 n_0 c}\frac{I(x,y,t')}{\hbar \omega_0} \hbar \omega_0 \\
     &= \frac{2 \hbar \omega_0}{\varepsilon_0 n_0 c} |\psi(x,y,t')|^2.
\end{aligned}
     \label{norm1}
\end{equation}
Here $|\psi(x,y,t')|^2$ is analogous to a density in photons.m$^{-2}$.s$^{-1}$. 
The conversion $t'\rightarrow \zeta'$  defined in Eq.~\ref{eq:rescaling} is then used to convert to   photons.m$^{-3}$ (in the compressed frame): 
\begin{equation}
    |\psi(x,y,t')|^2 = \sqrt{\frac{m_t}{m_{\perp}}} |\psi(x,y,\zeta')|^2.
    \label{norm2}
\end{equation}
We combine Eq.~\eqref{norm1} and Eq.~\eqref{norm2} and multiply  by $c$ to take into account the change of variable along $z$ and by  $\hbar$ to obtain: 
\begin{equation}
\begin{aligned}
    &\mathrm{i}\hbar\dfrac{\partial \psi (x,y,\zeta')}{\partial \tau} = \\&\left[-\frac{\hbar^2 }{2 \bar m}\nabla^2  - \mathrm{i}\bar{\gamma} + \bar{V}(\mathbf{r_\perp},\zeta') + \bar g_{3D}|\psi|^2\right]
    \psi (x,y,\zeta'),
    \label{eq:nlse_3d_gpe}
\end{aligned}
\end{equation}
\begin{equation}
    \text{with } \bar{g}_{3D}=\hbar c \times \frac{2 \hbar \omega_0}{\varepsilon_0 n_0 c} \times \sqrt{\frac{m_t}{m_{\perp}}} \times -k_0 \frac{n_2^E}{n_0}.
\end{equation}

\noindent This simplifies to 
\begin{equation}
    {\bar{g}_{3D}= -n_2^I\  (\hbar \omega_0)^2 \ \sqrt{\frac{m_t}{m_{\perp}}}.}
\end{equation}

\noindent The interaction energy in the GPE $E_{\text{int}}=g\rho$ is therefore obtain in the optics language by $E_{\text{int}}=\hbar \omega_0 \times -\Delta n $ which is simply the energy of one photon times the nonlinear index change.\\

We established a formal mapping of the dispersive NLSE to a 3D+1 GPE. 
Like in the previously treated 2D+1 case, we will now preferably use the native optics units of Eq.~\eqref{eq:nlse_paraxial_3d}.
As seen in Eq.~\eqref{eq:rescaling}, the conversion from the temporal units (relative to $t$ or $t'$ axis) to the spatial units along $\zeta'$ is done via a multiplication by the mass ratio $\sqrt{{m_t}/{m_\perp}}=1/\sqrt{-k_0D_0}$ in [m.s$^{-1}$].\\
In optics language, the ``momentum" of a perturbation along the co-moving $t'$-axis is then $\delta\omega=\omega-\omega_0$ as defined in Eq.~\eqref{eq:deltaomega} with $\omega_0$ the central frequency of the laser.
Then, the response of the system to a perturbation of momentum $\mathbf{k}=(k_x,k_y,\delta\omega)$ will be given by the ``3D" Bogoliubov dispersion relation:
\begin{equation}
    \Omega_\text{B}(k)=\sqrt{\left(\dfrac{k_\perp^2}{2k_0} +\dfrac{\delta\omega^2 |D_0|}{2}\right)\left(\dfrac{k_\perp^2}{2k_0} +\dfrac{\delta\omega^2 |D_0|}{2}+k_0|\Delta n|\right)} .
    \label{eq:3D_bogoliubov}
\end{equation}

The dispersion~\eqref{eq:3D_bogoliubov} is still linear at low $\mathbf{k}$, but $\mathbf{k}=(k_x,k_y,\delta\omega)$ is now three-dimensional.
An important consequence of this relation is that, for $\delta\omega \neq 0$, a gap opens in the dispersion along $k_{\perp}$ and the linear dispersion progressively vanishes. 
In other words, the dispersion of a weak excitation propagating along the $x$ direction is modified if its frequency differs from that of the driving laser $\omega_0$.  
Similarly, the temporal evolution of an excitation (in the co-moving frame) is influenced by its momentum in the transverse plane. 
In the following, we will estimate the various orders of magnitude to understand when this coupling starts to play a non-negligible role.

\subsection{Discussion on the dimensionality of the system}

The 3D+1 representation established here does not actually require a laser pulse to be correct; any slowly varying modulation of a continuous-wave (CW) laser is also well described in this way. 
However, in almost all experiments on paraxial fluids of light to date, only the two-dimensional dynamics in the transverse plane, described by a 2D+1 NLSE, have been explored (see Section~\ref{sec:expt}).
To know whether this picture is sufficient or if one should take into account the dynamics along the co-moving time axis $t'$, let us examine the different orders of magnitudes involved along the $t'$-axis.
We take again the example of a hot rubidium vapor cell, with $\Delta n \sim 10^{-5}$, $k_0\sim8\times10^6$~m$^{-1}$.
For the length of the medium, which typically ranges from 1 to 20~cm (see Section~\ref{sec:platforms}), we will take here $L=$10~cm.

In the 2D+1 experiments the system is only modulated in the transverse plane.
The detection is done by imaging the density (and the phase) on a camera with an exposure time $t_\text{exp}>$1~$\mu$s.
This exposure time is to be compared to the characteristic scale of superfluid dynamics along the $t'$ axis, which is $\xi_t=\sqrt{-D_0/2k_0\Delta n}$.
For hot rubidium vapor a typical value of $D_0$ near resonance is  $D_0\sim-10^{-18}$s$^2$m$^{-1}$, which gives $\xi_{t}\sim $ 0.5~ns $\ll t_\text{exp}$. 
So the typical time dynamics related to the interactions would be integrated out by the detection methods of experiments, giving access exclusively to the dynamics.

But, letting apart the detection method, is there \textit{any} dynamics taking place along the $t'$ axis in these experiments?
We can evaluate the smallest value of $\delta\omega$ giving a significant evolution of a weak excitation after propagation in a medium of length $L$: $\Omega_\text{B}(k_\perp=0,\delta\omega_\text{min})L\sim\pi$.
Then, $\delta \omega_\text{min}\sim2\pi \cdot \ 400$~MHz. 
This implies that any excitation below about $400$~MHz can be considered frozen out in the dynamics of fluids of light. 
At such frequencies, and even down to 1~MHz, modern CW lasers are typically shot-noise limited. As a result, unless temporal modulation or noise is deliberately added at frequencies above $\delta\omega_\text{min}$, the system contains only vacuum fluctuations above that frequency range.

The 2D+1 experiments do not impose any additional time perturbations, but time-excitations are generated from the quench of the interactions at the entrance of the nonlinear medium, which creates a distribution of pairs of correlated excitations of momentum ($\mathbf{k},-\mathbf{k}$) \citep{larre2015propagation}.
The signature of these excitations in the ($x$,$y$) plane was measured in the transverse noise spectrum \citep{steinhauer2022analogue}), and we expect a similar manifestation along the $t'$-axis.
However, these fluctuations average to zero, so we can neglect the effect of the quench in the mean-field data (in the absence of stimulation of the process). 
In the typical transverse experiments, we can thus safely neglect the time dynamics. 

Finally, we discuss the geometric aspect ratio in typical experiments. 
In the CW laser case, the fluid's dimensions are ($w_{0x},w_{0y},t_\text{coh}$) where $w_{0x,y}$ is the waist in the $x$, $y$ dimension and $t_\text{coh}$ is the coherence time of the laser.
The coherence time is the characteristic duration since it gives the maximal extension over which the fluids of light is coherent. After a time on the order of the coherent time, we could consider that it is a new fluid leading to a new ``run" of the experiment.
This size can be mapped to the axes ($x,y,\zeta'$), where the coherence time is converted to a length using the masses ratio of Eq.~\eqref{eq:rescaling}, such that a linewidth of 10~kHz gives a scale along $\zeta'$ larger than 10 m.
The beam waist is usually of a few millimeters, such that geometrically, typical experiments are highly cigar-shaped. There are however no excitations along the axis of the cigar, and the detection method integrates the fluid along this axis over the exposure time of the camera.
Hence, due to translational invariance along the $t'$ (or $\zeta'$ axis), the dynamics of the system is effectively 2D.
In the case of a pulse, $t_\text{coh}$ is replaced by the  pulse duration $t_\text{pulse}$ and the aspect ratio is modified.
It would nonetheless be interesting to use the 3D+1 picture, by adding a modulation of weak amplitude detuned by a  $\delta\omega\sim 1/\xi_t$, and resolving its dynamics in the spectrum of the fluid.

\section{Adimensionalization and comparison with cold atoms}
    
\subsection{Adimensional equation}
\label{sec:adim}

Following the previous discussion, we now only consider the 2D+1 NLSE.
To provide a quantitative comparison between paraxial fluids of light and ultracold atomic Bose gases, it is useful to adimensionalize the NLSE and the GPE.
We rescale the transverse dynamics by the healing length $\xi=1/k_0\sqrt{2\Delta n}$.
We also introduce another characteristic scale  in the longitudinal direction, the nonlinear length:  
\begin{equation}
    z_{\text{NL}}=\frac{1}{k_0|\Delta n|}.
\end{equation}
It is the propagation length (or effective evolution time) above which nonlinear effect, hence photon-photon interactions, become sizeable. 
One can note that it is also the inverse of the equivalent ``chemical potential" $k_0\Delta n$.
We make the following changes of variables:  $\tilde{\mathbf{r}}_\perp=\mathbf{r}_\perp/\xi$, $\tilde{z}=z/z_{\text{NL}}$, $\tilde{\mathcal{\psi}}=\mathcal{E}/\sqrt{2I_0/\varepsilon_0 c}$ where $I_0$ is the average intensity of the field.
Then, dropping the external potential and loss term for simplicity, Eq.~\eqref{eq:NLSE_2D} becomes
\begin{equation}
    \mathrm{i}\dfrac{\partial \tilde{\mathcal{\psi}}}{\partial \tilde{z}} = \left( -\tilde{\nabla}_\perp^2 + |\tilde{\psi}|^2\right)\tilde{\psi} , 
    \label{eq:NLSE_adim}
\end{equation}
where $\tilde{\nabla}_\perp$ indicates derivative with respect to the adimensional $\tilde{r}_\perp$.
All quantities in Eq.~\eqref{eq:NLSE_adim} are now dimensionless.
One can note that the new effective dimensionless ``time" is directly equal to the nonlinear phase accumulated during propagation over a distance $z$: $\tilde{z}=z/z_{\text{NL}}=k_0\Delta n z$.
Similar adimensionalization can be done for the GPE describing a uniform BEC by defining the healing length $\xi=\hbar/\sqrt{2m\mu}$, the nonlinear time $t_{\text{NL}}=\hbar/\mu$ and $\tilde{\psi}=\psi/\sqrt{\rho_0}$ where $\mu=g\rho_0$ and $\rho_0$ is the average density of the condensate and $\mu$ is the chemical potential.

\subsection{Comparison with cold atoms}
\label{comparison}

From the adimensional form of the NLSE and the GPE, we can compare the key quantities of the two platforms to gain insight on the observable dynamics in each of them.
Let us compare the adimensional quantities $z/z_{\text{NL}}$ (or $t/t_{\text{NL}}$ in the cold atom case), and $R/\xi$ where $R$ is the extension of the fluid.

In fluids of light, if we take the example of hot rubidium vapors, we typically measure $\Delta n \sim 10^{-5} $ \citep{piekarski2021measurement}, with maximal values up to~$10^{-4}$.
Then, with $\lambda_0$=780~nm, we get $\xi\sim 16$~µm and 
$k_\xi=1/\xi\sim $ 50 mm$^{-1}$.
Hence, the typical interaction value we measure gives an optically accessible $k_\xi$. 
Indeed, we can perturb the system with transverse waves of momentum $k_\perp$ orders of magnitudes smaller or bigger than $k_\xi$, for instance with a spatial light modulator (see Section~\ref{sec:SLM} for details).
Then, the minimal value of $k_{\perp,\text{min}}$ for which we can detect a change in energy is limited by the length of the nonlinear medium $L$.
Like in the previous section, we define as an order of magnitude the $k_{\perp,\text{min}}$ value for which the perturbation gets $\pi$ phase shift after propagation: $\Omega_\text{B}(k_{\perp,\text{min}})L=\pi$. 
We obtain $k_{\perp,\text{min}}\sim10$~rad/mm $<k_\xi$.
This means that we can probe the superfluid to normal-fluid transition, which happens at $k_\xi$.
The extension $R$ of the system is the waist $w_0$ of the beam, which is usually of a few millimeters. 
Typically, we get $R/\xi \sim$100.
Since superfluid features like vortices have a typical lengthscale of  $\xi$, this shows that we can observe a significant distribution of them and study their dynamics within the extent of the fluid.\\

In BECs, the interactions are quantified by the chemical potential $\mu=g\rho$, for which a typical order of magnitude is (for 2D BECs) $\mu \sim 10^{-30}$J  \citep{desbuquois_2DBEC}. 
Consequently, $\xi \sim $ 200~nm.
As a typical BEC size we take $R\sim$20 µm, which gives $R/\xi\sim$100, so a similar order of magnitude as paraxial fluids of light in rubidium vapor.

The relevant limiting time $t$ to consider for BECs is the coherence time, which ranges from milliseconds to seconds.
This gives $t/t_{\text{NL}}\sim$10$^3$-10$^6$.
For fluids of light, the nonlinear medium length $L$ (hence the effective evolution time) typically ranges from 1~cm to 20~cm, which gives $L/z_{\text{NL}}\sim$100. 
Hence we measure significantly shorter time dynamics than atomic BECs.
In addition, we can only detect the fluids after a fixed effective evolution time $\tau=L/c$, since we cannot image inside of the nonlinear medium.
We will see in Section~\ref{sec:techniques} techniques to circumvent these issues.

\section{Review of the experimental platforms}
\label{sec:platforms}
 At the heart of the physical phenomena underlying the fluid-like behavior of light is the nonlinear response of the propagation medium. 
Depending on the experimental platform, this nonlinearity can have multiple origins.
We will focus on the three most common experimental platforms: atomic clouds, photorefractive crystals and thermo-optic liquids.
    
\subsection{Atomic clouds}

Atomic media are well known for their strongly nonlinear response to a near resonance laser field. The field of nonlinear and quantum optics in hot atomic vapors is extremely broad \citep{glorieux2023hot} with applications from magnetometry to quantum memory.
While coherent excitation with multi-levels and multi-fields could open exciting perspectives \citep{lukin2000resonant}, most of the experimental realizations today focus on single-field near-resonant excitation.
In the following we provide the derivation of the $\chi^{(3)}$ nonlinearity in atomic vapor and present how to use this system for paraxial fluids of light.
We will follow the example of the $\mathrm{D}_2$ line of a rubidium 87 vapor.
In the case of atomic vapors, the $\chi^{(3)}$ response is derived from the optical Bloch equations \citep{grynberg2010introduction,glorieux2018quantum}.

  \subsubsection{$\chi^{(3)}$ nonlinearity for a two-level atom}
   
As a first approximation, we will study the case of a two-level atom.
For the rubidium $D_2$ line, these two levels are the $5S_{1/2}$ ground state $|g\rangle$ and $5P_{3/2}$ excited state $|e\rangle$.
We will note the energy of each level $E_g=\hbar\omega_g$ (resp. $E_e=\hbar\omega_e$) and $\omega_{eg}=\omega_e-\omega_g$ the resonant frequency of this transition.
The transition is driven by a monochromatic field polarized along a unit vector $\mathbf{u}$: $\mathbf{E}(t)=\dfrac{1}{2}\mathcal{E}e^{-\mathrm{i}\omega t}\mathbf{u} + \mathrm{c.c}.$.
We write the Hamiltonian  of an atom interacting with the electric field as $\hat{H} =\hat{H}_0+\hat{W}$,  with $\hat{H}_0$ the atomic Hamiltonian and $\hat{W}$ the interaction Hamiltonian:
\begin{equation}
\begin{split}
    \hat{H}_0 &= \hbar\omega_g \ket{g}\bra{g} + \hbar\omega_{e}\ket{e}\bra{e} \sim \hbar\omega_{eg}\ket{e}\bra{e} \\
    \hat{W}(t) &= -\mathbf{D}\cdot\mathbf{E}(t)=\left(\mathbf{d}^*\ket{g}\bra{e}  + \mathbf{d}\ket{e}\bra{g} \right)\cdot\mathbf{E}(t) \\
    & = \frac{\hbar}{2}\Big( \Omega^*\ket{g}\bra{e} + \Omega\ket{e}\bra{g} \Big)\times\Big( e^{-\mathrm{i}\omega t}+e^{\mathrm{i}\omega t}\Big) ,
    \label{eq:W}
\end{split}
\end{equation}
where $\mathbf{D}$ is the dipole operator and we defined the atomic Hamiltonian up to an energy constant.
In Eq.~\eqref{eq:W}, we have introduced the Rabi frequency $\Omega=-\frac{\mu_{eg}\mathcal{E}}{\hbar}$, and the transition dipole moment $\mu_{eg}=\bra{e}\mathbf{d}\cdot\mathbf{u}\ket{g}$.
The transition dipole moment contains the internal structure of the atom and encapsulates the selection rules depending on the polarization of the electric field.
We switch to the interaction picture with respect to the Hamiltonian $\hbar\omega \ket{e}\bra{e}$ to eliminate the time dependence of the interaction term, and within the rotating wave approximation,
we obtain:
\begin{equation}
    \hat{H} = \hbar
    \begin{pmatrix}
        0 & \frac{\Omega^*}{2} \\
        \frac{\Omega}{2} & -\Delta
    \end{pmatrix}
\end{equation}
with $\Delta=\omega-\omega_{eg}$.
\noindent Then, we calculate the evolution of the density operator $\hat{\rho}=\sum_{i,j\in[e,g]}\rho_{ij}\ket{i}\bra{j} $, which is given by the master equation:
\begin{equation}
\frac{\partial\hat{\rho}}{\partial t}=-\frac{\mathrm{i}}{\hbar}\left[\hat{H}, \hat{\rho}\right]
+\sum_{\nu}\left[\hat{L}_\nu \hat{\rho} \hat{L}_\nu^\dagger -
\frac{1}{2}\{\hat{L}_\nu^\dagger \hat{L}_\nu, \hat{\rho}\}\right],
\label{eq:2level_master}
\end{equation}
where  $\hat{L}_\nu$ are the jump operators.
Here, we include a jump operator to describe spontaneous emission: $\hat{L}_\text{sp} = \sqrt{\Gamma}\ket{g}\bra{e}$, where $\Gamma$ is the transition linewidth.
We also include decoherence with $\hat{L}_\text{dec} = \sqrt{\gamma}(\ket{e}\bra{e}-\ket{g}\bra{g})$, where $\gamma = \Gamma/2+\gamma_\text{col}$, where $\gamma_\text{col}$ is the collision rate.

\noindent We can now express the medium polarization as:
\begin{equation}
\mathbf{P}=\frac{N}{V}\langle\mathbf{D}\rangle = \frac{N}{V}Tr(\rho \mathbf{D}) = \frac{N}{V}\left(\frac{\mu_{eg}\rho_{eg}}{2}e^{-\mathrm{i}\omega t} + \textrm{c.c.}\right),
\end{equation}
where $\frac{N}{V}$ is the atomic density.
Writing $P = \frac{1}{2}\mathcal{P}e^{-\mathrm{i}\omega t} + \textrm{c.c.}$, we can find the expression for the complex envelope: $\mathcal{P} = \frac{N}{V}\mu_{ge}\rho_{eg}$.
The susceptibility $\chi$ of the atoms to the electric field then writes:
\begin{equation}
    \chi = \frac{N}{V}\frac{\mu_{ge}\rho_{eg}}{\varepsilon_0 \mathcal{E}}.
\end{equation}
Solving Eq.~\eqref{eq:2level_master} for the stationary state, using the fact that the total population $\rho_{ee}+\rho_{gg}$ is conserved, finally yields:
\begin{equation}
    \chi=\frac{\alpha(0)c}{\omega_{eg}}\frac{\mathrm{i}-\frac{\Delta}{\gamma}}{1+(\frac{\Delta}{\gamma})^2+|\frac{\mathcal{E}}{\mathcal{E}_s}|^2},
    \label{eq:chi_2level}
\end{equation}
where $\alpha(0)=\alpha(\mathcal{E}=0, \Delta=0) = \dfrac{\omega_{eg}}{c}\dfrac{N}{V}\dfrac{|\mu_{eg}|^2}{\varepsilon_0 \hbar\gamma}$ is the weak-probe line-center linear absorption coefficient and $\mathcal{E}_s=\sqrt{2}\gamma \hbar/\mu_{eg}$ is the line-center saturation field. 
In the limit where $|\frac{\mathcal{E}}{\mathcal{E}_s}|^2\ll 1+(\frac{\Delta}{\gamma})^2$, we expand the expression~\eqref{eq:chi_2level} in powers of ${\mathcal{E}}/{\mathcal{E}_s}$ up to the second order and obtain:
\begin{equation}\label{eq:susceptibility_analytical}
    \begin{split}
        \chi^{(1)}&=\frac{\alpha(0)c}{\omega_{eg}}\frac{\mathrm{i}-\frac{\Delta}{\gamma}}{1+(\frac{\Delta}{\gamma})^2}, \\
        \chi^{(3)}&=-\frac{\alpha(0)c}{|\mathcal{E}_s|^2\omega_{eg}}\frac{\mathrm{i}-\frac{\Delta}{\gamma}}{\Big[1+(\frac{\Delta}{\gamma})^2\Big]^2}.
    \end{split}
\end{equation}
As we can see from the expression of the susceptibility, the first order cancels out, as we predicted from a centro-symmetric medium (here we assumed isotropy which is stronger than centro-symmetry).

    \subsubsection{Three-level system: optical pumping and transit}
\label{beyondkerr}
The two-level closed system is a simple but unrealistic model \citep{phillips_no_two_level}. 
For a better understanding of the system, we have to consider the splitting of  $5S_{1/2}$ (previously defined as the ground state) into two hyperfine levels $F=1$ and $F=2$ (for $^{87}$Rb), which are separated by $\delta_0=2\pi\cdot$6.835~GHz \citep{steck2001rubidium}.
In the case of the excited state, we can neglect the hyperfine splitting as it spans over $2\pi \cdot$496~MHz, which is comparable to the Doppler-broadened atomic linewidth $\Gamma_D=k_0\sqrt{\frac{k_BT}{m}}\sim 2\pi \cdot250$MHz for a typical vapor temperature of 140°C ($\Gamma_D$ being $1/\sqrt{e}$ times the half-width of the broadened linewidth).
We also have to take into account the transit of the atoms through the beam.
Hence we model the atoms as an open three-level system. 
The two ground states $F=1$ and $F=2$ will be noted $\ket{1}$ and $\ket{2}$, and the excited state $\ket{3}$.

In the case where the laser field only drives the $F=1$ ground state, the atoms in the excited state will spontaneously decay to both hyperfine ground states, and thus will be pumped into the other ground state $F=2$ over time (which acts as a dark state).
With a continuous laser beam, this optical pumping effect would cause the medium to become transparent at a time scale of tens of \unit{\mu\second} \citep{kaiserlens}.

However, the atomic transit balances optical pumping out. Due to atomic motion,
 atoms exit the beam in a time $\tau_\mathrm{t} = \frac{w}{v} = \sqrt{\frac{\pi mw_0^2}{8k_B T}}$ \citep{Sagle_1996} and undergo relaxation processes. 
Here $w$ is the mean path taken by an atom inside the beam, $v$ is the mean speed given by the Boltzmann distribution, and $w_0$ is the beam waist ($w\leq w_0$). 
We assume that outside of the beam the gas behaves like a reservoir, and thus retains a stable mixture of atoms in the two ground states since the energy separation of the states is below $k_B T$. 
This means that the populations in the beam are constantly replenished by the reservoir, which counterbalances the optical pumping. 
The effectiveness of the effect of transit depends on the rate $\Gamma_\mathrm{t}={1}/{\tau_\mathrm{t}}$ (mostly depending on the beam waist, and marginally on the temperature), compared to the pumping rate (mostly depending on the laser detuning and intensity).

\begin{figure}
    \centering
    \includegraphics[width=1\linewidth]{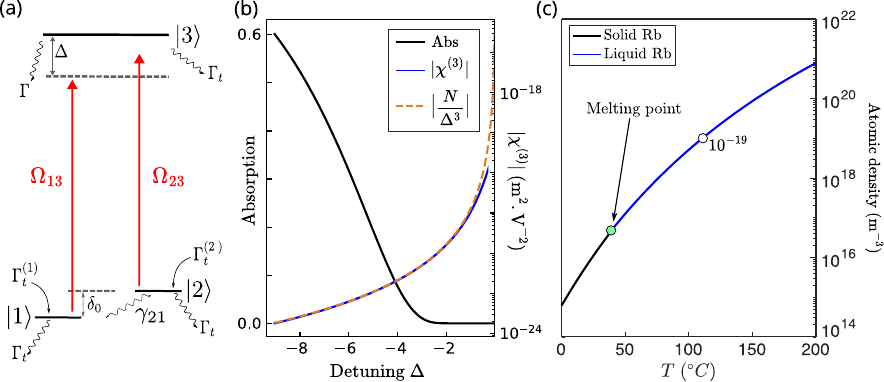}
    \caption{(a) Three-level system taken into consideration, with decay and influx rates due to spontaneous emission ($\Gamma$) and transit ($\Gamma_\mathrm{t}$).
    (b) Medium nonlinear susceptibility $\chi^{(3)}$ of the three-level model versus laser detuning $\Delta$ in blue, obtained within the three-level model.
    The dark curve shows the medium absorption spectrum for $10$ cm of propagation with an atomic density of $10^{19}\rm{\ m^{-3}}$.
    (c) Atomic density in a rubidium vapor as a function of temperature.
    }
    \label{fig:3levels}
\end{figure}

For a more complete picture, we can consider that the electric field also drives the transition from the other ground state $F=2$ to the excited state, at a detuning $\Delta+\delta_0$ as depicted in Fig.~\ref{fig:3levels}.
The hamiltonian in the interaction picture is now:
\begin{equation}
    \hat{H} = \hbar
    \begin{pmatrix}
        0 & 0 & \frac{\Omega_{13}^*}{2} \\
        0 & \delta_0 & \frac{\Omega_{23}^*}{2} \\
        \frac{\Omega_{13}}{2} & \frac{\Omega_{23}}{2} & -\Delta \\ 
    \end{pmatrix}
    ,
    \label{eq:3level_H}
\end{equation}
where $\Omega_{ij}=-\frac{\mu_{ij}\mathcal{E}_{ij}}{\hbar}$,  $\mu_{ij}$ and $\mathcal{E}_{ij}$ are respectively the Rabi frequency,  the dipole moment  and  the envelope of the driving field between the levels $\ket{i}$ and $\ket{j}$.
Spontaneous emission is modeled using Lindbladians and corresponds to relaxation of the excited population and a relaxation of the coherences:
\begin{align*}
 \hat{L}_\text{sp,32} &= \sqrt{\Gamma}\ket{3}\bra{2},\\ \hat{L}_\text{sp,31} &= \sqrt{\Gamma}\ket{3}\bra{1},\\
\hat{L}_\text{dec,32} &= \sqrt{\gamma}(\ket{3}\bra{3}-\ket{2}\bra{2}),\\ \hat{L}_\text{dec,31} &= \sqrt{\gamma}(\ket{3}\bra{3}-\ket{1}\bra{1}),
\end{align*}
where $\Gamma$ is the excited level linewidth (taken equal), and $\gamma = \frac{\Gamma}{2} + \gamma_\text{col}$ is the dephasing rate due to spontaneous emission and collisions.

Within the beam, the atoms of each level $\ket{i}$ exit the beam at a rate $\Gamma_t\rho_{ii}$ and go into the reservoir. 
We assume a memory-less process for simplicity, meaning that the atoms  in the reservoir are always taken in a statistical mixture of the two ground states.
Denoting $\rho_{11}^R$ and $\rho_{22}^R$ the populations of the states $\ket{1}$ and  $\ket{2}$ in the reservoir, the probability for an atom in state $\ket{1}$ (resp. $\ket{2}$) to enter inside the beam is $\Gamma_t^{(1)}=\Gamma_\mathrm{t}\rho_{11}^R$ (resp. $\Gamma_t^{(2)}=\Gamma_\mathrm{t}\rho_{22}^R$). 
In the absence of magnetic field, each hyperfine ground state $\ket{j}$ is degenerated $g_j=2F_j+1$ times (for $^{87}$Rb $F_1=1$, $F_2=2$) and the probability for an atom in the reservoir to be in state $\ket{j}$ is then $\rho_{jj}^R=\frac{g_j}{g_1+g_2}$.
In the example of $^{87}$Rb, $\rho_{11}^R=3/8$ and $\rho_{22}^R=5/8$.
The atoms in the beam and in the reservoir are at thermal equilibrium, and their populations are conserved: $\rho_{11}^R+\rho_{22}^R=\rho_{11}+\rho_{22}+\rho_{33}$=1.\\


Transit also causes relaxation of the coherences between the two ground states, as atoms potentially in a superposition of states $\ket{1}$ and $\ket{2}$ are replaced by atoms in a statistical mixture. This leads to the rate equations:
\begin{align*}
\left\{\frac{\partial \rho_{33}}{\partial t}\right\}_{\mathrm{transit}} &= -\Gamma_\mathrm{t}\rho_{33}, \\
\left\{\frac{\partial\rho_{22}}{\partial t}\right\}_\mathrm{transit} &= -\Gamma_\mathrm{t}\rho_{22} +\Gamma_\mathrm{t}^{(2)}, \\
\left\{\frac{\partial\rho_{11}}{\partial t}\right\}_\mathrm{transit} &= - \Gamma_\mathrm{t} \rho_{11} + \Gamma_\mathrm{t}^{(1)}, \\
\left\{\frac{\partial\rho_{12}}{\partial t}\right\}_\mathrm{transit} &= -\Gamma_\mathrm{t}\rho_{12}.
\end{align*}

\noindent The notation $\{...\}_\mathrm{transit}$ denotes the contribution of the transit in the rate equation. The full rate equation would include the $[H, \rho]$ commutator, as well as the Lindbladians.
The Rabi frequencies, decoherence and decay rates in the three-level system are represented in Figure~\ref{fig:3levels}(a).\\
We numerically solve the resulting equations to obtain $\rho_{13}$ in a steady state, to deduce $\chi = \frac{N}{V}\frac{\mu_{13}\rho_{13}}{\varepsilon_0 \mathcal{E}}$.
We fit $\chi$ with a polynomial expansion of $|\mathcal{E}/\mathcal{E}_s|$ and obtain $\chi^{(3)}$ as the coefficient of the second-order term.
The result is plotted in Figure~\ref{fig:3levels}(b), along with the linear absorption.

Interestingly, optical pumping can also be used to locally change the atomic population in the two ground states. Typically, this is done using an arrangement of two lasers: one on the $D_2$ line (the \textit{fluid} beam), and one localized one on the $D_1$ line (the \textit{potential} beam). 
The $D_1$ laser modifies locally the number of atoms visible by the $D_2$ line laser. This will lead to a local change of refractive index $\delta n(\mathbf{r})$ and therefore an effective potential for the fluid of light \citep{truscott1999optically} as shown in Section~\ref{sec:mapping}.
How to use this approach experimentally will be described in Section~\ref{sec:potential} and can be modeled by the master equations of a four-level system \citep{aladjidi2023full}.

\subsubsection{Corrections to the Kerr model}
A simple Kerr model with a real linear susceptibility and a nonlinear susceptibility restricted to a local \( \chi^{(3)}\) response does not capture the full picture of alkali vapors. 
Other important mechanisms have to be considered, which are notably influenced by the effect of atomic transit discussed previously.

\paragraph{Non-locality}
A phenomenon in hot atomic vapors that can significantly modify the effective nonlinearity is non-locality.
Due to atomic motion, the nonlinear response at a given point in the fluid is not solely determined by the local density but rather by the density within a characteristic interaction radius.
Taking into account a non-local optical response, the nonlinear term in Eq.~\eqref{eq:NLSE_2D} is modified into:
\begin{equation}
k_0\frac{n_2^E}{n_0}\int_S \mathrm{d}\mathbf{r'_\perp}  G(\mathbf{r_\perp} - \mathbf{r'_\perp})|\mathcal{E}(\mathbf{r'_\perp},z)|^2\mathcal{E}(\mathbf{r'_\perp},z),
   \label{eq:NLSE_nl_2D_rescaled}
\end{equation}
where $G$ is the non-local response function in real space, and $S$ is the beam surface.
This non-local response has been observed and identified as a crucial stabilizing factor in the study of collapse instabilities \citep{azam2021dissipation}. 
Two processes compete, the mean distance before atoms de-excite $l_\text{b}$, and the mean distance between atomic collisions  $l_\text{col}$. If $l_\text{b} \le l_\text{col}$, which is valid until $\sim$155°C for $^{87}$Rb  vapors \citep{skupin}, the atomic motion can be considered ballistic. This model leads to a non-local kernels in Fourier space:
\begin{align}
    G(k_\perp) = \frac{\sqrt{\pi}}{\Gamma}\frac{e^{\frac{1}{(k_\perp l_\text{b})^2}}}{k_\perp l_\text{b}} \text{Erfc} \left( \frac{1}{k_\perp l_\text{b}} \right),
    \label{eq:non-local_kernels}
\end{align}
where $l_b = \frac{1}{\Gamma}\sqrt{\frac{2k_bT}{m}} \sim \text{7.6}$ \unit{\micro \meter}. 
Accounting for the effects induced by non-locality amounts to convolving the electric field $\mathcal{E}(\mathbf{r_\perp},z)$ at each $z$ with the function $G$.  
In other terms, non-locality can be understood as a smoothing of the nonlinear interactions.
For example, we will see in Section~\ref{sec:thermo} that this effect is predominant in case of thermo-optic nonlinearity \citep{vocke2016role}.
Importantly, the presence of non-locality modifies the Bogoliubov dispersion relation as:
\begin{equation}
    \Omega_\text{B}(k_\perp) = \sqrt{ \frac{ k_\perp^2}{2k_0} \left( \frac{k_\perp^2}{2 k_0} + k_0 \Delta n \times G(k_\perp) \right) }.
    \label{eq:bogoliubov_non-local}
\end{equation}
It is actually possible to exploit this effect to observe an inflection in the dispersion relation and it has been predicted that this effect could lead to self-organization of superfluid light similar to supersolidity \citep{maucher2016self}. 

\begin{figure}
    \centering
    \includegraphics[width=1\linewidth]{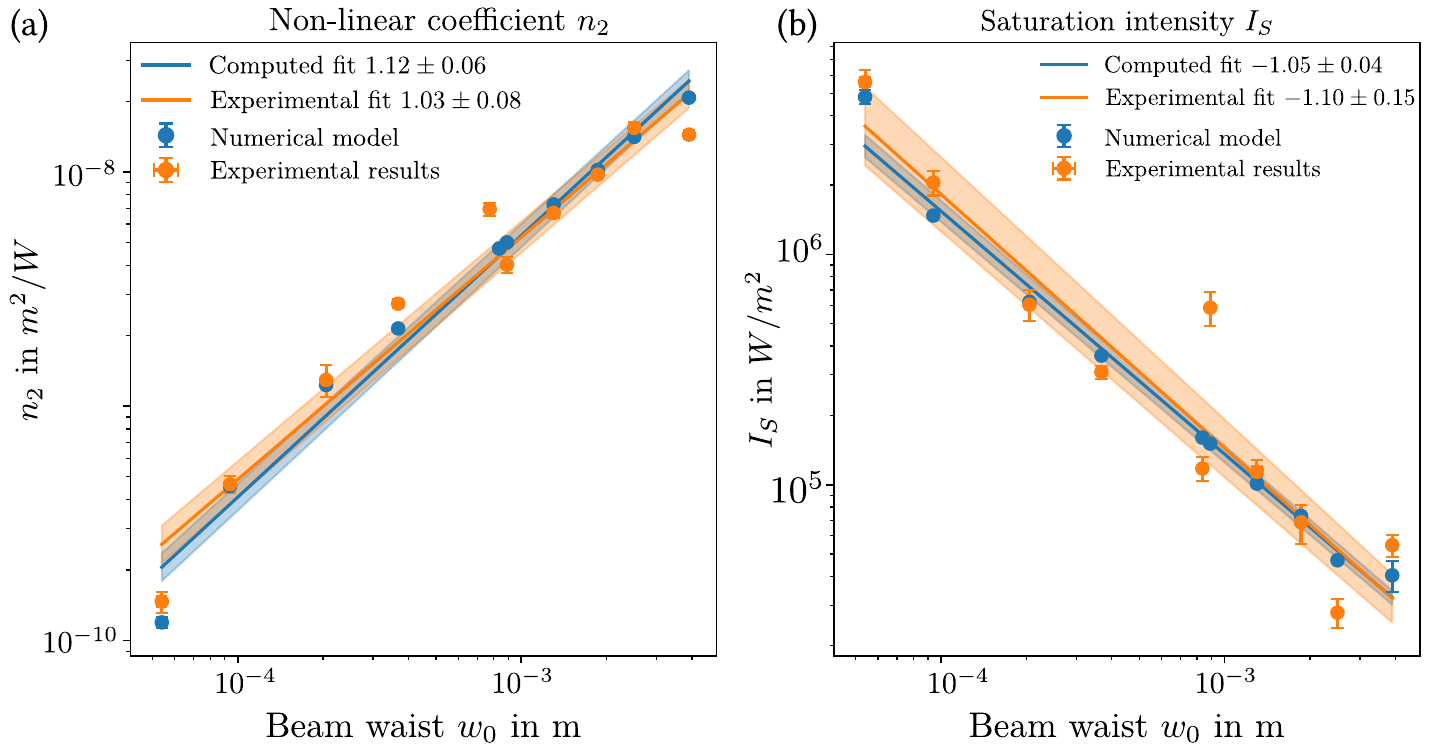}
    \caption{Dependency of $n_2$ (a) and $\Isat$ (b) as a function of the beam waist $w_0$. Orange dots are experimental data in rubidium vapor. Blue dots are numerical model using \cite{aladjidi2024nlse}. 
    The fits provide the scalings described in the main text.
    }
    \label{fig:n2_Isat_waist}
\end{figure}

\paragraph{Saturation}
At large optical intensity, the atomic medium becomes saturated and the nonlinear response is not of purely \( \chi^{(3)}\) type \citep{mccormick2004saturable}. 
A quintic term \( \chi^{(5)}\), with opposite sign, starts to play a role and reduces the nonlinearity.
This effect (actually accounting for all higher order terms) is known as saturation.
For a two-level system at resonance, the saturation intensity is expressed as
\begin{equation}
    I_\text{sat}(0)=\dfrac{\pi \hbar \omega_0 \Gamma}{3\lambda^3} ,
    \label{eq:Isat_resonant}
\end{equation}
where $\omega_0$, $\lambda$ and $\Gamma$ are respectively the pulsation, the wavelength and the linewidth of the atomic transition. 
For a field detuned by $\Delta=\omega-\omega_0$, the saturation intensity depends on the detuning as $I_\text{sat}(\Delta)=I_\text{sat}(0)(1+\frac{\Delta^2}{\Gamma^2})$.
Besides, due to optical pumping and to the transit time of the atoms in the beam, the beam size  plays a crucial role in this saturation intensity which can therefore be noted $I_\text{sat}(\Delta,w_0)$.
As described in \cite{aladjidi2023full} and shown in Fig.~\ref{fig:n2_Isat_waist}, $I_\text{sat}(\Delta,w_0)$ varies as $1/w_0$ while $n_2$ is proportional to $w_0$.
Instead of describing the refractive index $n$ as $n=n_0+n_2I$, a more realistic description incorporates the saturation term in the form:
\begin{equation}
    n=n_0+\dfrac{n_2(\Delta,w_0) }{n_0}\dfrac{I}{1+I/\Isat(\Delta,w_0)}.
\end{equation}
A direct consequence of this saturation term is that the speed of sound obtained within the Bogoliubov perturbation theory (see Section~\ref{sec:bogo}) gets modified as predicted by \cite{huynh2024}:
\begin{equation}
    c_\text{s}=\dfrac{\sqrt{n_2 I}}{1+{I}{/I_\text{sat}}} .
\end{equation}

\paragraph{Linear losses}
The linear susceptibility also includes an imaginary part, that leads to absorption. 
As discussed already, this is of critical importance when setting the trade-off on the laser detuning between high losses (near resonance) and low nonlinearity (far from resonance).
Within the two-level atoms model this is given by a linear imaginary term in the NLSE that scales as $N/\Delta^2$.
As we will see in Section~\ref{sec:atomic_vapors}, in the same model, $\chi^{(3)}$ scales as $N/\Delta^3$.
This would imply that for a constant transmission ($N/\Delta^2=$ constant), it is advantageous to reduce both $N$ and $|\Delta|$ to maximize \( \chi^{(3)}\). 
This simple argument is actually not true in practice for warm vapors since these scalings are only valid far detuned from resonance.
As already mentioned, a description of rubidium vapor should consider (at least) three atomic levels and account for the transit time of atoms through the laser beams, which affects the average atomic response due to optical pumping \citep{borde1976saturated}.
The optimal detuning and temperature are therefore dependent on the beam’s transverse size and should be optimized for each experiment.

\subsubsection{How to use alkali vapors ?}
\label{sec:atomic_vapors}
Hot atomic vapors provide a versatile and tunable platform for realizing paraxial fluids of light. As we will see, it has been used in many different experiments but it requires a careful implementation to take full benefit of this versatility.
Importantly, in the simple case of a single near resonant laser, the nonlinearity arises from the third-order susceptibility $\chi^{(3)}$, which, as derived in Eq.~\eqref{eq:susceptibility_analytical}, follows the scaling at large detuning:
\begin{equation}
\Re\left(\chi^{(3)}\right) \propto \frac{N}{\Delta^3}.
\end{equation}
This scaling is particularly favorable for experimental control. 

The atomic density is directly related to the vapor pressure of the atomic species, which follows the Clausius-Clapeyron relation. 
For rubidium, an accurate empirical model is provided by \cite{nesmeianov1963vapor}, giving the equilibrium vapor pressure \(P_{\text{Rb}}\) in torr as a function of temperature:
\begin{equation}
\log_{10} P_{\text{Rb}} = 15.88 - \frac{4529}{T} + 0.000586 \, T - 2.99 \log_{10} T,
\end{equation}
for temperatures satisfying $T \geq 312$ K (liquid phase).
Below the melting point, a different set of coefficients applies \citep{alcock1984vapour}, but for typical experimental conditions, this formula is sufficient.
The atomic density $N$ is then given by the ideal gas law:
\(
N = \frac{P_{\text{Rb}}}{k_B T},
\)
where $k_B$ is the Boltzmann constant. 
Fig. \ref{fig:3levels}(c) presents the atomic density of a rubidium vapor for typical experimental values.

To develop intuition about the atomic density, we provide reference values at two typical temperatures. 
At $T = 300$ K, the rubidium vapor pressure is approximately $P_{\text{Rb}} \sim 10^{-7}$ Pa, corresponding to an atomic density of $N$ below $10^{16}$ atoms/m$^3$. In contrast, at $T = 400$ K, the density increases dramatically to $N \sim 10^{19}$ atoms/m$^3$.  
This represents a change of more than three orders of magnitude over a temperature range of just 100 K, highlighting the strong exponential dependence of $N$ on $T$. 
This sensitivity makes temperature a highly effective control parameter for tuning the interaction strength in fluids-of-light experiments.\\

Conversely, the inverted cubic dependence on detuning (see Figure~\ref{fig:3levels}b) offers a flexible means to control both the magnitude and sign of the nonlinearity \citep{mccormick2003nonlinear}.
A red-detuned laser ($\Delta<0$) will lead to repulsive photon-photon interaction (self-defocusing), while a blue-detuned one ($\Delta>0$) will induce attractive photon-photon interaction (self-focusing).
However, at room temperature and above, Doppler broadening plays an important role.
At room temperature ($\sim 300$ K), the Doppler-broadened linewidth for rubidium is about 500 MHz. This broadening determines the range of detunings where nonlinear effects are accessible while keeping absorption low. 
In practice, detunings in the range of 1–10 GHz from resonance provide a good compromise between nonlinearity and transmission. A detuning too close to resonance ($|\Delta|< 2\pi \cdot 1$~GHz) leads to excessive absorption, limiting the effective propagation length, while a detuning too far from resonance ($ |\Delta|> 2\pi \cdot10$~GHz) reduces the nonlinear response significantly \citep{mccormick2004saturable}.

When setting up a fluid of light experiment in a rubidium vapor, a careful vapor cell design has to be taken into account. Standard glass cells can be used, but for high-temperature operation ($> 100^\circ$C), quartz cells with 2-sided-coated windows with optical contact are preferable.
A well designed oven is also required to ensure an homogeneous temperature to avoid rubidium deposition on the windows and air current convection outside of the cell that would distort the imaging system.\\

Among alkali vapors, rubidium (Rb) is widely used due to its convenient spectral properties and strong optical nonlinearities. 
Both isotopes, $^{87}$Rb and $^{85}$Rb, have well-characterized D1 and D2 transitions and exhibit relatively high vapor pressures at moderate temperatures, making them accessible for room-temperature experiments.
For fluids of light experiments, rubidium offers many advantages.
The transition dipole moment of rubidium ($\mu_{ge}$) is relatively large compared to heavier alkali, leading to a stronger $\chi^{(3)}$ response compared to cesium. 
Moreover, rubidium is chemically stable in standard sealed cells, unlike potassium or lithium, which require specialized handling due to their higher reactivity.
While sodium and cesium have also been considered for fluids of light and nonlinear optics, their absorption and vapor pressure characteristics often make them less practical. 
The lower vapor pressure of sodium requires higher temperatures to reach the same optical depth as rubidium, whereas cesium’s hyperfine structure complicates its resonance structure for single-laser experiments.

\subsection{Photorefractive crystals}

\begin{figure}
    \centering
    \includegraphics[width=1\linewidth]{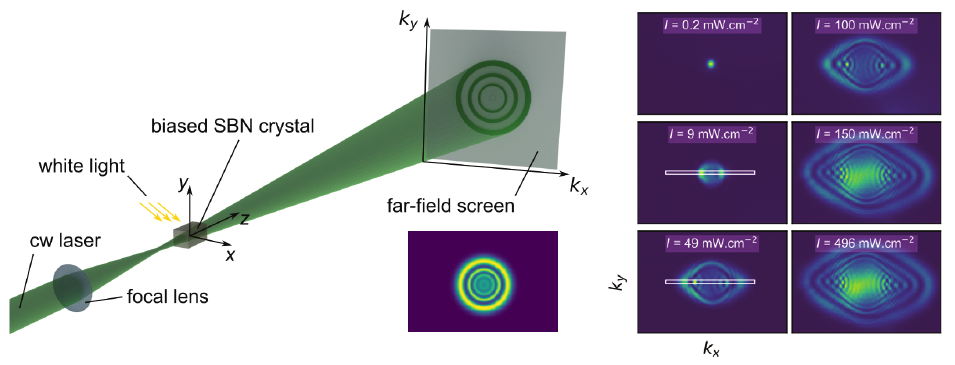}
    \caption{Adapted from  A. Boughad et al., Optics express \textbf{27}, 30360 (2019). Copyright by Optica Publishing Group (2019) \citep{boughdad2019anisotropic}. Typical photorefractive setup for the measurement of the nonlinearity using diffraction pattern in the far field. It can be seen that the response is strongly anistropic as described in \cite{boughdad2019anisotropic}.}
    \label{fig:photorefractive}
\end{figure}

Photorefractive crystals are historically the first platforms for studying fluid-like optical behavior thanks to their strong nonlinear response to optical fields. 
Unlike atomic vapors, where nonlinearity arises from resonant interactions with discrete atomic transitions, photorefractive crystals rely on charge transport mechanisms to induce a dynamic refractive index modulation. 
This fundamental difference affects their experimental accessibility, control parameters, and practical limitations.

The photorefractive effect is based on the redistribution of charge carriers in response to an incident optical field. 
When a laser beam illuminates the crystal, charge carriers (electrons or holes) are excited from donor states into the conduction band. 
These mobile carriers then migrate due to diffusion and drift effects, the latter being enhanced by an externally applied electric field. Eventually, the carriers become trapped in spatially displaced locations, leading to the formation of a space-charge field. 
This field modulates the refractive index of the material via the linear electro-optic (Pockels) effect, creating an intensity-dependent change in refractive index:

\begin{equation}
\Delta n(I) = -\frac{1}{2} n_0^3 r_{33} E_{\text{sc}}(I),
\end{equation}
where $n_0$ is the linear refractive index of the bulk crystal, $r_{33}$ is the electro-optic coefficient, and $E_{\text{sc}}(I)$ is the space-charge electric field, which depends on the local intensity $I$.
In a biased photorefractive crystal with an external electric field $E_{\text{ext}}$ applied along the c-axis of the crystal where the  nonlinear effects are the strongest \citep{lukasiewicz2008strontium}, the space-charge field can be expressed in terms of an electrostatic potential $\phi$ \citep{kukhtarev1978holographic}. 
The total potential consists of a light-induced electric potential $\phi_0$ and an external bias term $-|E_{\text{ext}}|x$, assuming that the c-axis is aligned with the propagation axis $z$  \citep{barsi2015model,boughdad2019anisotropic}. 
This results in an anisotropic description of the photorefractive effect, where the space-charge field is given by:
\begin{equation}
E_{\text{sc}} = E_{\text{ext}} - \nabla \phi_0.
\end{equation}
The term $E_{\text{sc}} - E_{\text{ext}} = -\nabla \phi_0$ is  referred to as the screening field, which is generated by the redistribution of charge carriers due to light excitation. 
Under steady-state conditions, where the photovoltaic effect is negligible and the drift effect dominates over diffusion in charge carrier migration, the potential equation for the electrostatic field $\phi_0$ is given by \cite{boughdad2019anisotropic}:
\begin{equation}
\nabla^2_\perp \phi_0 + \nabla_\perp \ln(1 + \tilde{I}) \cdot \nabla_\perp \phi_0 = |E_{\text{ext}}| \frac{\partial \ln(1 + \tilde{I})}{\partial_z } ,
\end{equation}
where $\nabla_\perp$ represents the gradient in the transverse plane, and $\tilde{I} = I/I_{\text{sat}}$ is the normalized intensity, accounting for the ratio between thermal and photoinduced excitations. In experimental setups, an incoherent white light background is typically used to introduce a controlled contribution to the thermal excitation, effectively modifying the saturation intensity and the overall nonlinear response of the system.

In the isotropic approximation, the space-charge field simplifies to:
\begin{equation}
E_{\text{sc}} = \frac{E_{\text{ext}}}{1 + \tilde{I}},
\end{equation}
and the nonlinear refractive index variation follows:
\begin{equation}
\Delta n(I) = \Delta n_{\text{max}} \frac{I}{I + I_{\text{sat}}},
\label{eq:photoref}
\end{equation}
where $\Delta n_{\text{max}} = \frac{1}{2} n_0^3 r_{33} E_{\text{ext}}$ is the maximum refractive index change.
This expression is identical to the one obtained for atomic vapor in the presence of saturation.
This means that once the choice of materials (e.g., SBN, LiNbO$_3$) has been made the only parameter controlling the nonlinear response is the external electric field $E_{\text{ext}}$ which enhances charge drift, increasing the nonlinear response.
This clearly simplifies the parameters space exploration compared to complex energy levels in atomic vapors but also limits the tunability.

In practice, the refractive index modification photo-induced in a biased nonlinear photorefractive
crystal can be accurately controlled by means of a background incoherent illumination and an external electric field.
One method to quantify this effect relies on measuring the diffraction patterns of the laser beam propagating through the medium and undergoing spatial self-phase modulation as shown in Fig. \ref{fig:photorefractive} (left).
It has been shown in \cite{boughdad2019anisotropic}, that the response is anisotropic in the stationary regime leading to asymmetric diffraction patterns as shown in Fig. \ref{fig:photorefractive} (right).

Finally, an important tool for investigating quantum fluids of light is the ability to imprint and manipulate external potentials within nonlinear optical media.
In photorefractive crystal, one effective approach involves optical induction, where an externally applied  potential is created semi-permanently using an extra laser beam.
The refractive index modulation induced by the optical induction persists due to the photorefractive screening nonlinearity, which stabilizes the imprinted potential under an applied electric field. 
This allows for creating a lattice potential for the light \citep{fleischer2003observation} or a localized potential such as a defect or a barrier \citep{michel2018superfluid} as proposed initially by \cite{PhysRevE.55.2835}.

\subsection{Thermo-optic media}
\label{sec:thermo}

Thermo-optic media exhibit a refractive index that depends on local temperature variations caused by the absorption of laser light.
When a beam propagates through such a medium, part of its energy is absorbed, creating local heating. 
This heating modifies the refractive index profile, resulting in nonlinear optical effects.
The induced refractive index change is expressed as:
\begin{equation}
\Delta n = \beta \Delta T,
\end{equation}
where \( \beta = \frac{\partial n}{\partial T} \) is the thermo-optic coefficient, characteristic of the material, and \( \Delta T \) is the induced temperature variation.

Under conditions of low absorption, heat transport predominantly occurs in the transverse plane (\( x,y \)), while the axial direction \( z \) can be neglected due to the relatively small temperature gradients along the propagation direction (\(\nabla_{\perp} T \gg \nabla_z T\)). In the steady-state approximation, the temperature distribution is governed by the two-dimensional heat diffusion equation:
\begin{equation}
\nabla^2_{\perp} \Delta T = -\frac{\alpha}{\kappa} I(\mathbf{r}_{\perp}),
\end{equation}
where \( \alpha \) is the linear absorption coefficient, \( \kappa \) is the thermal conductivity, and \( I(\mathbf{r}_{\perp}) \) is the optical intensity distribution in the transverse plane.

Given \( \Delta n = \beta \Delta T \), the heat diffusion equation transforms into a two-dimensional Poisson equation for the refractive index variation:
\begin{equation}
\nabla^2_{\perp} \Delta n = -\frac{\alpha \beta}{\kappa} I(\mathbf{r}_{\perp}).
\end{equation}
This equation can be solved using Green’s function methods, yielding the formal solution \citep{vocke2017analogue}:
\begin{equation}
\Delta n(\mathbf{r}_{\perp}) = \gamma \int \mathrm{d}^2 \mathbf{r}'_{\perp} R(\mathbf{r}_{\perp}-\mathbf{r}'_{\perp}) I(\mathbf{r}'_{\perp}),
\end{equation}
where \( R(\mathbf{r}) \) is the non-local response function related to the Green’s function of the heat diffusion problem and $\gamma$ is a constant that links the two quantities.
Physically, the function \( R(\mathbf{r}) \) represents how heat generated at one location spreads out spatially, smoothing intensity variations and producing a refractive index profile extending beyond the immediate area of absorption. 

\begin{figure}
    \centering
    \includegraphics[width=1\linewidth]{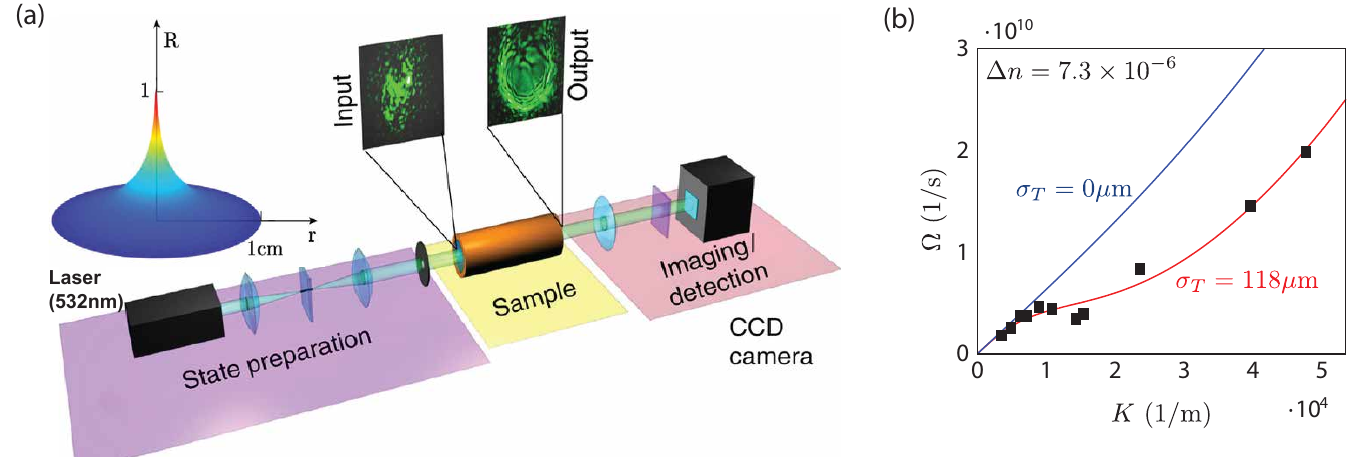}
    \caption{(a) Typical experimental set-up  for thermo-optic medium from \cite{xu2015coherent}.
    The beam from a CW laser is sent into the nonlinear material. The sample is a cylindrical tube filled with a solution of methanol and graphene nanoscale flakes. Top left: the typical non-local response of the medium obtained by numerical simulation of the diffusion equations from \cite{vocke2016role}.
    (b) Non-local Bogoliubov dispersion. Data are adapted from \cite{vocke2017analogue} and correspond to a non-local length of $\sigma_\mathrm{T}=118\mu$m (red). The local dispersion relation is plotted in blue for comparison. $K$ is the transverse wavevector $k_{\perp}$ and $\Omega$ is the Bogoliubov pulsation as defined in Eq.~\eqref{eq:bogo_disp} and applied for non-local nonlinearity following Eq.~\eqref{eq:bogoliubov_non-local}.
    $\Omega$ is converted from m$^{-1}$ to s$^{-1}$ with the rescaling used in Eq.~\eqref{eq:NLSE_2D_rescaled}.}
    \label{fig:faccio}
\end{figure}

To calculate the constant $\gamma$, a physical model should be defined for the system boundaries.
In a system with infinite boundaries, an exact solution for \( R(\mathbf{r}) \) is logarithmic and thus physically unrealistic since real thermal diffusion always occurs within finite dimensions \citep{vocke2017analogue}. 
Therefore, practical scenarios involve finite boundaries and typically result in exponentially decaying spatial profiles $R(\mathbf{r}) \propto e^{-\mathbf{r}/\sigma_\mathrm{T}}$ as in \cite{bar2013nonlocality} or Gaussian profiles $R(\mathbf{r})\propto e^{-\mathbf{r}^2/\sigma_\mathrm{T}^2}$ as in \cite{briedis2005ring}, with \( \sigma_\mathrm{T} \) defining the characteristic non-local length scale as shown in Fig.~\ref{fig:faccio}(a). 

Numerical calculations incorporating realistic beam profiles (such as Gaussian beams) and finite boundary conditions refine these approximations, providing insights into the spatial extent of thermal nonlinearities. The characteristic width \( \sigma_\mathrm{T} \) is typically estimated numerically, and its accurate determination is essential to describe nonlinear optical effects accurately.

Thermal nonlinearities have been studied in various liquids, with methanol and methanol-based solutions (such as methanol/graphene mixtures) being commonly used due to their strong negative thermo-optic coefficient. 
Unlike electronic Kerr nonlinearities, thermal nonlinearities exhibit response times on the millisecond to second scale. 
This allows to tune the effective photon-photon interaction as a function of time \citep{vocke2015experimental}.
The experimental setup typically involves a bulk liquid sample contained in a cylindrical cell with optical windows, through which a laser beam propagates, see Fig.~\ref{fig:faccio}(a). 
The absorption of light induces a temperature gradient, leading to a refractive index profile that mediates photon-photon interactions over macroscopic distances.
In this case the Bogoliubov dispersion relation is strongly modified by the non-locality. 
This has been measured by \cite{vocke2015experimental} and analyzed by \cite{fontaine2020interferences}. 
A typical dispersion relation is shown in Fig.~\ref{fig:faccio}(b).
\subsection{Comparison between the platforms}
The different platforms are compared in the Table~2. We have selected typical experimental values from \cite{aladjidi2023full}, \cite{boughdad2020fluids} and \cite{vocke2017analogue}.
\label{tab:nonlinear-systems}
\begin{table}[h]
\centering
\begin{tabular}{lccc}
\hline
\textbf{Quantity} & \textbf{Atom. Vap.} & \textbf{Phot. Ref.} & \textbf{Therm. Opt.} \\
\hline
$n$ & $\sim1$ & $\sim2.4$ & $\sim1.33$ \\
$\lambda$ (nm) & 780 & 633 & 532 \\

$n_2$ (m$^2$/W) & $\sim10^{-6}$ & $\sim10^{-7}$ & $\sim10^{-10}$ \\
$\Delta n_{\text{max}}$ & $5.10^{-5}$ & $10^{-4}$ & $10^{-6}$ \\

$L$ (mm) & 10–200 & 5–20 & 10–200 \\
Transverse size  & 5 mm& 2 mm& 10 mm\\
$\xi$ ($\mu$m) & 10-20 & 3-6 & 50-100 \\
$z_\text{NL}$ (mm) & 2-20 &  0.5-5 & 50-500  \\
$\tilde{z}=L/z_\text{NL}$  &  $\leq$100 & $\leq$40 & $\leq$5 \\
\hline
\end{tabular}
\caption{Typical parameters for the three nonlinear optical systems.  Values are adapted from from: \citep{aladjidi2023full}, \citep{boughdad2020fluids} and \citep{vocke2017analogue}.}
\label{tab:nonlinear-systems}
\end{table}

\section{Experimental and numerical techniques for fluids of light}
\label{sec:techniques}
This section aims to describe the  techniques used for paraxial fluids of light experiments and numerical simulations. 
Several techniques are derived from nonlinear and quantum optics while others have been adapted from ultracold atomic quantum gases.

\subsection{Arbitrary state generation: SLM's and DMD's.}
\label{sec:SLM}

The first step in fluid-of-light experiments is the preparation of the initial state by controlling both its intensity and phase.
A spatial light modulator (SLM) enables the generation of arbitrary classical optical states, providing independent manipulation of phase and amplitude. 
Typically, an SLM modifies the phase of an incoming optical field pixel by pixel. 
Amplitude shaping is achieved indirectly by encoding a suitable wavefront pattern, such as a blazed grating, allowing the desired intensity profile to be isolated through spatial filtering in the Fourier plane. 
This filtering is commonly implemented with an iris placed at the focal point of a 4f-system.
Several technical details can be found for SLM and digital micromirror devices (DMD) in the work of \cite{WavefrontshapingTutorials} and in the review written by \cite{gauthier2021dynamic}.

We note $\mathcal{E}(x,y) = a(x,y)e^{\mathrm{i}\phi(x,y)}$ the target field, with the normalized amplitude $a \in [0, 1]$.
The general problem is to find the phase function $\Psi[a,\phi](x,y)$ to be displayed on the SLM, that generates the target field after filtering in the Fourier plane.
After reflection on the SLM, the incident electric field is multiplied by the transfer function:
\begin{equation}
\begin{aligned}
       h[a,\phi](x,y) &= e^{\mathrm{i}\Psi[a,\phi](x,y)} \\
       &= \sum_j h_j[a,\phi](x,y) =  \sum_j c_j (a)e^{\mathrm{i}j\phi}, 
\end{aligned}
    \label{eq:slm_fourier_series}
\end{equation}
with $h_j[a, \phi]$ the $j^{th}$ term in the Fourier series expansion, and $c_k(a)$ the Fourier series coefficients. The $x$ and $y$ variables have been dropped in the interest of brevity.  With these definitions, the condition on $c_1$ to retrieve the target field in the first order of the Fourier plane is $c_1(a) \propto a$.
As shown by \cite{Arrizon:07}, $c_1(a) \leq a$,
which sets a maximum for the light efficiency of reconstruction $\eta = \frac{\|c_1\|^2}{\sum_j \|c_j\|^2} \leq \|a\|^2$, with $\|f\|^2 = \frac{1}{M}\sum_{x,y}|f(x,y)|^2$ the power of the signal $f$, and $M$ the number of pixels.

A solution for $\Psi(a, \phi)$ that fulfills the condition $c_1(a) = a$ and minimizes errors due to the discrete nature of the SLM is given by \cite{Arrizon:07}:

\begin{equation}
    \Psi[a, \phi] = \phi + J_0^{-1}(a)\sin{\phi},
\end{equation}
where $J_0^{-1}$ is the inverse of the Bessel function.
In Fourier space, the transfer function becomes $H[a,\phi](u,v) = \sum_jH_j[a,\phi](u,v)$ with $H_j[a,\phi](u,v) = \mathcal{F}\left[c_j(a)e^{\mathrm{i}j\phi}\right](u,v)$. 
There is no mathematical reason for $H_1$ to be spatially separated from the other $H_{j\neq 1}$ terms, and thus it cannot be spatially filtered in general. 
To improve filtering, it is required to add a phase gradient of spatial frequency $u_0$: $\Psi(a, \phi) \rightarrow \Psi(a, \phi + u_0x)$, such that the transfer function reads: 

\begin{equation}
\begin{aligned}
    h[a,\phi + u_0 x](x,y) 
    &= \sum_j h_j[a,\phi](x,y) \, e^{\mathrm{i} j u_0 x} \\
    \implies 
    H[a,\phi](u,v) 
    &= \sum_j H_j[a,\phi](u - j u_0).
\end{aligned}
\end{equation}
The phase gradient has the effect to spatially shift each $H_j$ term by $ju_0$, allowing the spatial filtering of the target signal present in $H_1$.
However, the SLM pixel density limits how large $u_0$ can be chosen, as the gradient's discretization error increases the fewer pixels are used to encode it.
The bandwidth of the signal must be smaller than $\frac{u_0}{2}$ to avoid any overlap between the diffraction orders, and ensure the correct reconstruction of the target field. Therefore the pixel density fixes the bandwidth of the target signal that can be reconstructed.

\begin{figure}[] \centering \includegraphics[width=1\linewidth]{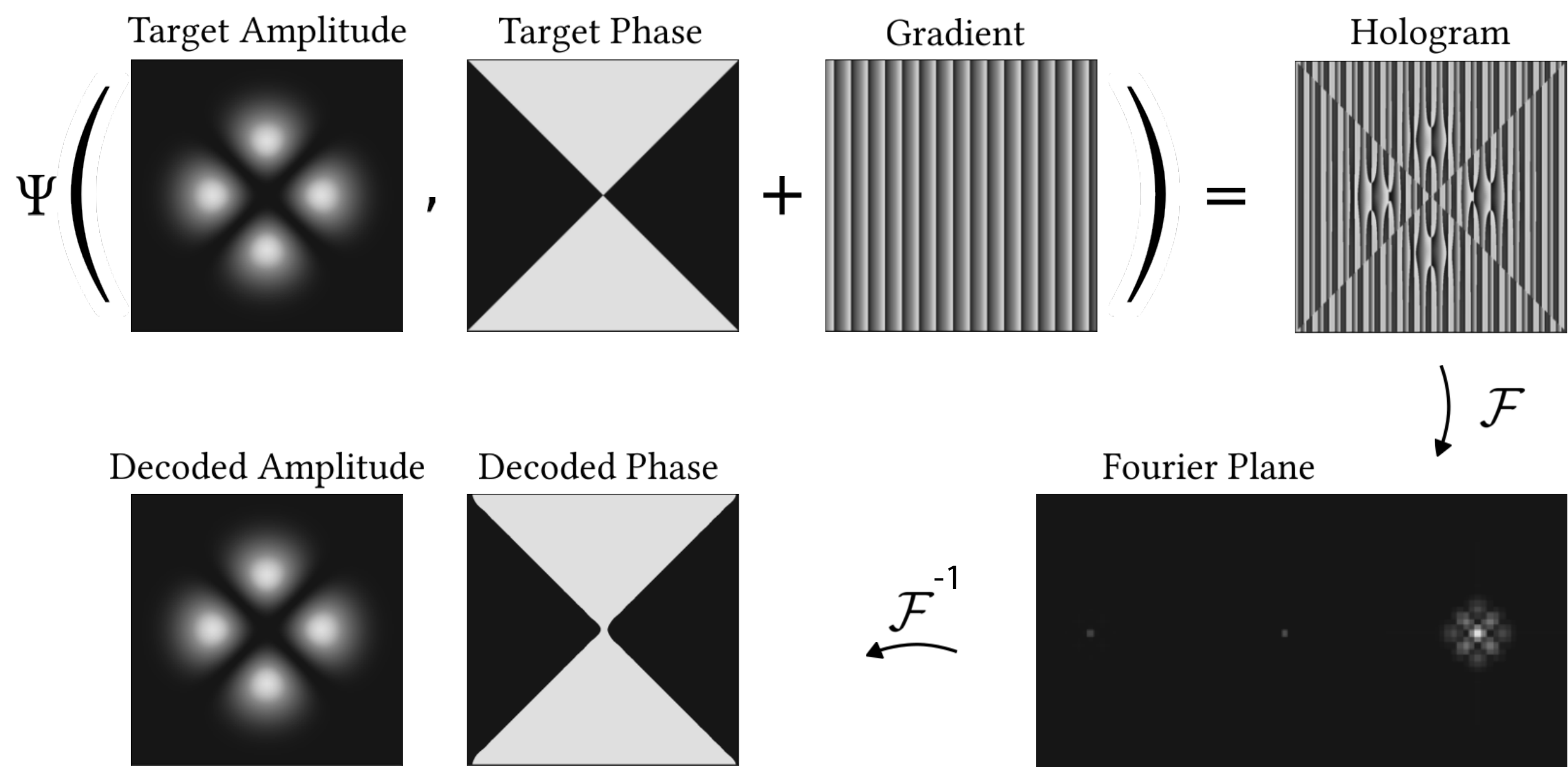} \caption{Laguerre Gauss field encoding and decoding using \cite{Arrizon:07} hologram.} \label{fig:slm_tuto} \end{figure}

As an example, a Laguerre Gauss field has been encoded and decoded using the aforementioned hologram in Fig.~\ref{fig:slm_tuto}. The phase of the target field has a high bandwidth, as it varies instantaneously from one quadrant to the next. As can be seen from the reconstructed field, the phase has been slightly distorted near the image center, a sign that the gradient is not steep enough to capture high spatial frequencies.

The main advantage of spatial light modulators lies in their ability to achieve high diffraction efficiencies, thanks to their relatively large modulation depth (8 to 16 bits). However, their refresh rates are limited, typically from $10$~Hz to a few hundred $\rm{Hz}$.

An alternative to SLMs is the Digital Micromirror Device (DMD). These devices, composed of arrays of reflective micromirrors, are traditionally used for amplitude-only modulation. However, various techniques have been developed to enable wavefront shaping by grouping micromirrors into macropixels \citep{Popoff2024,Gauthier:16}. While DMDs can reach refresh rates of several tens of kilohertz, this comes at the cost of significantly reduced light efficiency.

\subsection{Potential engineering}
  \label{sec:potential}

For the various platforms presented in this paper, it is possible to create an arbitrary potential  by locally modifying the linear refractive index. The ability to engineer such potentials enables the realization of complex potential landscapes. Here, we describe how attractive and repulsive potentials can be implemented in the three platforms.

In warm atomic vapors, one can exploit the hyperfine structure of the atomic levels to optically pump atoms into a \textit{dark} state, i.e. a state that does not interact with the probe field, and therefore reduce the effective density of atoms. 
By spatially shaping the intensity of a control beam, one can create spatially dependent optical pumping, thus inducing a local decrease in the optical susceptibility.
This results in a local change in the refractive index and therefore a potential seen by the fluid of light. 
A lower index creates a repulsive potential, while an increase can create an attractive one.
This idea has been implement in rubidium vapors to create optically written waveguides \citep{truscott1999optically, andersen2001light}.
More complicated schemes using the multi-level structure of alkali atoms have also been proposed \citep{zhang2013control}, in particular using electromagnetically induced transparency (EIT) \citep{sun2006optical, vudyasetu2009all, sheng2015observation}.
Interestingly, to keep the shape of the potential constant over $z$ while reaching a high resolution, non-diffractive Bessel beams have been used \citep{fontaine2019attenuation}.

In photorefractive crystals, the refractive index change arises from the spatial redistribution of photo-generated charge carriers. 
When the medium is illuminated with an intense beam, it becomes saturated as shown in Eq.~\eqref{eq:photoref}, and its response to additional light is reduced. 
A localized ``writing" beam can therefore induce a spatially varying saturation leading to a refractive index change determined by the local intensity distribution. 
Typically, in the self-defocusing regime (achievable in strontium-barium-niobate crystals with appropriate external field orientation), this results in a reduction of the refractive index in regions of high intensity, which can be conceptualized as the creation of repulsive optical potentials for light \citep{boughdad2019anisotropic,boughdad2020fluids}. 
Conversely, realizing attractive potentials (i.e., an increase in refractive index in bright regions) can be achieved through specific configurations such as inverting the applied external electric field in materials like SBN to enter a self-focusing regime, or by utilizing background incoherent illumination to manipulate the effective nonlinearity.

In thermo-optic liquid it is actually possible to include an object inside the medium (it is a liquid), therefore a macroscopic modification of the refractive index can be induced by choosing an object with a different index than the liquid.
This approach has been followed by \cite{vocke2016role}.

\subsection{Off-axis interferometry: phase measurement}
\label{sec:off-axis}

\begin{figure}[h]
    \centering
    \includegraphics[width=0.9\linewidth]{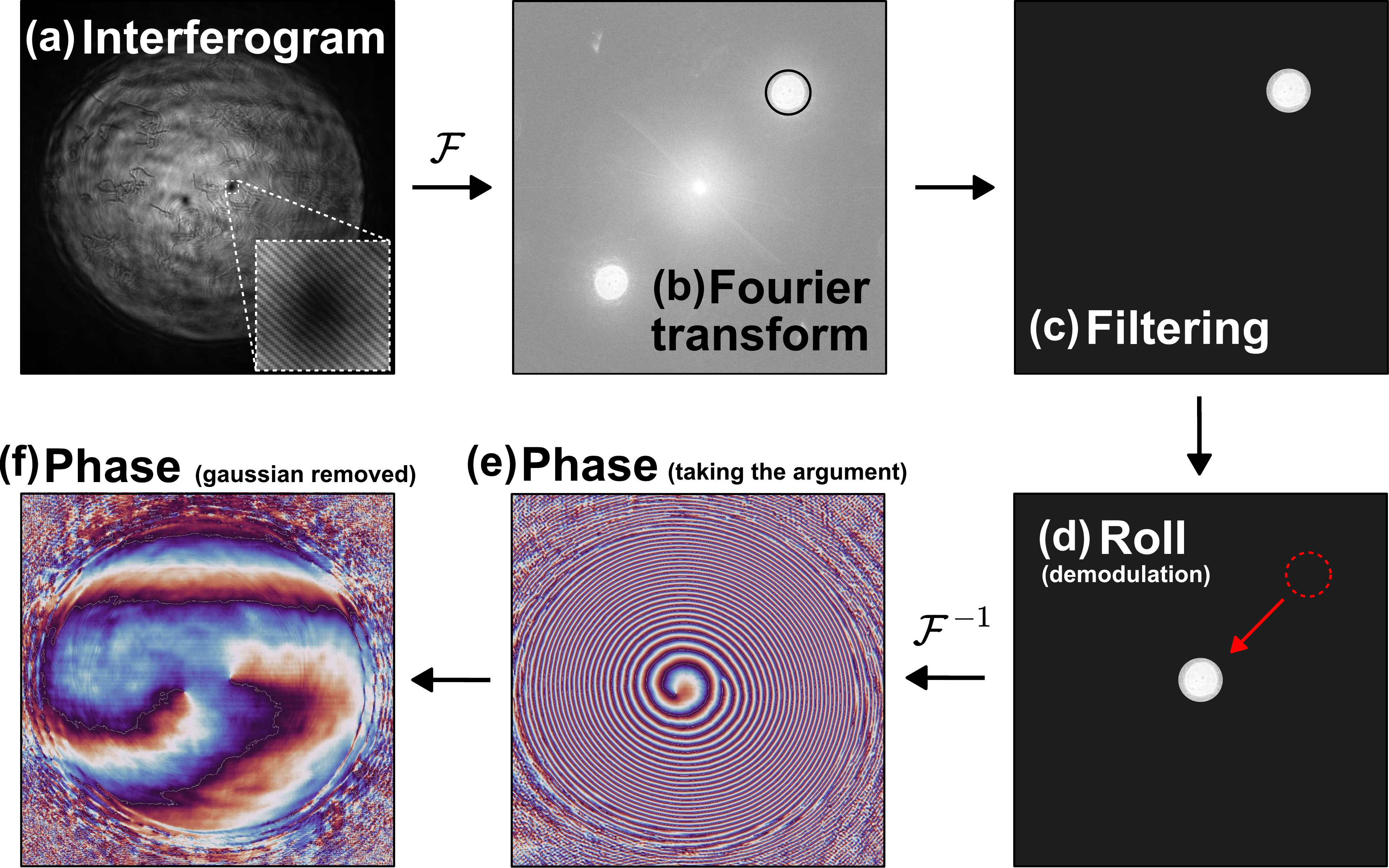}
    \caption{Off-axis interferometry. (a) Intensity measured by a camera. (b) Real part of the 2D Fast Fourier transform. (c) Selection of one of the sidebands. (d) Demodulation to remove the $\mathbf{k}'_\perp$ component. (e) Argument of the 2D Inverse Fast Fourier transform. (f) Extra-step to flatten the phase. }
    \label{fig:phase_tuto}
\end{figure}
A considerable advantage of fluids of light compared to ultracold atomic gases is the simple access to the phase of the field, using interferometric method like off-axis interferometry \citep{fienup1982phase,cuche1999simultaneous}.
This technique is widely in optics since it gives fast reconstruction (possibly above 30~Hz) with high resolution. \\
The signal beam $E_s$, which propagated through the nonlinear medium, is overlapped with a reference beam $E_r$.
The resulting interference pattern is imaged on a camera and can be expressed as follow:

\begin{equation}
  \begin{split}
    &I_{\rm{cam}}(\textbf{r}_\perp)\propto|E_s(\textbf{r}_\perp)
    e^{\mathrm{i}\phi(\textbf{r}_\perp)}+E_r(\textbf{r}_\perp)e^{\mathrm{i}\textbf{k}'_\perp\textbf{r}_\perp}|^2=\\
    &
    \underbrace{I_s(\textbf{r}_\perp)+I_r(\textbf{r}_\perp)}_{\textrm{DC part}}+
    \underbrace{E_s^*E_re^{-i(\phi-\textbf{k}'_\perp\textbf{r}_\perp)}+E_sE_r^*e^{i(\phi-\textbf{k}'_\perp\textbf{r}_\perp)}}_{\textrm{modulated part}}
  \end{split}
\end{equation}
where  $\textbf{k}'_\perp$ is the transverse wavevector of the reference beam with respect to the signal.
The phase of $E_r$ is ignored as it is set to not vary with $r_\perp$.
In order to demodulate the signal, we take the Fourier transform of this expression, giving:
\begin{equation}
  \begin{split}
    &\tilde{I}_{\rm{cam}} (\mathbf{k}_\perp) = \underbrace{\tilde{I}_\text{s} (\mathbf{k}_\perp) +
     \tilde{I}_\text{r} (\mathbf{k}_\perp)}_{\textrm{DC part}} +\\
    &
    \mathcal{F} \left[E_s e^{\mathrm{i}\phi (\mathbf{r}_\perp)} \right] (\mathbf{k}_\perp) *
    \left\{\tilde{E}_r(\mathbf{k}_\perp - \mathbf{k}'_\perp) \right\} +\\
    &\mathcal{F} \left[E_s e^{-\mathrm{i}\phi (\mathbf{r}_\perp)} \right] (\mathbf{k}_\perp) * \left\{\tilde{E}_r(\mathbf{k}_\perp + \mathbf{k}'_\perp) \right\}.
  \end{split}
\end{equation}
The phase can be extracted from the sidebands, but the difficulty is that the convolution product is hard to invert.
However, if one uses a very large reference beam compared to the signal beam, its Fourier transform will be much narrower than the one of the signal.
We can then approximate the Fourier transform of the reference by a Dirac function, meaning that the convolution product will simply shift the signal by $\textbf{k}'_\perp$ in the Fourier plane. 
\noindent In this scenario, the camera intensity in the Fourier domain becomes:
\begin{equation}
  \begin{split}
    &\tilde{I}_\text{cam} (\mathbf{k}_\perp) \simeq \underbrace{\tilde{I}_\text{s} (\mathbf{k}_\perp) +
     \tilde{I}_\text{r} (\mathbf{k}_\perp)}_{\textrm{DC part}} +\\
    &\underbrace{
    \mathcal{F} \left[E_s e^{\mathrm{i}\phi (\mathbf{r}_\perp)} \right] (\mathbf{k}_\perp+\mathbf{k}'_\perp)+
    \mathcal{F} \left[E_s^* e^{-\mathrm{i}\phi (\mathbf{r})} \right] (\mathbf{k}_\perp-\mathbf{k}'_\perp)}_{\textrm{modulated part (sidebands)}}.
  \end{split}
\end{equation}

We then spatially filter the Fourier plane in order to recover the information carried by the  sidebands.
This is done with a band-pass filter $\tilde{T}(\mathbf{k}_\perp)$  around one of the first sideband term, and by shifting the signal in Fourier domain by  $-\mathbf{k'}_\perp$ to get rid of the off-axis term.
By applying an inverse Fourier transform, we recover:
\begin{equation}
  E_s e^{\mathrm{i}\phi (\mathbf{r}_\perp)} * \mathcal{F}^{-1}\left[\tilde{T}\right](\mathbf{r}_\perp).
\end{equation}
If the band-pass filter $\tilde{T}(\mathbf{k}_\perp)$ has a circular shape, its inverse Fourier transform, $\mathcal{F}^{-1}\left[\tilde{T} \right]$, corresponds to an Airy function. 
This implies that the recovery process is not perfect, as it degrades the spatial resolution of the reconstructed field.
We select the largest possible region around the satellite peak in the Fourier plane which still excludes the zeroth order.
This fixes the choice of the relative angle between signal and reference 
 $\mathbf{k'}_\perp$ so that the satellite peak is (approximately) positioned at the center of a quadrant of the Fourier plane \citep{carpenter2022digholohighspeedlibrary}.
This phase reconstruction process is depicted in Fig.~\ref{fig:phase_tuto}. For this example, we used as an illustration $\mathbf{k'}_\perp=\frac{\pi}{d} \left( \frac{1}{\sqrt{2}} , \frac{1}{\sqrt{2}}\right)$, where $d$ is the effective camera pixel pitch.
Using adequate numerical implementations, this process can run at several hundreds of $\unit{Hz}$ for high resolution images, allowing for fast data acquisition, as well as live imaging of the full field and its derived quantities.

\subsection{Velocity decomposition and kinetic energy spectrum}

\begin{figure}[h]
    \centering
    \includegraphics[width=1\linewidth]{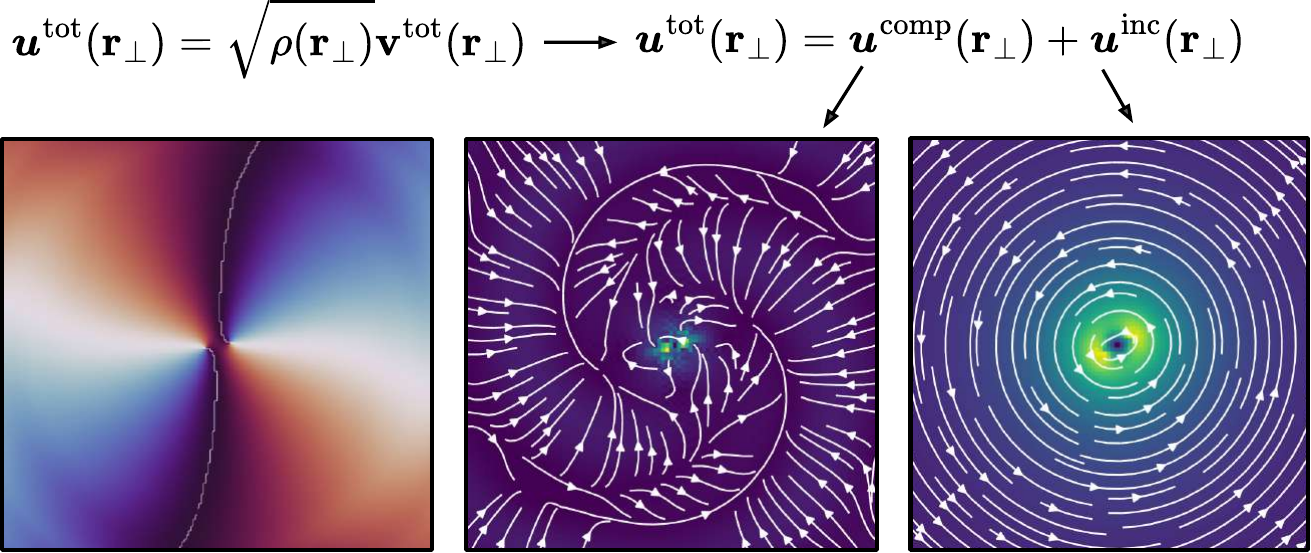}
    \caption{Velocity decomposition for two vortices of same sign. 
    The phase map is used to compute the total velocity $\mathbf{v}^{\mathbf{tot}}$ in order to achieve the Helmholtz decomposition. 
    Each component gives the velocity contribution of the acoustic waves (compressible) and vortices (incompressible).
    }
    \label{fig:helmholtz_tuto}
\end{figure}

As explained in Section~\ref{sec:hydro}, the velocity field of the paraxial fluid of light is defined as the phase gradient in the transverse plane $\mathbf{v}=\dfrac{c}{k_0}\mathbf{\nabla}_\perp \phi$.  
As we have shown in the previous section, the spatially resolved phase can be obtained using interferometric techniques.

To analyze these velocity fields, we need to reconstruct the total velocity from the experimentally measured phase map. 
The total velocity field $\mathbf{v}^{\text{tot}}(\mathbf{r})$ is derived from the spatial gradients of the phase after careful unwrapping along both spatial axes to avoid phase discontinuities to appear. 
The unwrapped phases along the $x$ and $y$ directions are noted $\phi'_x$ and $\phi'_y$, respectively. 
The total velocity field is then expressed as the combination of the gradient components along each spatial axis:

\begin{equation}\label{eq:v_tot}
\mathbf{v}^{\text{tot}}(\mathbf{r}_\perp)
= \frac{c}{k_0}\left(\frac{\partial \phi'_x(x,y)}{\partial x}\ ,\  \frac{\partial \phi'_y(x,y)}{\partial y}\right).
\end{equation}

In the absence of phase singularities, this effectively gives a zero curl: $\nabla\times \mathbf{v} = 0$. 
However this property does not hold in the case of the density-weighted velocity $\boldsymbol{u}^{\text{tot}}(\mathbf{r}_\perp) = \sqrt{\rho(\mathbf{r}_\perp)}\mathbf{v}^{\text{tot}}(\mathbf{r}_\perp)$.
By introducing this quantity, it becomes possible to distinguishing between the compressible (irrotational) and the incompressible (rotational) components of the velocity field. 
These components are defined by the Helmholtz decomposition:
\begin{equation}
    \label{u_tot}
    \boldsymbol{u^{tot}}(\mathbf{r}_\perp)=\underbrace{\boldsymbol{\nabla}\theta(\mathbf{r}_\perp)}_{\textrm{compressible}} + \underbrace{\boldsymbol{\nabla}\times\textbf{A}(\mathbf{r}_\perp)}_{\textrm{incompressible}},
\end{equation}
where $\theta(\mathbf{r}_\perp)$ is a scalar potential and $\textbf{A}(\mathbf{r}_\perp)$ a vector potential.
In momentum space, the Helmholtz decomposition of the velocity field is given by:
\begin{equation}
\boldsymbol{\tilde{u}}^{\text{tot}}(\mathbf{k}_\perp)
= \mathrm{i}\mathbf{k}_\perp \tilde{u}_{\theta}(\mathbf{k}_\perp)
+ \mathrm{i}\mathbf{k}_\perp\times\boldsymbol{\tilde{u}}_{A}(\mathbf{k}_\perp),
\end{equation}
with
\begin{equation}
\tilde{u}_{\theta}(\mathbf{k}_\perp) = \frac{\mathbf{k}_\perp\cdot\boldsymbol{\tilde{u}}^{\text{tot}}(\mathbf{k}_\perp)}{\mathrm{i}|\mathbf{k}_\perp|^2},
\quad
\boldsymbol{\tilde{u}}_{A}(\mathbf{k}_\perp) = \frac{\mathrm{i}\mathbf{k}_\perp\times\boldsymbol{\tilde{u}}^{\text{tot}}(\mathbf{k}_\perp)}{|\mathbf{k}_\perp|^2}.
\end{equation}
By performing inverse Fourier transforms, these components are retrieved in real space:
\begin{equation}
\begin{split}
\boldsymbol{\nabla}\theta(\mathbf{r}_\perp) &= \mathcal{F}^{-1}\left[\mathrm{i}\mathbf{k}_\perp \tilde{u}_{\theta}(\mathbf{k}_\perp)\right],\\
\boldsymbol{\nabla}\times\boldsymbol{A}(\mathbf{r}_\perp) &= \mathcal{F}^{-1}\left[\mathrm{i}\mathbf{k}_\perp \times \boldsymbol{\tilde{u}}_A(\mathbf{k}_\perp )\right].
\end{split}
\end{equation}
To simplify the computation, the incompressible velocity component is often obtained by directly subtracting the compressible part from the total velocity \citep{panico2023onset}. 
This approach allows for a clear separation of the compressible and incompressible contributions to the fluid’s kinetic energy \citep{baker2023turbulent}.
An example of this velocity decomposition is shown in Fig.\ref{fig:helmholtz_tuto} for a vortex pair of same sign.

 \subsection{Vortices and topological charge detection}

 \begin{figure}[h]
    \centering
    \includegraphics[width=1\linewidth]{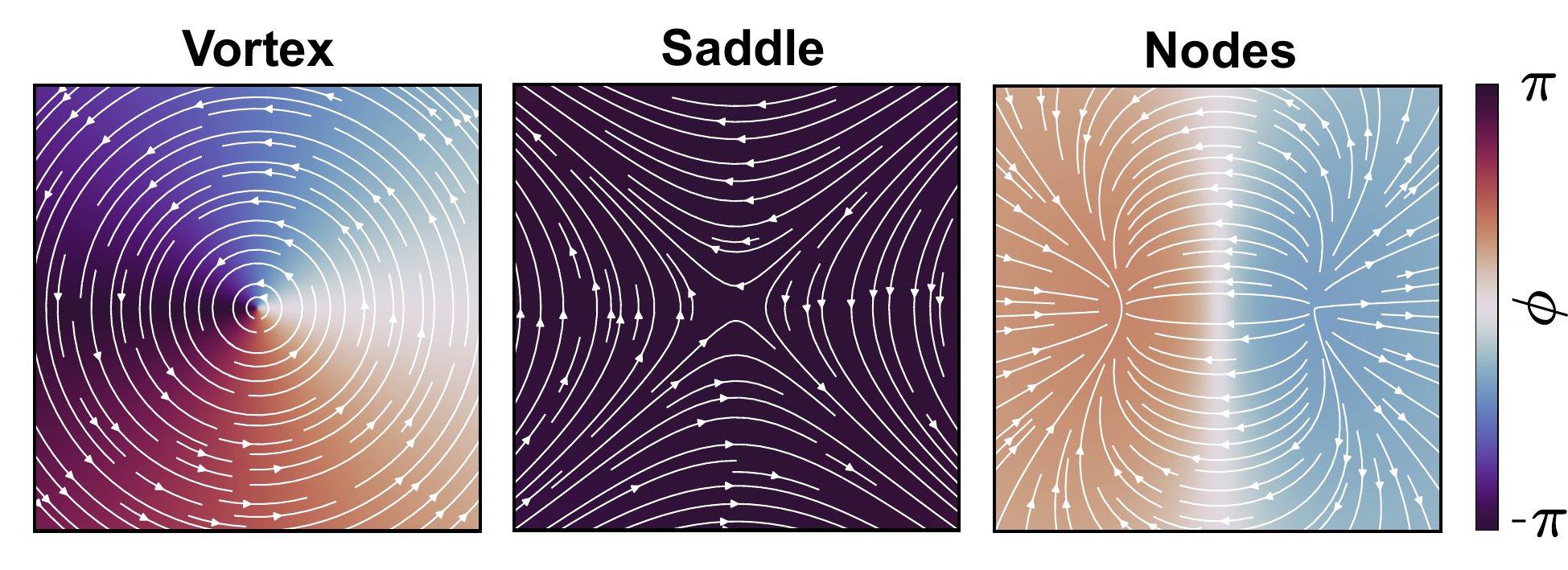}
    \caption{Topological charge zoology.
    Phase and associated velocity streamplot for a vortex (left), a saddle point (middle) and two nodes (right).}
    \label{fig:topological_charges}
\end{figure}

Having a direct access to the phase in fluids of light allows for unambiguous identification of topological objects.
Vortices are defined by a variation $\Delta \phi$ of the phase of the wavefunction (or the phase of the electric field envelope) by a multiple of $2\pi$ due to the uniqueness of the condensate wavefunction (or electric field envelope) \citep{pethick2008bose}:
\begin{equation}
    \Delta \phi = \oint \nabla\phi\cdot \textrm{d}l = 2\pi \ell,
\end{equation}
with $\ell$ an integer.
This formula gives the circulation $\Gamma_c$ around a closed contour $\mathcal{C}$:
\begin{equation}
    \Gamma_c = \oint_\mathcal{C} \mathbf{v}\cdot\textrm{d}l = \frac{\hbar}{m}2\pi\ell = \frac{h}{m}\ell.
\end{equation}
The quantization of the circulation (and thus of the velocity) in units of $h/m$ was first proposed by \cite{onsager_statistical_1949} in superfluid liquid $^4\textrm{He}$.
We will use these properties to identify numerically vortices and anti-vortices (opposite rotation or $\ell<0$) in a fluid of light

\begin{figure}[]
    \centering
    \includegraphics[width=0.85\linewidth]{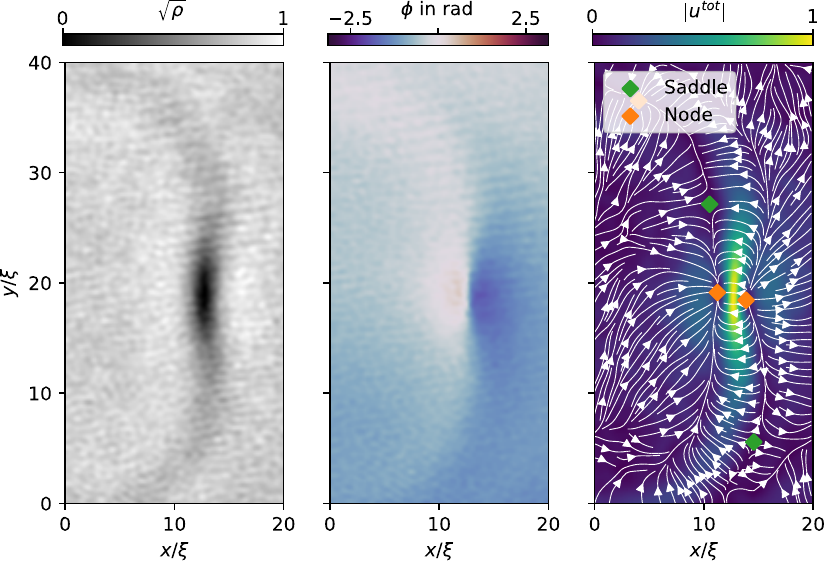}
    \caption{Two opposite sign vortices have merged into a Jones-Roberts soliton.
    Its density (left) and phase (middle) give the signature of a localized soliton.
    The associated total velocity field (right) allows to detect the remaining topological charges, given by the Poincaré index computed with the circulation of polar angle $\mathbf{v^{tot}}$.}
    \label{fig:jrs}
\end{figure}

There is no known analytical formula to describe a generic vortex in a quantum gas. 
However, in the approximation of an infinite medium without an external potential, \cite{bradley_energy_2012} demonstrated that the structure of a charge $\ell = 1$ vortex can be determined by finding the function that minimizes the system’s energy.
In this case, one can approach the vortex profile using:
\begin{equation}
\Psi_v (\mathbf{r}) = \sqrt{\rho_0} \frac{\mathbf{r_\perp} e^{-\mathrm{i} \theta}}{\sqrt{\mathbf{r_\perp}^2 + (\Lambda^{-1} \xi)^2}},
    \label{vortex_solution}
\end{equation}
where $\mathbf{r_\perp}$ is the radial distance, $\xi$ the healing length, $\rho_0$ the average density and $\Lambda \sim 0.8249 $ is a constant determined numerically.

Experimentally we use the quantization of the circulation to detect and track the topological charge of single charged vortices
as shown in Fig.\ref{fig:topological_charges} 
by numerically computing the circulation of the phase map:
\begin{equation}
    \rm{C_v} = \frac{1}{2\pi}\oint_\mathcal{C} d\phi = 0,\pm 1.
\end{equation}

Moreover, recent studies \citep{congy2024, panico2024} have observed the presence of other topological points, such as saddle points and nodes, characterized by their Poincaré index \citep{poincare1988}:  
\begin{equation}\rm{C_P} = \frac{1}{2\pi}\oint d\theta_v= 0,\pm 1,\end{equation}
where $\theta_v$ is the polar angle of $\mathbf{v^{tot}}$, given by $\theta_v = \rm{atan}(\mathbf{v_y}/\mathbf{v_x})$.
All these topological points are schematically described on Fig.~\ref{fig:topological_charges}

As an experimental illustration, in the scenario depicted in Fig.~\ref{fig:jrs}, where two vortices of opposite sign have merged into a Jones-Roberts soliton. 
\cite{baker2025observation} demonstrated that the opposite phase circulations vanish, leaving only topological features such as saddles and nodes in the velocity field, as described by \cite{congy2024}.

\subsection{Static structure factor }
\label{sec:ssf}

A typical observable in order to characterize the density response function \citep{dalfovo1999theory} of a system is the static structure factor.
It describes the density fluctuations distribution in momentum space and gives insight into the correlation properties of the system.
It has been used in atomic quantum gases to study the effect of interaction quenches in multiple experiments 
\citep{hung2011extracting,chen2021observation,landig2015measuring}.

In order to describe the fluctuations of the system, we introduce the annihilation operator $\hat{a}_\mathbf{k}$ for the photon modes $\mathbf{k}$ and we define the density fluctuation operator as:
\begin{equation}\label{eq:delta_rho_hat_definition}
  \delta\hat{\rho}_{\mathbf{k}_{\perp}}(z) = \int\mathrm{d}\mathbf{q}
  \hat{a}_{\mathbf{k}_{\perp}+\mathbf{q}}^\dagger\hat{a}_\mathbf{k}.
\end{equation}

\noindent The definition of the static structure factor is then given by:
\begin{equation}\label{eq:s_k_definition}
  S(\mathbf{k}_{\perp}) = \frac{1}{N}\langle{\delta\hat{\rho}_{\mathbf{k}_{\perp}}^2 -
    |\langle\delta\hat{\rho}_{\mathbf{k}_{\perp}}\rangle|^2}\rangle,
\end{equation}
where $N$ is the photon flux depending on laser power, beam cross-section and integration time.
The static structure factor is thus simply the variance of density fluctuations in momentum space.

In order to measure it experimentally, we adopt the following approach. We send light through the cell, integrate the intensity on a camera over a given time and then measure the shot to shot fluctuations and compute the variance.
In order to limit losses, there should be as little optical elements between the output of the cell and the camera sensor, whose quantum efficiency is already limited.
To this extent, it is required to send short pulses of light in order to avoid camera saturation.

Data is then processed as follows.
We first compute the average density $\langle{\rho(r_{\perp})}\rangle_\mathrm{N_\text{rep}}$ and then compute the density fluctuations $\delta\rho(r_{\perp})=\rho(r_{\perp})-\langle{\rho(r_{\perp})}\rangle_\mathrm{N_\text{rep}}$.
From there, we calculate the density fluctuations spectrum as $\delta\rho(k_{\perp})=\mathcal{F}\left(\rho(r_{\perp})-\langle{\rho(r_{\perp})}\rangle_\mathrm{N_\text{rep}}\right)$
and finally, compute the variance
        $S(k_{\perp})=\frac{1}{\mathrm{N_\text{rep}}}\mathrm{Var}_\mathrm{N_\text{rep}}\left[\delta\rho(k_{\perp})\right]$, 
where the $\mathrm{N_{\rm rep}}$ subscript represents the averaging over the experiment realizations.
In case of cylindrical symmetry of the system, it is possible to improve the signal further by  calculating an azimuthal average of the signal.
Finally, the static structure factor could then be Fourier transformed back into a real space density-density correlation and expressed as the second order correlation function $g^{(2)}(\mathbf{r_\perp}, \mathbf{r'_\perp}, z)$.

\subsection{Analogue Bragg spectroscopy}

A typical experimental technique to probe the excitation spectrum and density fluctuations in quantum gases is Bragg spectroscopy~\citep{stamper1999excitation,steinhauer2002excitation}. 
In atomic systems, this method relies on inducing density modulations through short optical pulses and detecting scattered particles.
This technique have been extended to paraxial fluids of light, using direct inspiration from the atomic quantum gases approach.

\begin{figure}[]
    \centering
    \includegraphics[width=1\linewidth]{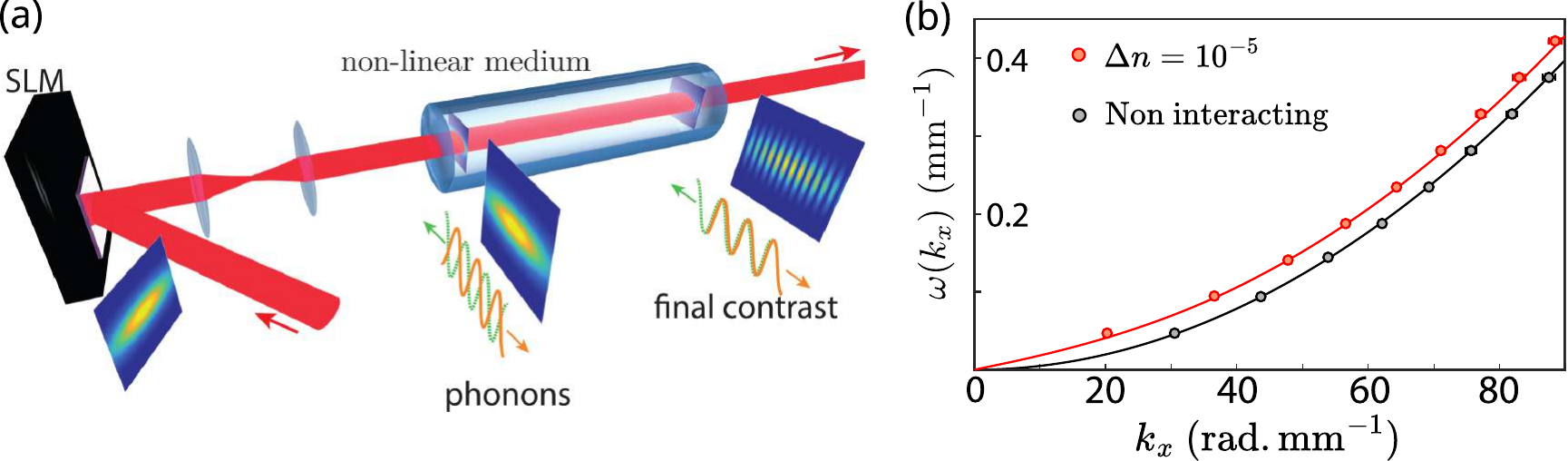}
    \caption{(a) Sketch of the Analogue Bragg spectroscopy setup. Counter propagating phonons are injected along the $x$ axis at the input plane $z=0$ using a SLM. Contrast is measured in the final plane at $z=L$ for various values of the phonon wavevector.
    (b) Experimental dispersion relation for the non-interacting and nonlinear case, respectively shown by the dark and red curves.
    More details are in \cite{piekarski2021measurement}.
    }
    \label{fig:Bragg}
\end{figure}

The implementation of analogue Bragg spectroscopy in a fluid of light involves imprinting a sinusoidal phase modulation onto the photon fluid using wavefront shaping via a spatial light modulator (SLM)~\citep{piekarski2021measurement}. 
The analogue of a short Bragg pulse is a pair of counterpropagating phonon-like excitations characterized by opposite wave vectors $\pm k_x$. 
These excitations are imprinted at the input of the medium and then interfere during propagation through the nonlinear medium, producing a measurable standing wave pattern in the photon density as shown in Fig. \ref{fig:Bragg}(a). 
Experimentally, this modulation depth is kept sufficiently small to prevent altering the medium’s nonlinear refractive index and thus stay in the Bogolioubov perturbative regime.

A central application of this technique is the high-resolution measurements of the dispersion relation.
Experimentally, the contrast of the perturbation density  at the exit plane of the medium is recorded as a function of transverse wave vector $k_x$.
At the input of the medium, the initial contrast is set to maximum for all $k_x$.
Thus, by identifying the wavevectors corresponding to successive extrema of contrast at the end of the medium and knowing the medium's length $L$, the dispersion relation $\omega(k_x)$ can be extracted through the relation $\omega(k_x)=p\pi/L$, where $p$ is an integer corresponding to all successive extrema as shown in Fig. \ref{fig:Bragg}(b).
Additionally, this technique allows for the measurement of the static structure factor~\citep{piekarski2021measurement}.

This technique can be extended to two fluids Bragg spectroscopy in presence of miscible binary mixture and to temporal domain by phase modulation in time using an electro-optic modulator rather than in space with the SLM.

\subsection{Effective time propagation}

A key limitation of paraxial fluids of light lies in the fixed length of the nonlinear medium, which imposes a fixed evolution time since it is not possible to image within in a nonlinear medium (except in the specific configuration of \cite{ford2024measurement}).

In Eq.~\eqref{eq:NLSE_adim}, we introduced a dimensionless form of the NLSE by rescaling the transverse coordinates with the healing length $\xi = \dfrac{1}{k_0\sqrt{2|\Delta n|}}$ and the propagation direction with the nonlinear length $z_{\text{NL}} = \dfrac{1}{k_0|\Delta n|}$. 
In this framework, changing $\tilde{z} = z/z_{\text{NL}}$ can be interpreted as an effective temporal evolution. 
However, since the physical propagation distance $z$ is fixed by the length $L$ of the medium, direct control over $\tilde{z}$ via $z$ is not possible.

Fortunately, $z_{\text{NL}}$ depends on the optical intensity $I$ and the nonlinear refractive index $n_2$, both of which can be tuned. 
By adjusting $I$ or $n_2$, one can then effectively control the evolution time $\tilde{z}$ while still imaging the output plane of the medium. 
To remain consistent within the dimensionless framework, the transverse coordinates must also be rescaled by the healing length, which itself varies with $I$ and $n_2$.

\begin{figure}
    \centering
    \includegraphics[width=1\linewidth]{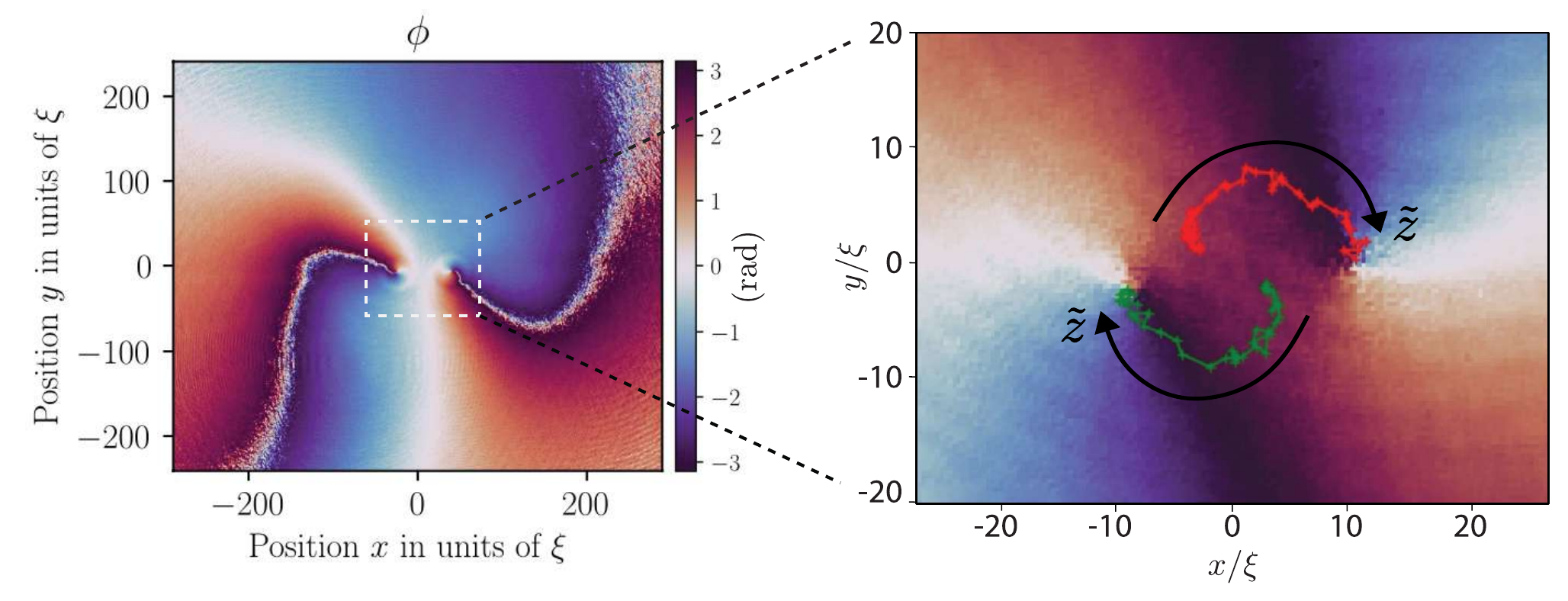}
    \caption{Left: Phase of the initial state containing two same-sign vortices. Right: Zoomed-in view of the central region at $\tilde{z}=120$.
    The red and green curves represent the trajectories of the two vortices as a function of $\tilde{z}$. The data were obtained in a rubidium vapor.}
    \label{fig:vortex}
\end{figure}

Figure~\ref{fig:vortex} illustrates this approach. 
The left panel shows the phase of the initial state, featuring two same-sign vortices. The right panel presents a zoomed-in view of the central region at the effective time $\tilde{z} = 120$. The red and green lines are the trajectories of the vortices as a function of $\tilde{z}$.

While this technique allows to probe different values of $\tilde{z}$, it remains limited to relatively short-time dynamics, typically $\tilde{z} < 160$. 
In Section~\ref{sec:feedback}, we will explore how to extend this method by incorporating an electronic feedback loop, enabling access to longer evolution times.

\label{sec:effective}

 \subsection{Electronic feedback loop}
\label{sec:feedback}

As discussed in Section~\ref{sec:effective}, using the dimensionless form of the NLSE enables us to interpret light propagation in a nonlinear medium as an effective temporal evolution. 
However, the maximum evolution $\tilde{z} = z / z_{\text{NL}}$ remains limited by the physical length $L$ of the medium and the value of the nonlinear length $z_{\text{NL}}$. 
While, in principle, one could extend $L$ to increase the effective propagation time, a more fundamental constraint arises from absorption: as light travels through the medium, its intensity decreases exponentially, eventually limiting the length $L$.

To overcome these limitations, \citet{fleischer2003observation} proposed a digital feedback loop technique. The concept involves capturing both the intensity and phase of the optical field at the output of the medium, and using a SLM, to recreate this exact same field at the input (using the technique described in Subsection~\ref{sec:SLM}. 
A pair of SLMs can also be used for independent control of amplitude and phase (see for instance \citet{ferreira2024digital}).
By iterating this process, one can artificially extend the effective propagation length, enabling the observation of longer dynamics without being worried by absorption.

\begin{figure}[h]
    \centering
    \includegraphics[width=1\linewidth]{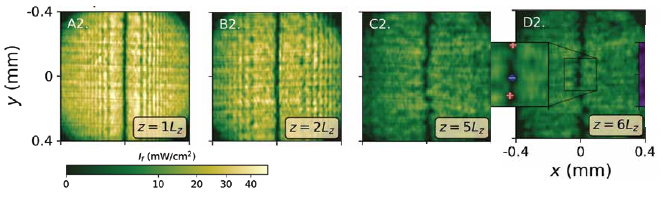}
    \caption{Figure adapted from \cite{ferreira2024digital}. Experimental feedback loop to observe the snake instability in a dark soliton stripe in 2D. From left to right: intensity images after 1,2,5,6 loops. Copyright (2025) by the American Physical Society.
}
    \label{fig:loop_f}
\end{figure}

This technique was recently implemented successfully by \citet{ferreira2024digital} to study the evolution of a planar dark soliton. As shown in Fig.~\ref{fig:loop_f}, the feedback loop was repeated six times, effectively multiplying the observable evolution time by a factor of six compared to the propagation in a single-pass photorefractive crystal.
A similar technique is used in optical fiber with a recirculating loop instead of a SLM \citep{copie2023space}.

However, the electronic feedback approach faces a major challenge due to the accumulation of noise in the measured and reconstructed fields (see Section~\ref{sec:SLM} and Section~\ref{sec:off-axis}).
Since each loop reuses the previous output, any imperfections or measurement noise are reinjected and amplified with each iteration. This accumulation quickly leads to instabilities in the dynamics, currently limiting the method to fewer than ten iterations in practice.

Despite this limitation, the electronic feedback loop represents an interesting tool to explore nonlinear optical dynamics beyond the physical constraints of the medium. 
For example, it could be used to simulate a periodic modulation of the system by changing the parameters of the nonlinearity between each iteration and therefore simulate Floquet-like physics.

\subsection{Modern tools for numerical simulations of the NLSE}

Numerous effects that are observed in the experiments go beyond the perturbative analytical treatment described before. 
This limits the available analytical tools for understanding the experimental results even in the mean field approximation. 
Progress with optimized solver on the CPU \citep{stagg2016numerical} or more modern specialized graphical processing units (GPU) \citep{aladjidi2024nlse} allows for an efficient implementation of numerical schemes.

To numerically solve the NLSE a typical approach is to use split-step spectral methods \citep{javanainen2006symbolic}.
Space and time are discretized, with uniform adimensional spacing $\delta x$, $\delta y$ and $\delta z$,
with the integrated equation reading:
\begin{equation}
    E(z + \delta z) = E(z) e^{\mathrm{i}\delta z (\mathcal{D} + \mathcal{N})E(z)}.
\end{equation}
\begin{itemize}
    \item
$\mathcal{D}$ is the kinetic energy operator, best applied in Fourier space where $\nabla^2$ transforms into a multiplication by $-(k\delta r)^2$ (or $-\sin(k \delta r)^2$, see \cite{Sunaina_2018}).
    \item 
$\mathcal{N}$ is the potential operator. It also contains the nonlinear term in case of the NLSE.
\end{itemize}

The $\mathcal{D}$ and $\mathcal{N}$ operators do not commute, and it is not possible to multiply the two exponentials to get an exact result. Instead an approximate solution can be used \citep{javanainen2006symbolic}:

\begin{equation}
\label{eq:splitstep_linear}
e^{\mathrm{i}\delta z(\mathcal{D}+\mathcal{N})}= e^{\mathrm{i}\frac{\delta z\mathcal{N}}{2}}e^{\mathrm{i}\delta z\mathcal{D}}e^{\mathrm{i}\frac{\delta z\mathcal{N}}{2}} + \mathcal{O}(\delta z^3).
\end{equation}

In the case of the NLSE, $\mathcal{N}$ depends on the field, and the above approximation breaks down. There is multiple choices for when to sample the field, i.e. before or after the first and second exponential multiplication. 
The following sampling choice conserves the convergence properties of \eqref{eq:splitstep_linear} \citep{javanainen2006symbolic}:
\begin{align}
E_1 &= e^{\frac{\mathrm{i} \delta z}{2} g|E_0|^2}E_0 \\
E_2 &= e^{\mathrm{i} \delta z \mathcal{D}}E_1 \\
E_3 &= e^{\frac{\mathrm{i} \delta z}{2} g|E_2|^2}E_2,
\end{align}
where $\mathcal{N}(E) = V + g|E|^2$.
This gives one step of the algorithm, we then loop over the $\frac{L}{\delta z}$ times until the end of the nonlinear medium cell to get the output field.

This method converges with $\mathcal{O}(\delta z^3)$ temporal accuracy and $\mathcal{O}(\delta r^2)$ spatial accuracy. 
As explain in \cite{splitstep_weideman}, the stability condition is $\delta z \leq \frac{\delta r^2}{\pi}$. The strategy to balance performance and precision is to choose $\delta r$ such that $\delta r < \xi$, and adjust $\delta z$ to satisfy the stability condition.

In the case of the simple Euler scheme presented above, all of the steps are \textit{diagonal} in real and Fourier space. 
This means that all calculations can be carried out element-wise over the real and Fourier space grids.
This makes it particularly suited for GPU acceleration as this type of hardware is heavily optimized for such matrix calculations, as well as packing a lot of computing power at an accessible price. 
Compared to CPU-based implementations, GPU-based solvers offer important speedups depending on problem size and hardware \citep{aladjidi2024nlse}. 
This makes it possible to model large optical systems with high spatial and temporal resolution.
Open source packages are available to implement this approach without any pre-existing knowledge of GPU programming \citep{aladjidi2024nlse}.

\section{Recent experimental advances}
\label{sec:expt}
During the last decade, experiments with fluids of light have moved towards more and more connections with ultracold atomic quantum gases. 
Using similar vocabulary and formalism, this has opened a new era for the field with three objectives in mind: i) validate the paraxial fluid of light systems as a potential platform to study quantum gases; ii) improve our understanding of quantum gases with new observables or better resolution using fluids of light with respect to ultracold atomic systems; iii) invent novel optics experiments directly inspired by ultracold quantum gases.
These three goals are somewhat chronological. 
The fluid of light community made a clear effort initially to settle this system as a solid and rigorous equivalent of ultracold quantum gases at the mean-field level, observing superfluidity, Bogoliubov dispersion, vortex nucleation, etc.
In a second step, extending the knowledge of quantum gases thanks to the specific tools available for fluids of light has shown great results in turbulence, Bose-Bose mixtures and out-of-equilibrium dynamics.
The third step is not clearly established yet. While several original optics experiments have been conducted, no truly novel nonlinear or quantum optics phenomena has been discovered yet. We will propose a few ideas along this direction at the end of this review.

\subsection{Hydrodynamics and nonlinear dynamics}

Hydrodynamics is one of the main testbed for fluids of light.
Compared to atomic BECs, fluids of light gives access to direct imaging of the beam intensity and off-axis digital holography allows to reconstruct the phase of the light field and thereafter the velocity field. 
Historically, these studies were targeting the observation of bright and dark solitons to demonstrate the nonlinear nature of these systems, where the balance between diffraction and nonlinearity allows for the formation of self-confined waves \citep{fleischer2003observation}.
Subsequently, study of jetlike tunneling \citep{PhysRevA.88.043833}, optical analogue of the Laval nozzle \citep{Fouxon_2010} and Rayleigh-Taylor instabilities \citep{jia2012rayleigh} have reinforced the hydrodynamics interpretation.
In the past years, several other topics have emerged. 
Several studies have focused on the formation of dispersive shock waves, analogous to those observed in other fluid systems, including atomic BECs \citep{wan2007dispersive,bienaime2021quantitative,abuzarli2021blast,azam2021dissipation,xu2015coherent,copie2020physics,walczak2015optical, dieli2024observation}. 
These shock waves manifest when perturbations propagate at different speeds due to dispersion and nonlinearity, leading to oscillatory structures. 

Notably, the dynamics of shock waves in fluids of light have been quantitatively analyzed using Whitham modulation theory, showing good agreement with experimental observations \citep{bienaime2021quantitative,azam2021dissipation}. 
Blast waves, characterized by increased pressure and flow followed by a negative pressure wind, have also been experimentally generated and studied in fluids of light, providing insights into compressible fluid dynamics in different spatial dimensions \citep{abuzarli2021blast}. 
Similarly, dam break experiments have been conducted in fluids of light, where the interaction of dispersive shock waves propagating in orthogonal directions gives rise to a 2D ensemble of solitons \citep{dieli2024observation}. 
These experiments involved measuring analogous physical properties like hydrostatic pressure, density, and particle velocity, and showed good agreement with hydrodynamical models.\\

Very tightly connected is the study of topological defect formation, particularly vortices. Experiments have demonstrated the spontaneous generation of vortices following perturbations \citep{aladjidi2023full}, such as the snake instability of solitons arising from an elliptical dark stripe \citep{ferreira2024digital} or perturbation by a  moving obstacle \citep{azam2022vortex}.
The observation of vortex annihilation, often accompanied by radiative losses, offers crucial insights into non-equilibrium dynamics. Recent research has focused on topological constraints governing vortex formation, highlighting the role of initial density and velocity distributions in determining nucleation pathways \citep{congy2024}. 

Specifically, vortex-antivortex pairs emerge from phase extrema (node collisions), while phase saddles contribute to vortex annihilation. The node collision mechanism is particularly effective in compressible, non-stationary light-fluid flows.
A recent experiment has evidenced Jones-Roberts solitons (JRS). This provides a direct link between soliton dynamics and vortex formation, since this structure experiences a transition from a vortex-antivortex dipole to a rarefaction pulse \citep{baker2025observation}. 
This illustrates a coherent mechanism for vortex annihilation, where incompressible vortex flow is converted into compressible wave excitations.\\

As the number of vortices in the system increases, the study naturally extends to turbulence in quantum fluids of light \citep{barenghi2014introduction,panico2023onset,alperin2019quantum}. 
Turbulence, a highly non-equilibrium regime characterized by energy transfer across multiple length scales, remains an open challenge in fluid physics. 
Fluids of light provide a unique platform for investigating this phenomenon due to the precise control over initial conditions and direct access to phase information \citep{eloy2021experimental}.

Recent studies have examined the creation and annihilation of vortex-antivortex pairs, as well as the role of hydrodynamic instabilities such as the Kelvin-Helmholtz instability in triggering turbulence \citep{ferreira2024exploring}. 
In counter-propagating geometries, where two fluids of light collide, kinetic energy spectrum analysis reveals characteristic power laws, including the inverse energy cascade with a  $k^{-5/3}$  scaling \citep{baker2023turbulent}. 
Advances in numerical techniques now enable higher-resolution energy spectra, allowing turbulence dynamics to be explored across a broader range of length scales \citep{bradley2022spectral}.

Furthermore, the study of quantum-like turbulence has been extended to two-component paraxial fluids of light, where orthogonally polarized components interact. This setup enables the observation of both direct and inverse turbulent cascades, with tunable properties controlled by the relative angle of incidence between the components \citep{silva2021exploring}. These investigations pave the way for a deeper understanding of turbulence in quantum fluids and its underlying mechanisms.

\subsection{Superfluidity}

\begin{figure}[h]
    \centering
    \includegraphics[width=0.96\linewidth]{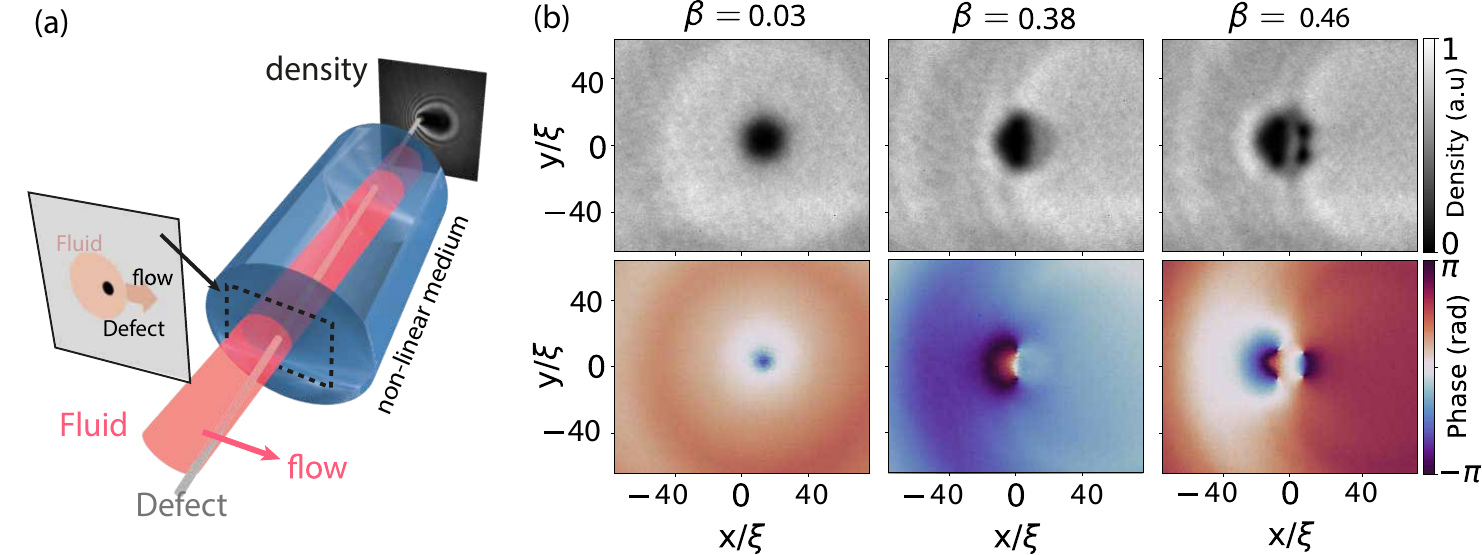}
    \caption{(a) Sketch of an experiment with a moving defect (or a moving fluid). The defect is imprinted using a local modification of the refractive index (constant along $z$). The relative velocity is given by the angle between the fluid beam and the defect.
    (b) Experimental observation of a moving impurity of finite size and finite mass in a fluid of light. Experimental data are taken in rubidium vapor. From left to right, the velocity of the moving impurity (normalized by the speed of sound) is increased from $\beta=0.03$ to  $\beta=0.46$. Top: intensity. Bottom: phase. More details are in \cite{aladjidi2023full}.
    }
    \label{fig:defect_expt}
\end{figure}

Hydrodynamics in quantum fluids directly leads to the study of superfluidity.
Superfluidity is the frictionless flow of a fluid below a critical velocity and it is one of the most striking manifestations of macroscopic quantum physics.
Fluids of light have provided a novel experimental platform for studying various aspects of this phenomenon.

Pioneering experiments have revealed indirect signatures of superfluidity, notably through the measurement of the Bogoliubov dispersion relation for elementary excitations \citep{fontaine2018observation,vocke2015experimental,piekarski2021measurement}.
This dispersion exhibits a linear (phononic) regime at low wave vectors \citep{fontaine2020interferences} and a quadratic (particule-like) regime at high wave vectors, mirroring the behavior of atomic superfluids or liquid helium, as described in Section~\ref{sec:bogo}.
Observations of the Bogoliubov dispersion have been reported in both local \citep{fontaine2018observation} and non-local light fluids \citep{vocke2015experimental}.

Beyond these indirect signatures, direct evidence of superfluidity in light has also been obtained.
Experiments have demonstrated dissipation-less flow around obstacles below a critical velocity, a hallmark of superfluidity, as the fluid overcomes imperfections without energy loss \citep{ferreira2018superfluidity,ferreira2024digital, amo2009superfluidity}.
The suppression of the drag force on a slowly moving defect further reinforces this interpretation \citep{michel2018superfluid}.
At supersonic velocities, the nucleation of quantized vortices has been extensively studied, with investigations characterizing their core size, circulation, and role in the breakdown of superfluidity \citep{aladjidi2023full}.

Moreover, the study of superfluidity in rotating system \citep{silva2017persistent} could draw analogy with astrophysical phenomena, such as rotating black holes \citep{vocke2018rotating,marino2016emergent}. 
For example, the observation of amplified scattered waves from a rotating fluid of light has provided an experimental demonstration of Penrose super-radiance in an analogue system \citep{braidotti2022measurement}. 
Typical experiments consist in creating a rotating superfluid by imposing a rotating phase to the fluid of light and observing the scattering of a plane wave on both side of the center of rotation.
Under specific conditions, well described by analogue gravity model of ergoregion instability \citep{giacomelli2020ergoregion}, the plane wave is amplified extracting energy from the rotating superfluid.

\subsection{Out-of-equilibrium dynamics and quenches}
The study of out-of-equilibrium dynamics in fluids of light, particularly their response to sudden parameter changes (quenches) is especially relevant in this system since there is an interaction quench directly embedded in the geometry of the platform.
Indeed, when light enters the nonlinear medium at the initial time (input face), it experiences a drastic change of the photon-photon interaction from virtually zero (in air) to a finite value in the medium. A second quench occurs when the light exits the medium (output face) and the interaction is again set to zero.
These quenches have been used to study the non-equilibrium pre-condensation  of classical waves \citep{vsantic2018nonequilibrium}, the emergence of coherence \citep{abuzarli2022nonequilibrium,fusaro2017emergence}, and the excitation of elementary modes such as Bogoliubov phonons \citep{steinhauer2022analogue,fontaine2020interferences}.

Condensation and pre-condensation of classical waves has long been an important topic for fluids of light \citep{vsantic2018nonequilibrium,connaughton2005condensation,sun2012observation,aschieri2011condensation}.
Recently, a different approach has been studied with a focus on the dynamical evolution of non-equilibrium states, particularly the phenomenon of pre-thermalization following an interaction quench and the emergence of a quasi-equilibrium state. 

In this framework, the emergence of pre-thermal states has been observed \citep{abuzarli2022nonequilibrium}. These states retain partial memory of their initial conditions while resembling their thermal counterparts. In this study, the emergence of long-range algebraic correlations spreading within a light cone—a signature of a quasi-steady state resembling a 2D thermal superfluid—was proposed in \cite{bardon2020classical} and observed through direct measurements of the first-order correlation function in \cite{abuzarli2022nonequilibrium}.
This approach enables the study of transitions between different correlation regimes within pre-thermal states. 
A controlled increase in fluid fluctuations revealed a crossover from algebraic to short-range (exponential) correlations, drawing a direct analogy to the Kosterlitz-Thouless transition in thermal equilibrium \citep{hadzibabic2006berezinskii,situ2020dynamics}. 
These findings suggest the presence of non-equilibrium precursors to thermodynamic phase transitions, providing an interesting direction for applying fluids of light to study quantum phase transitions and many-body physics far from equilibrium.

The intrinsic quenches embedded in the experimental platform leads to the excitation of elementary modes and complex relaxation dynamics.
One direct consequence of these quenches is the spontaneous generation of Bogoliubov phonons, which propagate through the fluid and imprint a characteristic oscillatory structure in both real and momentum space. 
The static structure factor $S(k)$ has been measured in fluids of light as described in Section~\ref{sec:ssf} and reveals the formation of ring-like interference patterns, indicating a well-defined phase relation between excitations. 
These oscillations provide a direct probe of the system’s coherence and response to sudden interaction changes.
Correlated quasiparticle pairs emitted due to the quench can be tracked using density correlation measurements. The emergence of expanding spatial modulations in the fluid’s density fluctuations reflects the system’s attempt to relax toward a new steady state. By analyzing the Fourier transform of the correlation function, experiments have revealed oscillatory structures that encode information about pair production and collective excitations following the quench \citep{steinhauer2022analogue}.
So far, these measurements have been limited to classical noise above the standard quantum limit. 
In these experiments, quantum noise is expected to be dominant, as the modes at $\mathbf{k}_{\perp} \neq 0$ are assumed to be in a vacuum state or similarly that the spatial intensity noise of the laser is at the shot noise limit. In principle, quantum depletion \cite{chang2016momentum}, or a non-equilibrium response with an envelope linked to the quantum depletion should be visible.
However, low-frequency technical noise in the temporal domain has so far prevented the observation of quantum correlations.

\subsection{Photonic lattices and analogies with condensed matter systems}
\begin{figure}[]
    \centering
    \includegraphics[width=1\linewidth]{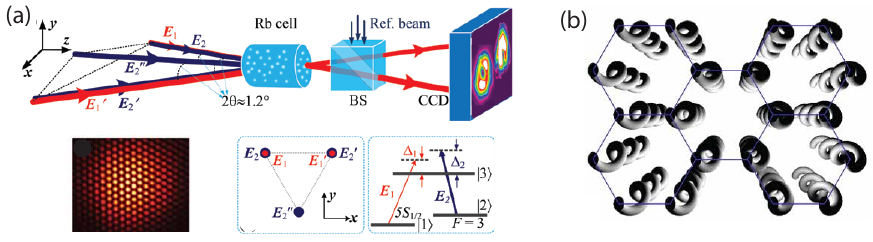}
    \caption{(a) Figure adapted from \cite{zhang2019particlelike}. Creation of a lattice mimicking graphene using a modulation of the refractive index in a rubidium vapor. Below the simplified setup, the typical lattice is shown with the position of the beams in the Fourier plane and the required atomic structure.
    (b) Figure adapted from \cite{rechtsman2013photonic}. Schematic diagram of an helical waveguides in a photorefractive crystal.
    \label{fig:graphene}}
\end{figure}

Fluids of light have also been explored to study analogies with condensed matter systems. 
Photonic lattices provide a powerful tool for exploring wave dynamics in structured optical media as illustrated in Fig.~\ref{fig:graphene}.

In this approach, a periodic potential is introduced through a transverse modulation of the refractive index, creating an effective lattice for photons. The evolution of light within such lattices can be either static or dynamically modulated along the propagation axis $z$, effectively simulating time-dependent Hamiltonians. 
These systems allow for the study of fundamental effects such as Bloch oscillations, topological transport, and the emergence of photonic Landau levels.

In the linear regime, photonic lattices enable the study of Bloch oscillations and Zener tunneling, which are optical analogues of electronic transport phenomena in solids \citep{zhang2017optical}.
These effects have been demonstrated in waveguide arrays and photonic crystals, where light undergoes periodic oscillations under an external force or tunnels across band gaps between Bloch states. 
Such dynamics have been used to engineer optical beam splitters and robust interconnects for photonic circuits \citep{zhang2017optical}.
Additionally, strain-induced modifications to the lattice can create pseudo-magnetic fields, leading to the formation of photonic Landau levels—quantized modes that mimic the behavior of electrons in a magnetic field \citep{barsukova2024direct}. 

The study of photonic lattices has also revealed rich topological properties. When time-reversal symmetry is broken, such as through waveguide helicity (see Fig.~\ref{fig:graphene} b) or gyromagnetic materials, topologically protected edge states emerge, allowing light to propagate without back-scattering \citep{rechtsman2013photonic,mukherjee2020observation,barsukova2024direct}. 
These photonic analogues of the quantum Hall effect have been realized in honeycomb lattices of waveguides, where light propagates along the lattice edges in a robust manner, immune to disorder and defects.
Furthermore, the Klein tunneling effect—where photons traverse a potential barrier without reflection—has been observed in photonic graphene (see Fig.~\ref{fig:graphene} a), demonstrating an angular dependence in agreement with Dirac-point physics \citep{zhang2022angular}.

Beyond linear effects, photonic lattices serve as an ideal setting for studying nonlinear wave dynamics and topological defects \citep{zhang2019particlelike,fleischer2003observation}.
In highly nonlinear regimes, optical vortices form and interact similarly to quantum vortices in superfluid systems.
These optical defects exhibit well-defined dynamical laws, including Magnus forces and mutual vortex interactions, suggesting a deeper connection between photonic and quantum fluids.
The introduction of controlled nonlinearity further enables the observation of solitons in topological bands, where light remains self-localized while moving through the lattice.

Another recent development involves the study of non-Hermitian photonic lattices, where gain and loss are spatially engineered to break parity-time (PT) symmetry \citep{zhang2018non}.
These systems exhibit exotic phase transitions and enable new functionalities such as unidirectional invisibility and exceptional-point-based sensors.
Experiments with PT-symmetric photonic lattices have demonstrated abrupt phase transitions where the eigenvalues of the system shift from entirely real to complex, leading to a breakdown of conventional wave dynamics.
A detailed review of these effects can be found in \cite{zhang2018non}.

Overall, photonic lattices provide a versatile testbed for studying fundamental physics, from solid-state-inspired transport phenomena to nonlinear wave interactions and topological effects thanks to the ability to engineer periodic and quasi-periodic potentials, combined with the intrinsic tunability of optical systems.

\section{Future directions and perspectives}
\label{sec:future}
In this review, we have presented a historical overview of paraxial fluids of light and described recent advances in the field. In this final section, we propose three research directions that we envision as particularly promising for the coming years.

First, exploring the degrees of freedom available in Bose-Bose mixtures could significantly broaden the scope of the field. Multiple experimental configurations could allow for this exploration. In warm atomic vapors, three promising approaches include using the polarization degree of freedom, employing two lasers tuned to the D1 and D2 lines, or addressing the F=1 and F=2 transitions of rubidium 87. Among these, only the polarization degree of freedom has begun to be experimentally explored as we will describe.
Second, identifying nonlinear media with enhanced control and larger nonlinear indices ($n_2$) represents another important direction. We propose utilizing cold atomic clouds trapped in magneto-optical traps and discuss the advantages offered by such systems.
Finally, we consider the potential of using fluids of light as platforms to investigate many-body physics. While achieving this requires transitioning to different interaction regimes, we outline several promising directions for future studies.

\subsection{Two-component mixture}
The recent realization of two-component fluids of light by \cite{piekarski2024spin} has extended the field to multi-component interactions, collective excitations \citep{fava2018observation}, and novel instability regimes. 
These experiments have demonstrated the existence of spin and density modes in a miscible fluid of light \citep{piekarski2024spin}.
Many open questions remain regarding the transition to immiscibility \citep{Papp2008}, the dynamics of phase separation, and the formation of composite topological structures such as massive vortices \citep{richaud2020vortices,richaud2021dynamics}. 
This section discusses these potential directions, focusing on the interplay between nonlinear interactions, inter-component coupling, and hydrodynamic instabilities.\\

A two-component quantum fluid of light consists of two interacting optical fields that propagate through a nonlinear medium, behaving analogously to a binary Bose-Einstein condensate \citep{Baroni2024}.
One possible implementation is to use the polarization degree of freedom.
Indeed, when deriving the NLSE in Section~\ref{sec:nlse_derivation}, we considered a linearly polarized field hence neglecting the tensorial nature of the third-order susceptibility.
However, if we consider the more general case of an elliptically polarized beam, still propagating in an isotropic medium, then the atomic polarization is given by \citep{boyd_nl_optics}:
\begin{equation}
    \mathbf{P}=6\varepsilon_0\chi_{1122}(\mathbf{E}\cdot\mathbf{E}^*)\mathbf{E}+3\varepsilon_0\chi_{1221}(\mathbf{E}\cdot \mathbf{E})\mathbf{E}^* .
\end{equation}

We decompose the electric field in the circular polarization basis: $\mathbf{E}=E_+\hat{\mathbf{\sigma}}_+ + E_-\hat{\mathbf{\sigma}}_-$, where $\hat{\mathbf{\sigma}}_+$ (resp.  $\hat{\mathbf{\sigma}}_-$) is the unitary circular left (resp. right) vector. Then the nonlinear  susceptibility can be rewritten as 
\begin{equation}
    \mathbf{P}=P_+\hat{\mathbf{\sigma}}_+ + P_-\hat{\mathbf{\sigma}}_-
\end{equation}
where $P\pm = \varepsilon_0\chi_\pm E_\pm $, with
\begin{align}
    \chi_\pm = A|E_\pm|^2 + (A+B)|E_\mp|^2 ,
\end{align}
with $A=6\chi_{1122}$ and $B=6\chi_{1221}$.
Injecting this expression into the Helmholtz propagation equation, one finds that
in the paraxial approximation, the evolution of the envelopes of the two circular polarization components, $\mathcal{E}_+$ and $\mathcal{E}_-$, is described by two coupled nonlinear Schr\"odinger equations (CNLSE):

\begin{equation}
    \mathrm{i}\partial_z \mathcal{E}_\pm = \left( -\frac{\nabla_\perp^2}{2k_0} + g |\mathcal{E}_\pm|^2 + g_{12} |\mathcal{E}_\mp|^2 \right) \mathcal{E}_\pm,
    \label{eq:nlse}
\end{equation}
where $g=-k_0A$ is the intra-component interaction coefficient, $g_{12}=-k_0(A+B)$ is the inter-component interaction coefficient. 
We have neglected absorption out of  simplicity.
Similarly as for the single-component case, these equations are formally equivalent to the coupled Gross-Pitaevskii equations governing binary Bose mixtures, with the propagation coordinate $z$ playing the role of time \citep{pethick2008bose}.

The sign and relative weight of $g$ and $g_{12}$ determine the accessible interaction regime  of the system \citep{pitaevskij_bose-einstein_2016}.
$g>0$ (resp. $g<0$) implies repulsive (resp. attractive) intra-component interaction, and similarly, $g_{12}>0$ (resp. $g_{12}<0$) implies repulsive (resp. attractive) inter-component interaction.
The fluid is stable against collapse as long as the total resulting interaction is repulsive.
Then, if $g > |g_{12}| > 0$, the two components are miscible and they remain homogeneously mixed.
If $ g_{12} > |g| > 0$, the system  is non-miscible and undergoes phase separation, leading to spontaneous domain formation.

\subsubsection{Spin and density modes in the miscible regime}

\begin{figure}
    \centering
    \includegraphics[width=1\linewidth]{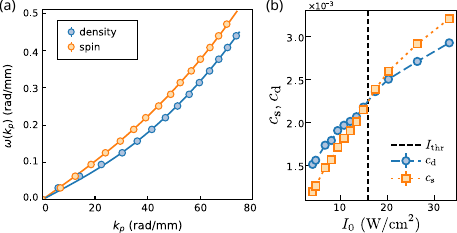}
    \caption{(a) Dispersion relation for density and spin modes in a two-component fluid of light. Data are taken in a rubidium vapor using two circular polarizations. 
    (b) Spin (blue circles) and density (orange squares) speeds of sound measured in a saturable nonlinear medium, as a function of the laser intensity. At low intensity, $c_s < c_d$, while in the saturated regime $c_s > c_d$.
    More details are to be found in~\cite{piekarski2024spin}.
    }
    \label{fig:2fluids}
\end{figure}

In the miscible regime, it has been predicted that two-component fluids of light support two distinct collective modes \citep{martone2021spin,martone2023time}, the density and the spin mode, just like two-component BECs \citep{pethick2008bose, cominotti_2fluids_dispersion}.
The density mode corresponds to excitations of the sum of the densities of the two components, and the spin mode to excitations of the difference of the densities of the two components.
They correspond to two distinct branches of dispersion, both having a Bogolioubov shape but with different speeds of sound.
Recent experiments have demonstrated a miscible two-component fluid of light  using a red-detuned laser from the D1 line of rubidium 87, and have observed the two branches of dispersions (see Fig.~\ref{fig:2fluids}), with the density speed of sound
\begin{equation}
    c_d^2 = \frac{g + g_{12}}{2k_0} |\mathcal{E}|^2,
    \label{eq:density_mode}
\end{equation}
and the spin speed of sound
\begin{equation}
    c_s^2 = \frac{g - g_{12}}{2k_0} |\mathcal{E}|^2,
    \label{eq:spin_mode}
\end{equation}
where $|\mathcal{E}|^2=|\mathcal{E}_+|^2 + |\mathcal{E}_-|^2$.
Interestingly, two-component fluids of light have also provided a novel effect compared to two-component BECs: an inversion of the relative value of the two sound velocities due to interaction terms beyond two-body contact interactions. 
As described in Section~\ref{beyondkerr}, higher order terms in the nonlinear atomic response will lead to a saturation and a decrease of the effective photon-photon interactions. 
In two-component fluids this leads to an unconventional inversion of the two speeds of sound, which mimics attractive interactions between the two species.
This novel degree of freedom should open interesting directions in the study of  fluid-mixture hydrodynamics with light.

\subsubsection{Non-miscible regime and coarsening dynamics}

\begin{figure}
    \centering
    \includegraphics[width=0.95\linewidth]{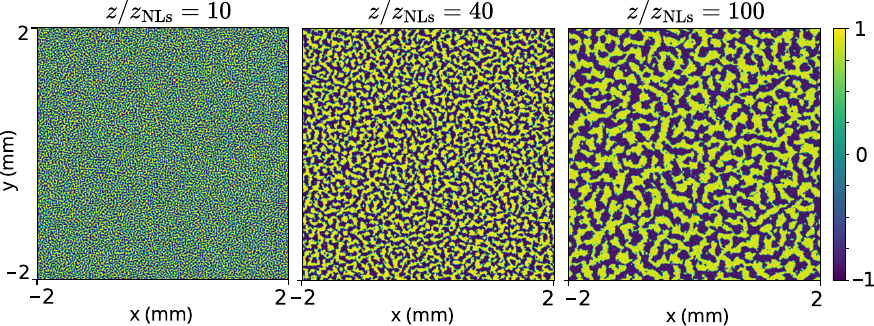}
    \caption{Coarsening simulation. Evolution of the magnetization $m=(|E_+|^2-|E_-|^2)/(|E_+|^2+|E_-|^2)$, after injection of a weak Gaussian speckle on a linearly polarized background, with $g_{12}/g=1.8$.
    Each image is taken at different transverse plane during propagation from $z/z_\text{NL,s} = 10,40$ and $100$, where $z_\text{NL,s}=1/\sqrt{k_0(g-g_{12})}$ is the spin nonlinear length.}
    \label{fig:coarsening}
\end{figure}
When $g_{12} > g$, the spin mode becomes imaginary, leading to an instability that drives the separation of the two components. 
This transition gives rise to a coarsening dynamics, a topic that remains largely unexplored in photonic fluids.
In Bose mixtures, coarsening dynamics refers to the evolution of an initially unstable mixture into separated domains, governed by defect formation and hydrodynamic interactions \citep{goo2022universal}. 
In the case of fluids of light, similar mechanisms could be studied by tuning $g$ and $g_{12}$ into the regime $ g_{12} > g > 0$. This is actually possible when working on the red side of the D2 line of rubidium.

Recent studies in atomic Bose-Einstein condensates suggest that the early coarsening stage plays a critical role in determining the final defect density. The Kibble-Zurek mechanism predicts that defect formation follows a power-law scaling with the quench rate, but recent experiments have shown that early coarsening can introduce corrections to this scaling, leading to defect density saturation \citep{zeng2023universal}.
This suggests that fluids of light could be a very relevant platform to study these effects thanks to the quench dynamics, the ability to detect vortices with high resolution and the control of the initial kinetic energy spectrum.

Interesting effects are also predicted for point-like vortices in one component that will become massive quasiparticules due to the second component.
Theoretical models suggest that the dynamics of these vortices can be described using a Lagrangian formulation similar to that of charged particles in a magnetic field, where the vortex cores behave as massive objects undergoing precession \citep{richaud2020vortices,richaud2021dynamics}.
This mass induces novel inertial effects, such as radial oscillations and modifications to the equilibrium vortex spacing, which have been confirmed in numerical simulations of coupled Gross-Pitaevskii equations and fluids of light could allow to observe this experimentally.

\subsection{Nonlinear media with cold atoms}
Using cold atoms as a nonlinear medium represents an exciting direction for experiments involving paraxial fluids of light. 
Even more than traditional nonlinear optical media described in this review, laser-cooled atomic ensembles could provide a deeper control over optical nonlinearities and the possibility to exploit atomic coherence.

The major advantage of using cold atoms lies in their narrow resonance lines, which drastically reduces Doppler broadening. 
As the optical nonlinearity in atomic media typically scales as $N/\Delta^3$, this narrow spectral width allows experiments to approach resonance closely without encountering the significant absorption losses inherent to hot atomic vapors.
From a typical detuning of $\sim 5$~GHz in hot atomic vapor experiments, it becomes relevant to set the laser frequency detuned by the natural linewidth, typically $\sim6$~MHz for alkali atoms.
It is then possible to win three orders of magnitude on $\Delta$ and therefore nine orders of magnitude on $1/\Delta^3$.
This is however compensated by the atomic density $N$. 
In warm vapors, we have seen that the density is typically on the order of $10^{13}$ at.cm$^{-3}$, while in dense cold atomic cloud it is on the order of $10^{11}$ at.cm$^{-3}$, which is only 2 orders of magnitude smaller \citep{camara2014scaling}.
Overall, it appears advantageous to move to that regime in order to reach similar $\Delta n$ with much weaker intensity. 

Moreover, it is possible to refine the excitation scheme using coherent optical effects such as electromagnetically induced transparency (EIT), coherent population trapping (CPT), or electromagnetically induced absorption (EIA) \citep{lukin2000resonant}. 
These phenomena are capable of strongly increasing the effective photon-photon interaction strength.
Specifically, coherent nonlinear optics can generate interactions whose strength, sign (attractive or repulsive), and spatial range are externally controllable, extending the range of accessible quantum hydrodynamic regimes.
So far, coherent scheme have not been used in the nonlinear regime to create fluids of light. 
Orders of magnitude tell that the maximum nonlinear phase shift will not overpass the one observed in warm vapors, however it will be reached at a much lower number of photons.
This leads to substantially enhanced photon-photon interactions and potentially many-body physics with light.

This idea is not new in the field of nonlinear quantum optics \citep{chang2014quantum,firstenberg2013attractive,firstenberg2016nonlinear,roy2017colloquium}, but it has mainly been tried along $t$ in the 1D+1 configuration.
However, a significant limitation in the temporal $t$ dimension is that the effective photon mass, arising from the group velocity dispersion, tends to be relatively large for typical parameters, severely constraining achievable nonlinear dynamics. 
An interesting extension of nonlinear quantum optics therefore lies in considering dynamics in the transverse $(x,y)$ plane.
In this transverse propagation approach, the effective mass of the photons, determined by diffraction rather than dispersion, is significantly smaller for typical experimental parameters.
Lighter mass means lower kinetic energy and faster dynamics, such that the requirements on the interaction term are drastically lower than in the case of previous experiments.
As a consequence, we believe that we could use this mass imbalance to observe quantum phase transition in 2D along the transverse direction.

In particular, we propose to study the superfluid to Mott insulator transition in 2D \citep{spielman2007mott,kohl2004superfluid}.
Since structured optical potentials are feasible with cold atoms, such as optical lattices formed by standing waves of laser beams, this enables to implement a lattice potential and tunable on-site interactions.
A fluid of light will be injected at the input of the medium in a superfluid (coherent) state, with an average of $N_t$ photons. If the on-site energy becomes larger than the tunnel rate, it becomes favorable for the photons to equilibrate between the sites.
If the number of sites is equal to $N_t$, the predicted final state at the output of the medium is an array of single photons (one per site).
Achieving this photonic analogue of the Mott insulator phase transition would represent a breakthrough, providing a direct route to generate and control single-photon states. 
Even though a perfect single-photon source is likely hard to achieve, this approach is a promising new way to create non-classical (squeezed) light inspired by the quantum gases phenomenology \citep{greiner2002quantum}.

\subsection{Quantum effects and beyond mean-field}
\label{sec:quantum}
As we just saw, going beyond the mean-field description of fluids of light represents an exciting frontier in the study of optical quantum fluids. 
While current experiments largely operate within a classical, mean-field regime described accurately by the Gross-Pitaevskii equation, introducing quantum fluctuations and correlations into these systems opens important perspectives both for fundamental physics and applications in quantum technologies.\\

A natural next step in fluids of light experiments, since it relies on well established quantum optics techniques, involves exploring genuinely quantum phenomena such as entanglement and photon number squeezing. 

One experiment has tried to go into this direction \citep{steinhauer2022analogue}, but currently, the major difficulty is to find a way to avoid the low frequency noise coming from the exciting laser source \citep{marino2012extracting,corzo2013rotation}.
Indeed, for paraxial fluids of light in the transverse ($x,y$) plane, cameras are used to record the integrated intensity over time $t$.
If the camera exposure time is $T_{\mathrm{cam}}$, all laser noises on timescales longer than $T_{\mathrm{cam}}$ will add fluctuations to our data when recording a series of images.
When computing the noise variance, this low frequency noise will artificially increase the variance above the standard quantum limit and hide purely quantum effects \citep{clark2012imaging}.
We propose two strategies to circumvent this current limitation. 

The laser beam propagating in the nonlinear medium could be seen as the macroscopic occupation of the ground state similar to an atomic BEC.
This laser is at $k_x=0,k_y=0$. 
Taking into account time $t$ as a third spatial dimension, we could define the laser frequency as $k_t=0$.
The first idea is to filter spatially this ground state to only look at fluctuations in $k_i\neq0$ modes, where $i=x,y \text{ or } t$.
By putting a spatial filter in the Fourier plane, it is possible to remove the $k_x=0,k_y=0$ components. 
After the filter only remains the fluctuations that could be detected using standard homodyne detection \citep{vogl2014advanced,glorieux2012generation,agha2011time}.

The second approach follows a similar idea but might be more practical. 
It requires to introduce a local oscillator to select the fluctuations that we want to detect.
We propose a modified version of the off-axis interferometry.
As explained earlier, off-axis interferometry consists in creating an interference pattern with a tilted reference beam with respect to the signal beam and filter the detected intensity after numerical Fourier transform to reconstruct both amplitude and phase of the signal beam. 
However, this technique still integrates all the low frequency noise  on the camera as explained above.
Nevertheless, if we set the local oscillator to a  different laser frequency, only fluctuations at this specific local oscillator frequency will create a non-zero signal on the camera. 
Following then the same procedure as off-axis interferometry, this version of non-resonant off-axis interferometry will allow us to measure only the noise variance around the local oscillator frequency getting rid of the low frequency noise. This technique could be compared to a nonlinear implementation of similar ideas to gain temporal resolution in phase measurement \citep{tikan2018single}.\\

As we have seen, several novel experimental techniques are required to bring fluids of light to the truly quantum optics regime. 
Once this is achieved, it will become possible to study quantum depletion in a fluid of light with potentially quantum correlations within depleted pairs \citep{chang2016momentum,lopes2017quantum}; the quantum response to an interaction quench \citep{larre2015propagation}; the non-classical statistics across phase transitions \citep{larre2018postquench} and analogue of spontaneous Hawking radiation \citep{jacquet2022analogue}.
Such quantum effects could be realized by carefully exploiting nonlinearities enhanced by coherent atomic effects, as described previously. 
These quantum states would offer a testbed for examining entanglement generation, quantum squeezing, and non-local correlations in propagating optical fields.

\section{Conclusion}
Paraxial fluids of light offer a compelling and highly controllable platform for exploring quantum hydrodynamic phenomena.
Through the analogy between nonlinear optical propagation and the Gross-Pitaevskii equation governing ultracold atomic quantum gases, optical systems have enabled detailed experimental and theoretical investigations of hydrodynamics, superfluidity, and non-equilibrium dynamics. 
We have presented recent experiments and highlighted the potential of fluids of light for probing phenomena such as vortex dynamics, turbulence, quantum correlations, and dimensional crossover effects.
We have also described the significant challenges that remain, particularly in achieving quantum regimes and overcoming practical limitations such as intrinsic classical noise sources.
To conclude this review, we have presented what we believe to be future directions of the field including the exploration of two-component mixtures, coherent nonlinear media with laser-cooled atomic systems, and studies of quantum effects beyond mean-field approximations.

\section*{Acknowledgements}
The authors thank Devang Naik,  Simon Lepleux, Nicolas Cherroret, Romain Dubessy and Claire Michel for useful discussions.
The authors want to specifically acknowledge Hélène Perrin for her several rounds of review of the manuscript, and for her detailed and precise comments. Her contributions have substantially enhanced the clarity and quality of this work.
Quentin Glorieux thanks Institut Universitaire de France.
This work is supported by the ERC Consolidator Grant MistiQ-Light and the ANR FOLIO and QuantumSOPHA.

\section*{Contributions of the authors}
QG wrote Sections 1, 5, 7 and 8 and coordinated the review.
CP wrote Section 4 and 8.1.
TA, QS and CP wrote Section 5.1.
QS, CP and QG wrote Section 2 and 3.
MBR edited the figures.
All authors contributed to Section 6 and corrected the manuscript.
\bibliography{biblio}

\end{document}